\documentclass[12pt]{iopart}
\makeatletter
\def\set@curr@file#1{%
  \begingroup
    \escapechar\m@ne
    \xdef\@curr@file{\expandafter\string\csname #1\endcsname}%
  \endgroup
}
\def\quote@name#1{"\quote@@name#1\@gobble""}
\def\quote@@name#1"{#1\quote@@name}
\def\unquote@name#1{\quote@@name#1\@gobble"}
\makeatother

\usepackage{graphicx}
\usepackage{float}
\usepackage{wrapfig}
\expandafter\let\csname equation*\endcsname\relax
\expandafter\let\csname endequation*\endcsname\relax
\usepackage{color,soul}
\usepackage{amsmath}
\usepackage{listings}
\usepackage{amssymb}
\usepackage{relsize}
\usepackage{tikz}
\usepackage{siunitx}
\usepackage[linktocpage=true]{hyperref}
\hypersetup{colorlinks=true,citecolor=red,urlcolor=blue}
\usepackage[retainorgcmds]{IEEEtrantools}
\usepackage{pgfplots}
\usepackage{tabulary}
\usepackage{xfrac}
\usepackage{ctable}
\usepackage{xr-hyper}
\usepackage{gensymb}
\usetikzlibrary{patterns}
\pgfplotsset{width=10cm, compat=1.10}
\usepgfplotslibrary{fillbetween}
\pgfmathdeclarefunction{surf1}{0}{\pgfmathparse{0.2*sin(x)}}
\pgfmathdeclarefunction{surf2}{0}{\pgfmathparse{-4+0.2*sin(x)}}
\newcommand{\ket}[1]{| #1 \rangle}
\renewcommand*{\contentsname}{ELGAR}

\begin{document}

\title[ELGAR]{ELGAR - a European Laboratory for Gravitation and Atom-interferometric Research}

\author{B.~Canuel$^1$, S. Abend$^2$, P. Amaro-Seoane$^{3,4,5,6,7}$, F~Badaracco$^{8,9}$, Q.~Beaufils$^{10}$, A.~Bertoldi$^1$, K. Bongs$^{11}$, P. Bouyer$^1$, C.~Braxmaier$^{12,13}$, W. Chaibi$^{14}$, N.~Christensen$^{14}$, F.~Fitzek$^{2,15}$, G.~Flouris$^{16}$, N.~Gaaloul$^2$, S.~Gaffet$^{17}$, C.~L.~Garrido Alzar$^{10}$,  R.~Geiger$^{10}$, S.~Guellati-Khelifa$^{18}$, K.~Hammerer$^{15}$, J.~Harms$^{8,9}$, J.~Hinderer$^{19}$, J.~Junca$^1$, S.~Katsanevas$^{20}$, C.~Klempt$^2$, C.~Kozanitis$^{16}$, M.~Krutzik$^{21}$, A.~Landragin$^{10}$, I.~L{\`a}zaro~Roche$^{17}$, B.~Leykauf$^{21}$, Y-H.~Lien$^{11}$, S.~Loriani$^2$, S.~Merlet$^{10}$, M.~Merzougui$^{14}$, M.~Nofrarias$^{3,4}$, P.~Papadakos$^{16,22}$, F.~Pereira$^{10}$, A.~Peters$^{21}$, D.~Plexousakis$^{16,22}$, M.~Prevedelli$^{23}$, E.~Rasel$^2$, Y.~Rogister$^{19}$, S.~Rosat$^{19}$, A.~Roura$^{24}$, D.~O.~Sabulsky$^1$, V.~Schkolnik$^{21}$, D.~Schlippert$^2$, C.~Schubert$^2$, L.~Sidorenkov$^{10}$, J.-N.~Siem{\ss}$^{2,15}$,  C.~F.~Sopuerta$^{3,4}$, F.~Sorrentino$^{25}$, C.~Struckmann$^2$, G.M.~Tino$^{26}$, G.~Tsagkatakis$^{16,22}$, A.~Vicer{\'e}$^{27}$, W.~von~Klitzing$^{28}$, L.~Woerner$^{12,13}$, X.~Zou$^1$}

\address{$^1$ LP2N, Laboratoire Photonique, Num{\'e}rique et Nanosciences, Universit{\'e} Bordeaux--IOGS--CNRS:UMR 5298, rue F. Mitterrand, F--33400 Talence, France}
\address{$^2$ Leibniz Universit{\"a}t Hannover, Institut f{\"u}r Quantenoptik, Welfengarten 1, D-30167 Hannover, Germany}
\address{$^3$ Institute of Space Sciences (ICE, CSIC), Campus UAB, Carrer de Can Magrans s/n, 08193 Cerdanyola del Vall\`es (Barcelona), Spain}
\address{$^4$ Institute of Space Studies of Catalonia (IEEC), Carrer del Gran Capit\`a, 2-4, Edifici Nexus, despatx 201, 08034 Barcelona, Spain}
\address{$^5$ Kavli Institute for Astronomy and Astrophysics, Beijing 100871, China}
\address{$^6$ Institute of Applied Mathematics, Academy of Mathematics and Systems Science, CAS, Beijing 100190, China}
\address{$^7$ Zentrum f{\"u}r Astronomie und Astrophysik, TU Berlin, Hardenbergstra{\ss}e 36, 10623 Berlin, Germany}
\address{$^8$ Gran Sasso Science Institute (GSSI), I-67100 L'Aquila, Italy}
\address{$^9$ INFN, Laboratori Nazionali del Gran Sasso, I-67100 Assergi, Italy}
\address{$^{10}$ LNE--SYRTE, Observatoire de Paris, Universit{\'e} PSL, CNRS, Sorbonne Universit{\'e}, 61, avenue de l'Observatoire, F--75014 PARIS, France}
\address{$^{11}$ Midlands Ultracold Atom Research Centre, School of Physics and Astronomy, University of Birmingham, Birmingham, B15 2TT, United Kingdom}
\address{$^{12}$ ZARM, Unversity of Bremen, Am Fallturm 2, 28359 Bremen, Germany}
\address{$^{13}$ DLR, German Aerospace Center, Linzer Strasse 1, 28359 Bremen, Germany}
\address{$^{14}$ ARTEMIS,  Universit{\'e} C{\^o}te d'Azur, Observatoire de la C{\^o}te d'Azur, CNRS, F--06304 Nice, France}
\address{$^{15}$ Institute for Theoretical Physics and Institute for Gravitational Physics (Albert-Einstein-Institute), Leibniz University Hannover, Appelstrasse 2, 30167 Hannover, Germany}
\address{$^{16}$ Institute of Computer Science, Foundation for Research and Technology - Hellas, 70013, Heraklion, Greece}
\address{$^{17}$ LSBB, Laboratoire Souterrain Bas Bruit,  CNRS, Avignon University, University Nice Sophia-Antipolis - La grande combe, 84400 Rustrel, France}
\address{$^{18}$ Laboratoire Kastler Brossel, Sorbonne Universit\'e, CNRS, ENS-PSL Research University, Coll\`ege de France, 4 place Jussieu, 75005 Paris, France}
\address{$^{19}$ Institut de Physique du Globe de Strasbourg, UMR 7516, Universit{\'e} de Strasbourg/EOST, CNRS, 5 rue Descartes, 67084 Strasbourg, France}
\address{$^{20}$ European Gravitational Observatory (EGO), I-56021 Cascina (Pi), Italy}
\address{$^{21}$ Humboldt-Universit{\"a}t zu Berlin, Institute of Physics, Newtonstrasse 15, 12489 Berlin, Germany}
\address{$^{22}$ Computer Science Department, University of Crete, 70013, Heraklion, Greece}
\address{$^{23}$ Dept. of Physics and Astronomy, Univ. of Bologna, Via Berti-Pichat 6/2, I-40126 Bologna, Italy}
\address{$^{24}$ Institute of Quantum Technologies, German Aerospace Center (DLR), S\"oflinger Str.~100, 89077 Ulm, Germany}
\address{$^{25}$ Istituto Nazionale di Fisica Nucleare (INFN) Sezione di Genova, via  Dodecaneso 33,  Genova, Italy} 
\address{$^{26}$ Dipartimento di Fisica e Astronomia and LENS Laboratory, Universit\`a di Firenze and INFN-Sezione di Firenze, via  Sansone 1,  Sesto Fiorentino, Italy} 
\address{$^{27}$ Universit\`a degli Studi di Urbino ``Carlo Bo'', I-61029 Urbino, Italy and INFN, Sezione di Firenze, I-50019 Sesto Fiorentino, Firenze, Italy}
\address{$^{28}$ Institute of Electronic Structure and Laser, Foundation for Research and Technology - Hellas, 70013, Heraklion, Greece}

\ead{benjamin.canuel@institutoptique.fr}

\begin{abstract}
Gravitational Waves (GWs) were observed for the first time in 2015, one century after Einstein predicted their existence. 
There is now growing interest to extend the detection bandwidth to low frequency.
The scientific potential of multi-frequency GW astronomy is enormous as it would enable to obtain a more complete picture of cosmic events and mechanisms. This is a unique and entirely new opportunity for the future of astronomy, the success of which depends upon the decisions being made on existing and new infrastructures. The prospect of combining observations from the future space-based instrument LISA together with third generation ground based detectors will open the way towards multi-band GW astronomy, but will leave the infrasound (0.1 Hz to 10 Hz) band uncovered.
GW detectors based on matter wave interferometry promise to fill such a sensitivity gap.
We propose the European Laboratory for Gravitation and Atom-interferometric Research (ELGAR), an underground infrastructure based on the latest progress in atomic physics, to study space-time and gravitation with the primary goal of detecting GWs in the infrasound band.
%
ELGAR will directly inherit from large research facilities now being built in Europe for the study of large scale atom interferometry and will drive new pan-European synergies from top research centers developing quantum sensors. 
ELGAR will measure GW radiation in the infrasound band with a peak strain sensitivity of $4.1 \times 10^{-22}/\sqrt{\text{Hz}}$ at 1.7 Hz.
The antenna will have an impact on diverse fundamental and applied research fields beyond GW astronomy, including gravitation, general relativity, and geology.
\end{abstract}

\maketitle

\tableofcontents

\section*{Introduction}

The first confirmed observation of gravitational waves (GWs)~\cite{Abbott2016} opened a new window into the study of the universe by accessing signals and revealing events hidden to standard observatories, i.e. electromagnetic \cite{Abramovici1992} and neutrino~\cite{Ando2013} detectors. Since then, several violent cosmological events have been reported, in detail ten binary black hole mergers and a binary neutron star inspiral \cite{LIGO2019}. Moreover, the complimentary information provided by GW astronomy could, for example, bring new insight for the study of dark matter or the exploration of the early universe, where light propagation was impossible. Expected sources of GWs range from well understood phenomena, such as the merging of neutron stars or black holes~\cite{Mandel2018}, to more speculative ones, such as cosmic strings~\cite{Nielsen1973} or early universe phase transitions~\cite{Weir:2017wfa}.

The era of GW astronomy was opened by the ``second" generation of laser interferometers LIGO~\cite{Aasi2015} and VIRGO~\cite{Acernese2014} that operate in a frequency band ranging from 10 Hz to 10 kHz. Other instruments operating in different frequency ranges are now required to expand the breadth of GW astronomy. Exploring the universe with GWs from low to high frequencies (mHz to kHz) can render possible the discovery of new sources of GWs. This is a unique opportunity to expand our knowledge of the laws of nature, cosmology, and astrophysics~\cite{Sesana2016}. The success of future GW astronomy depends on the choice of low frequency GW detectors. The proposal to construct the space-based Laser Interferometer Space Antenna (LISA) ~\cite{Jennrich2009} to investigate GWs sources at very low frequency, combined with the planned third generation ground-based laser interferometer (Einstein Telescope - ET) ~\cite{Punturo2010} will contribute to ``multiband GW astronomy", but will leave the infrasound (0.1 Hz-10 Hz) band uncovered. An infrasound GW detector is critical to the completion of available and considered observation windows in GW astronomy \cite{Mandel2018,Kuns2019,ArcaSedda2019,AEDGE2019}; such instrument could help answer long-standing questions of cosmology involving dark energy, the equivalence principle, cosmic inflation, and the grand unified theory.

The European Laboratory for Gravitation and Atom-interferometric Research (ELGAR) proposes matter-wave interferometry to fill the sensitivity gap in this mid-band. One century after the discovery of Quantum Mechanics and General Relativity, advanced concepts have resulted in dramatic progress in our ability to control matter at the quantum level. Manipulating atoms at a level of coherence that allows for precise measurement has led to the development of extremely sensitive inertial sensing devices that measure, with high accuracy and precision, accelerations~\cite{Peters1999,Freier2016}, rotations~\cite{Gustavson1997,Stockton2011,Savoie2018}, and even the tidal force induced by the spacetime curvature~\cite{Asenbaum2017}. The outstanding performances of inertial atom sensors motivate the surge of AI experiments both in fundamental physics - e.g., to measure fundamental constants \cite{Rosi2014,Bouchendira2011}, test general relativity \cite{Dimopoulos2007,Aguilera2014,Zhou2015,Rosi2017,Overstreet2018}, and set limit on the dark energy forces \cite{Burrage2015,Jaffe2017,Sabulsky2019} - and in applied contexts - e.g., space geodesy \cite{Trimeche2019}, geophysics \cite{deAngelis2009,Mnoret2018,Bidel2018}, and inertial navigation \cite{KasevichPatent2006,Barrett2019}. Triggered by the latest progress in this field, ELGAR will use a large scale, multidimensional array of correlated Atom Interferometers (AIs) in free fall~\cite{Chaibi2016}. In such a scheme, the GW signal is obtained by a a set of differential measurements between the different matter wave interferometers, providing a strong immunity to seismic noise and an important rejection of Newtonian Noise, i.e. two of the most important effects impacting the performances of infrasound GW detectors.  

The future infrastructure will not only fill the gap of infrasound GW observation, but could also have applications in other research domains including fundamental physics, gravitation, general relativity, and geology. ELGAR will monitor the evolution of the earth's gravitational field and rotation rate in three dimensions, thus improving our understanding of geophysical fluctuations of the Earth's local gravitational field, as well as our knowledge of slow variations in gravity gradients and rotations. The data produced by ELGAR could allow empirical tests of fundamental theories of physics with unprecedented precision. For example, precise time-mapping of the fluctuation of gravitational forces could provide limits on the violations of the Lorentz invariance~\cite{Chung2009} leading to improved understanding of quantum gravity~\cite{Disi1987,Penrose1996,Pikovski2015}. The precise monitoring of the earth’s rotation could also shed light on the Lense-Thirring effect~\cite{Pfister2007}, one of the many effects predicted by general relativity~\cite{Dimopoulos2007}. 

This paper is organized as follows: Sec.~\ref{sec_Detector_configuration} introduces the measurement concept of large-scale atom interferometry, presents the ELGAR geometry and derives its sensitivity to GWs and main noise sources. Sec.~\ref{sec_Detector_site} then presents different installation sites for the infrastructure in Italy and France, which are evaluated in terms of ambient noise. Sec.~\ref{sec_Atom_optics} and Sec.~\ref{sec_Atom_source} details the realization of the matter wave beam splitters and of the atom source. Sec.~\ref{sec_seis} gives insights of the suspension system required for the interrogation optics of the interferometer. Sec.~\ref{sec_NN_reduction} presents the sensitivity of ELGAR to different sources of Newtonian Noise and presents its mitigation strategy. Sec.~\ref{sec_other_noise_couplings} then gives a complete view of the metrology of the instrument, identifying and projecting the different noise sources in terms of equivalent strain. The ELGAR sensitivity curve and its complementary with other projects is then described in Sec.~\ref{sec_Sensitivity_curves}. The new possibility offered by ELGAR for astrophysics, gravity and fundamental physics are then studied in Sec.~\ref{sec_Sources_of_GWs}.

\section{Detector configuration and signal extraction}\label{sec_Detector_configuration}

\subsection{Atom interferometry}
\par An atom interferometer utilizes matter-wave beam splitters and mirrors to create a quantum mechanical analog to an optical interferometer~\cite{Cronin09}. 
Atom interferometric techniques take advantage of a fundamental property of quantum mechanics, interference, to offer unparalleled sensitivity to changes in space-time.
Here, we briefly introduce atom interferometry before delving into more details in later sections. 
\par
\begin{figure}[hbt!]
\centering
\includegraphics[width=.5\linewidth]{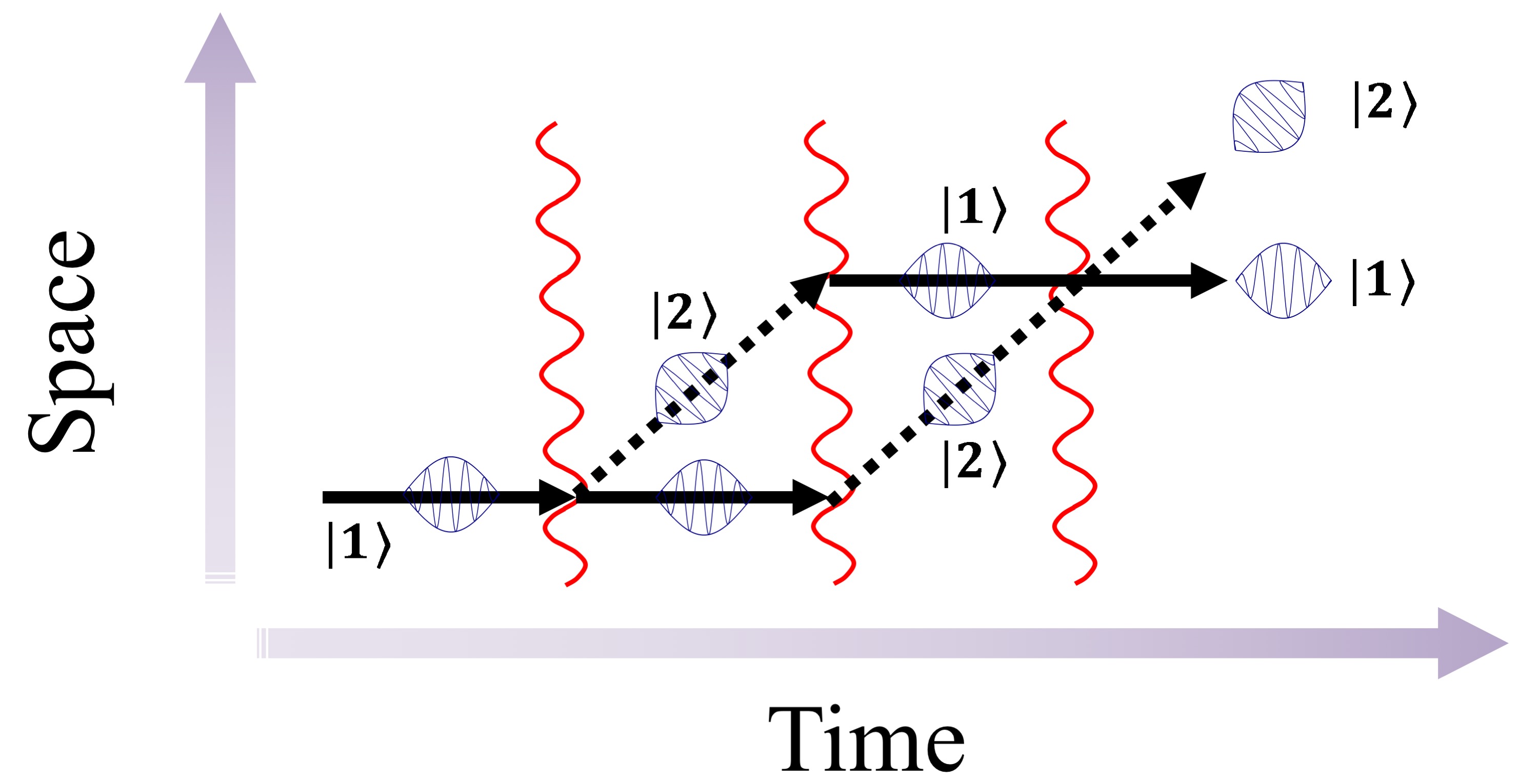}
\caption{
Space-time diagram schematic of an atom interferometer using light pulses. The atomic trajectories are represented in black: the solid lines refer to the propagation in state $\ket{1}$, the dashed ones in state $\ket{2}$. The propagation paths are represented as straight lines, whereas uniform gravity makes them parabolic. The two states have a momentum separation given by the two-photon momentum exchange imparted by the interferometric pulses, represented by the sinusoidal red lines. \label{fig:spacetime}} 
\end{figure}

 The atomic wave-function needs to be split, deflected, and recombined in order to observe interference, just like an electromagnetic wave in an optical interferometer.
After splitting, the atomic wave-packet follows a superposition of two different path and the interference pattern obtained after its recombination is a function of the relative phase shift accumulated along the paths. This phase shift is the result of free evolution of the atomic wave-function along each path~\cite{Storey94}.
We focus our attention on light pulse atom interferometers, where the interrogation of the atoms for splitting, deflecting, and recombining is accomplished using coherent pulses of laser light~\cite{Kasevich1991}. The space-time diagram in Fig.~\ref{fig:spacetime} shows a schematic version of this process for a single atom. 
An atom interferometer of this type measures the motion of the atomic wave-function relative to the reference frame defined by the laser phase fronts.
This has made light pulse atom interferometry a platform for inertial sensors that offers unparalleled precision and accuracy~\cite{peters_high-precision_2001}. 
Sensitivity to GWs is intrinsically linked to the response of an atom interferometer to the local phase of the manipulating optical field: the GW induces a variation of the travel time of photons between the atom and the laser~\cite{Dimopoulos2008}. 
\par The measurement of the atom interferometer phase requires a number of steps including preparation of the atomic sample, coherent manipulation of the matter waves which defines the instrument geometry and sensitivity, and finally detection of the output ports. 
Restricting our discussion to atom interferometers using cold atom in free-fall~\cite{Kasevich1991}, sample preparation requires collecting a dilute cloud of cold atoms, prepared with standard laser cooling and trapping techniques~\cite{RevModPhys.71.S253}.
Using ensembles with a small spread of momenta about their center-of-mass velocity ensures that atoms travel along the intended trajectory and avoids introducing spurious signals or reducing the interferometric contrast. 
After the cooling phase, these ensembles are transferred into the interferometer region by launching them onto a ballistic trajectory, accomplished via a moving molasses~\cite{Bertoldi2006}, coherent momentum transfer from laser light~\cite{Denschlag2002,Dickerson2013}, or by simply dropping them. This transfer allows for the separation of the interferometric region from the atomic source, which allows for optimization of several parameters like vacuum pressure and optical access, independently.
%
%
%
In the interferometer zone, a sequence of light pulses is applied to the atomic ensemble, to coherently divide, deflect, and finally recombine their wave-function. The light pulses are functionally made into beam splitters or mirrors based on the amount of time in which they interrogate the atomic ensembles. While illuminating the atoms, the resonant electromagnetic field introduces coherent transitions between different atomic states, so-called Rabi oscillations.
A beam splitter is realized for the pulse time corresponding to the creation of a superposition of states with equal probability, obtained at a fourth of a Rabi period and thus called a $\pi/2$ pulse. In a similar way, a $\pi$ pulse corresponds to a flipping of the atom states and to the realization of a mirror for the matter-waves. The interrogation sequence - defining the succession of $\pi$ and $\pi/2$ pulses and their distance in time - together with the direction of light with respect to the atom trajectory will define the sensitivity of the atom interferometer.
%
%
%
\par
\begin{figure}
\centering
\includegraphics[width=1\linewidth]{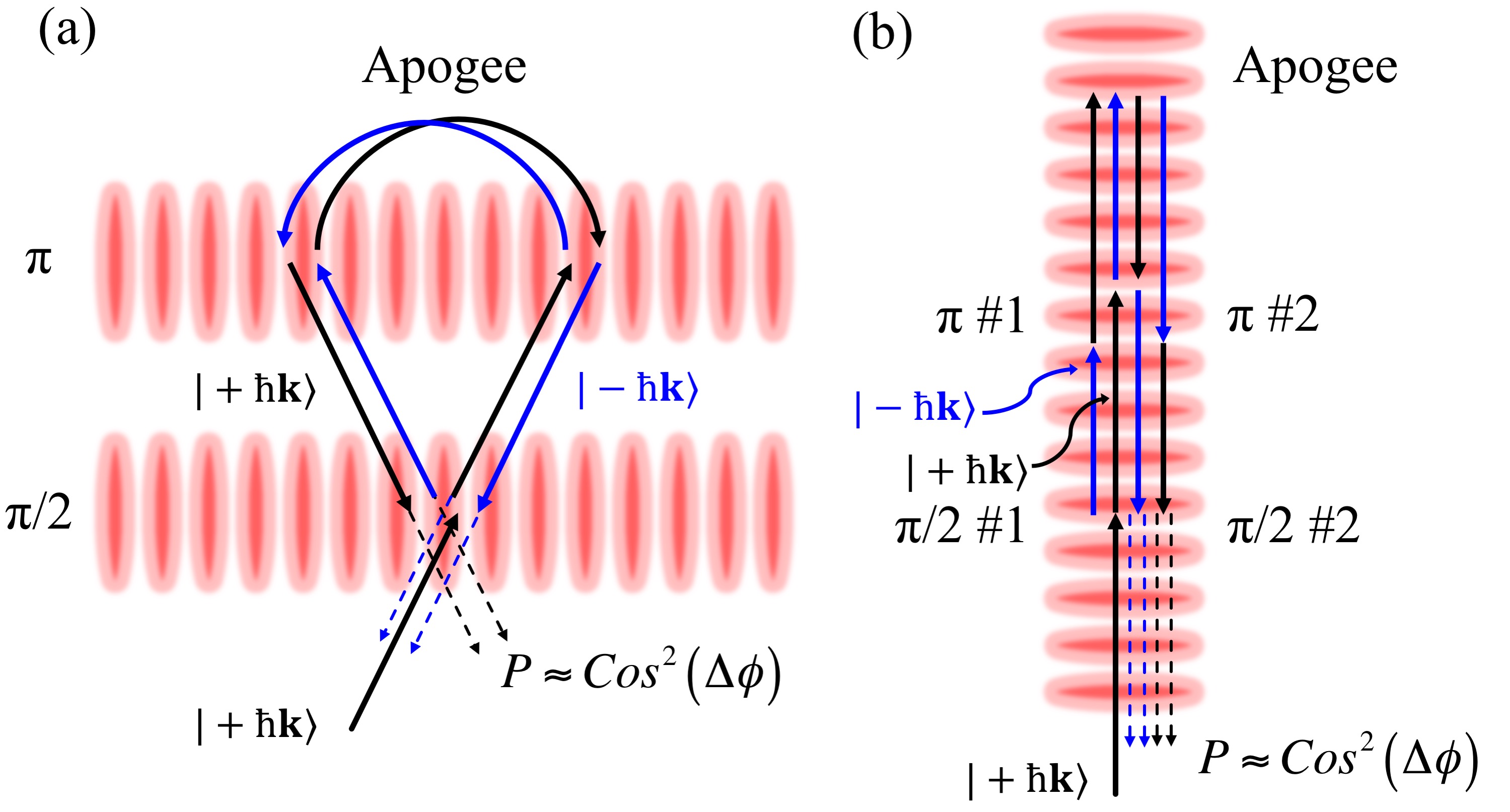}
\caption{Example schematic of a four-pulse atom interferometer. 
(a) Horizontal setup. 
A cloud with some initial momentum is interrogated by a lower beam to the tune of a $\pi/2$ pulse while an upper beam, where the apogee is, operates as a $\pi$ pulse. 
The atoms begin to fall with gravity and they are recombined with another $\pi/2$ pulse when they fall through lower interrogation beam. 
After, the population imbalance is measured which is a direct measurement of the atomic phase.
(b) Vertical setup. 
The same interferometer geometry, under constant gravitational acceleration. 
}
\label{fig:expsch}
\end{figure}

We now focus on techniques suitable to the ELGAR project. The antenna uses various laser cooling techniques for an all-optical production of atom ensembles with a 3D kinetic temperature below 1 $\mu$K, while maintaining a density dilute enough to mitigate atom-atom interactions - see section Sec.~\ref{sec_Atom_source}.
After cooling, atom are launched on a vertical parabolic trajectory into the interrogation region, where the interferometer is created in a symmetric way around the apogee using a set of two horizontal laser beam - see Fig.~\ref{fig:expsch}. Different interferometer sequences can be used for ELGAR; we focus our attention on the four-pulse ``butterfly"~\cite{dutta} configuration, whose geometry is shown in Fig.~\ref{fig:expsch}, which consists of a sequence of \mbox{$\pi/2$-$\pi$-$\pi$-$\pi/2$} pulses separated in time by \mbox{$T$-$2T$-$T$}.
This configuration~\cite{Canuel2006}, first proposed to measure gravity gradients, shows now sensitivity to DC accelerations and offers robustness against spurious phase terms.
The first interferometer pulse is a beam splitter, putting the atomic ensemble into a superposition of states. The second and third pulses deflect the states, and create a folded geometry. At the location of the second beamsplitter, the trajectories overlap and the two output ports are measured.
The details of the interrogation process will be treated in Sec.~\ref{sec_Atom_optics}. In brief, among the multiple techniques for the exchange of momentum between atoms and photons, the ELGAR project will focus on Bragg diffraction and Bloch oscillations~\cite{PhysRevA.88.053620}, based on their scalability and demonstrated efficacy~\cite{LKBgroup} in highly sensitive atom interferometer setups. 
%
%
At the conclusion of the interferometer, each atom of the ensemble is in a superposition of the output states.
For detection, we measure one observable of this quantum system, the occupancy of the states.
This operation is typically accomplished using a variety of destructive readout techniques, such as fluorescence and absorption~\cite{Rocco14}, to obtain the probability that an atom will be found in a particular state. 
This probability is a function of the relative phase acquired along the paths of the interferometer, which depends upon the variation of the interrogation laser phase during the time of the interferometer, where such variations may arise from the effect of incident GWs. 

Based on the horizontal interferometer geometry presented here, we now consider the sensitivity to GWs obtained from a gradiometric configuration using two spatially separated AIs, the basis of the ELGAR detector. 
\subsection{Gravitational wave signal from an atom gradiometer}
\begin{figure}[ht!]
\centering
\includegraphics[width=.5\linewidth]{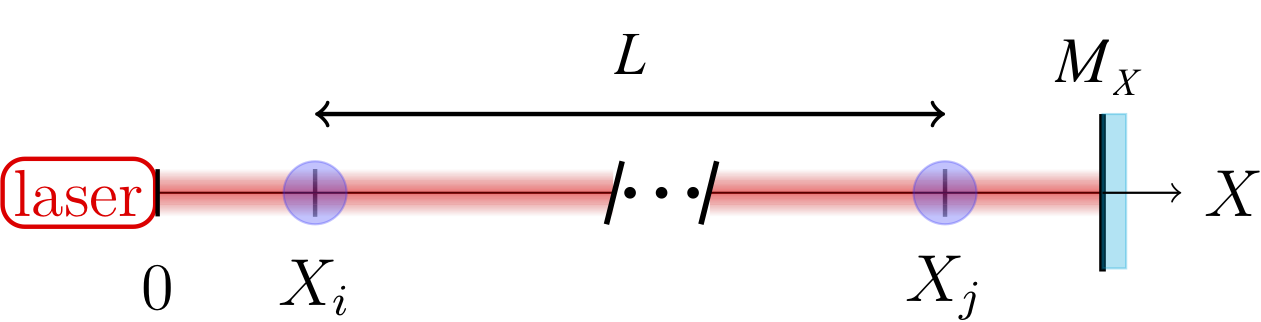}
\caption{Gravity gradiometer schematic diagram. Two AIs placed at $X_{i,j}$ are separated by a distance $L$ and coherently manipulated by a common laser retro-reflected by a mirror placed at position $M_X$.
\label{fig:AIgradio}}
\end{figure}

Here we present a schematic description of how the ELGAR detector is sensitive to gravitational waves. 
As shown in Fig.~\ref{fig:AIgradio}, we consider an atom gradiometer using two free-falling atom interferometers placed at positions $X_{i,j}$ along the $x$-axis and interrogated by a common laser beam which is retro-reflected by a mirror placed at position $M_X$. The geometry of each AI is the four-pulse ($\pi /2$--$\pi$--$ \pi$--$ \pi/2$) presented in the previous section.
The interferometric signal is read out as a ground state population imbalance that depends upon the phase difference $\Delta\varphi_{las}$ between the two counter-propagating beams.
Considering large momentum transfer techniques where $2n$ photons are coherently exchanged during the interrogation process, the output phase $\Delta \phi(X_i,t)$ of the atom interferometer at position $X_{i}$ and time $t$ is:
\begin{equation}
\Delta \phi(X_i,t)=n \int^{\infty}_{-\infty}\Delta\varphi_{las}(X_i,\tau)g'(\tau-t)d\tau+\epsilon(X_i,t) \, ,
\end{equation}
where $g'$ is the time derivative of the sensitivity function of the AI~\cite{Cheinet08, leveque} and $\epsilon(X_i,t)$ is the detection noise related to the projection of the atomic wave-function during the measurement process.
Accounting for the effects of laser frequency noise $\delta\nu(\tau)$, vibration of the retro-reflecting mirror $\delta x_{M_X}(\tau)$, gravitational wave strain variation $h(\tau)$, and fluctuation of the mean trajectory of the atoms along the laser beam direction induced by the fluctuating local gravity field $\delta x_{at}(X_i,\tau)$, the last equation can be written as~\cite{Canuel2018, Chaibi2016}:
\begin{align}\label{eq:Deltaphix}
\Delta \phi(X_i,t)&=\int^{\infty}_{-\infty}2nk_l\Big[\Big(\frac{\delta\nu(\tau)}{\nu}+\frac{h(\tau)}{2}\Big)(L-X_i)\nonumber\\
&+\Big[\delta x_{M_X} (\tau)-\frac{L-X_{i}}{c}\delta x'_{M_X}(\tau)\Big]-\delta x_{at}(X_i,\tau)\Big] g'(\tau-t) d\tau \nonumber\\
&+\epsilon(X_i,t) \, ,
\end{align}
where ${k_l=\frac{2\pi\nu}{c}}$ is the wave number of the interrogation laser, and $L=X_j-X_i$. 
It should be noted that seismic condition does not only impact movement of the retro-reflector, linked to the term $[\delta x_{M_X} (\tau)-\frac{L-X_{i}}{c}\delta x'_{M_X}(\tau)]$, but also creates frequency noise from movement of the input optics, which is included in the term $\frac{\delta\nu(\tau)}{\nu}$.
By simultaneously interrogating two atom interferometers with the same laser, one can cancel the sensitivity to position of the retro-reflecting mirror, a common-mode noise.
The resulting differential phase $\psi(X_i,X_j,t)$ is~\cite{Canuel2018}:
\begin{align}\label{eq:diffatphi}
\psi(X_i,X_j,t)=\Delta \phi(X_i,t)-\Delta \phi(X_j,t)\nonumber=\int^{\infty}_{-\infty}2n k_l\Big[\Big(\frac{\delta\nu(\tau)}{\nu}+\frac{h(\tau)}{2} - \frac{\delta x'_{M_X}}{c}\Big)L \nonumber\\
+\delta x_{at}(X_j,\tau)-\delta x_{at}(X_i,\tau)\Big]g'(\tau-t) d\tau +\epsilon(X_i,t)-\epsilon(X_j,t) \, .
\end{align}
With the assumption that the detection noise is spatially uncorrelated, we write the power spectral density of the differential interferometric phase as:
\begin{equation}\label{eq:Spsi}
S_{\psi}(\omega)=(2nk_l)^2 \Big[\Big(\frac{S_{\delta \nu}(\omega)}{\nu^{2}}+\frac{S_h(\omega)}{4}+\frac{\omega^{2}}{c^{2}}S_{\delta x_{M_X}}(\omega)\Big)L^2+S_{NN_1}(\omega)\Big]|\omega G(\omega)|^2+2 S_{\epsilon}(\omega) \, ,
\end{equation}
where $S_u$ denotes the Power Spectral Density (PSD) of a given time function $u\left(t\right)$.
The term $G(\omega)$ represents the Fourier transform of the sensitivity function of the interferometer to phase variations, which for the four-pulse configuration is~\cite{leveque}:
\begin{equation}
|\omega G(\omega)|^{2}=64 \sin^2\left(\omega T\right)\sin^{4}\left(\frac{\omega T}{2}\right) \, ,
\end{equation}
In Eq.~(\ref{eq:Spsi}) the term $S_{NN_1}(\omega)$ is the PSD of the relative displacement of the atom test masses with respect to the interrogation laser:
\begin{equation}
NN_1(t)=\delta x_{at}(X_j,t)-\delta x_{at}(X_i,t)\, ,
\end{equation}  
which is related to the difference of the local gravity field between the points $X_i$ and $X_j$ projected along the gradiometer direction, so-called Newtonian Noise (NN), i.e. terrestrial gravity perturbations of various origins, which we treat in detail in Sec. \ref{sec_NN_reduction}. 
This perturbation introduces an atomic phase variation that is indistinguishable from the signal produced by an incident GW, as shown in Eq.~(\ref{eq:Spsi}), and constitutes a limit for the detector that sums with other contributions.
\par Taking the gravitational wave term as the signal of interest in Eq.~(\ref{eq:Spsi}) and dividing it by the other terms, we obtain the signal to noise ratio (SNR) of the detector.
Setting the limit of detection as an SNR of 1, we define the strain sensitivity of the gradiometer as the sum:
\begin{align}\label{eq:Sh}
S_{h}&=\frac{4S_{\delta \nu}(\omega)}{\nu^{2}}+\frac{4S_{NN_1}(\omega)}{L^2}+\frac{4\omega^{2}S_{\delta x_{M_X}}(\omega)}{c^{2}}+\frac{8S_{\epsilon}(\omega)}{(2nk_l)^2L^{2}|\omega G(\omega)|^{2}}.
\end{align}
Here, we have derived the sensitivity of an atom gradiometer to changes in space-time strain, a configuration which is the basis of the ELGAR detector. We now present the full instrument geometry which is configured to optimize the sensitivity to the different noise term listed in Eq. (\ref{eq:Sh}).

\subsection{The ELGAR detector}
\subsubsection{ELGAR structure.}
\par
\begin{figure}[htp!]
\centering
\includegraphics[width=1\linewidth]{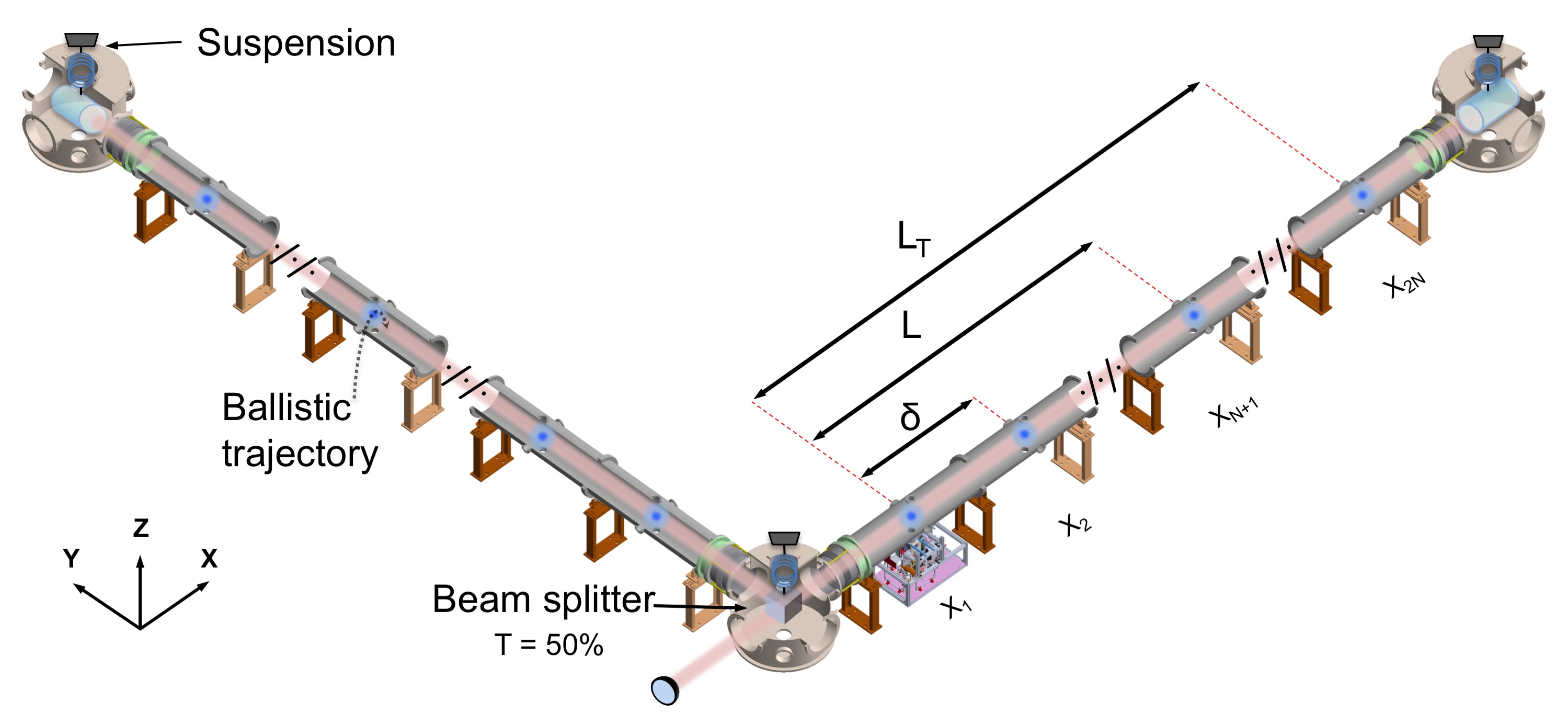}
\caption{Schematic diagram of the proposed ELGAR infrastructure. 
Each atom interferometer is separated from the next nearest interferometer by distance $\delta$, with the shortest gradiometric length $L$ on a baseline $L_{T}$.
There are a total of $N$ two-point gradiometers of minimum separation $L$ which comprise $2N$ total atom interferometers. 
The retro-reflection mirrors and the beam splitter are placed on suspension systems.
The entire detector is inside a large steel vacuum vessel that maintains a base pressure of $\leq 1 \times 10^{-9}$ mbar.
A vertical arm could be located with a horizontal detector, going below the beam splitter, or equally be located at a distributed site. 
\label{fig:ElgarGeo}}
\end{figure}

 In order to manage over the different terms limiting the strain sensitivity of a single atom gradiometer and given in Eq. \ref{eq:Sh}, we propose for ELGAR a detector structure shown Fig.~\ref{fig:ElgarGeo}.
The distinct feature of this geometry is the use of a 2D-array of atom interferometers, interrogated by a common laser beam in order to reduce sensitivity to Gravity-Gradient noise. Such a noise source is expected to be one of the main limitations of the sensitivity at low frequency of the next generation GW detectors based on optical interferometry, like the Einstein Telescope~\cite{Punturo2010}. Detectors relying on single atom gradiometers will be strongly limited in their performances by GGN in a large portion of their sensitivity window, and it will be crucial to mitigate its impact. To this end, we use a sensor geometry of an optimised array to statistically average GGN~\cite{Chaibi2016}, and bring its contribution below the target sensitivity of the instrument.

In this geometry the laser field is divided by a beam-splitter and retro-reflected by two end mirrors in order to obtain two symmetric and perpendicular arms in gradiometric configuration. Using parameters from~\cite{Chaibi2016}, each arm of ELGAR is composed by $N=$ 80 atom gradiometers of baseline $L=$ 16.3~km, regularly spaced by a distance $\delta=$ 200~m, leading to a total arm length of $L_T=(N-1)\delta+L=$ 32.1~km.
 
The whole detector is placed under ultra high vacuum with a residual total pressure less than $\leq 1 \times 10^{-9}$ mbar in order for gas collisions from the environment to play a marginal role in the operation of the individual atom interferometers. This would make the ELGAR vacuum vessel similar in terms of size of performances as the one of existing large experiments such as VIRGO/LIGO. Such vacuum performances could be reached with a series of pumping stations distributed along the arms,  containing oil-free backing pumps and large turbo-molecular pumps but also non-evaporable getters and ion pumps used after initial evacuation to high vacuum conditions. The individual atom sources are encased inside a magnetic shield sufficient to attenuate the Earth's magnetic field by a factor 1000 and the interrogation region is placed in a magnetic shield that covers the vacuum vessel in the few meters around each atom source; an example of such system is the magnetic shielding of the MIGA demonstrator~\cite{Canuel2016}).

\subsubsection{ELGAR GW signal extraction and strain sensitivity.}

To extract the GW signal, we consider the difference between the average signal of the gradiometers of each arms: 
\begin{equation}
\label{average}
H_N\left(t\right) =H_N^X\left(t\right)-H_N^Y\left(t\right)=\frac{1}{N}\sum^{N}_{i=1} \psi(X_i,X_{N+i},t)-\psi(Y_i,Y_{N+i},t). 
\end{equation}
Using Eq.~\ref{eq:diffatphi} we obtain:
\begin{align}\label{}
H_N\left(t\right)\nonumber=\frac{1}{N}\sum^{N}_{i=1}\int^{\infty}_{-\infty}2n k_l\Big[\Big(-\frac{\delta\nu_{Bs}(\tau)}{\nu}+h(\tau) - \frac{\delta x'_{M_X}}{c}+\frac{\delta x'_{M_Y}}{c}\Big)L \nonumber\\
+\delta x_{at}(X_j,\tau)-\delta x_{at}(X_i,\tau)-\delta x_{at}(Y_j,\tau)+\delta x_{at}(Y_i,\tau)\Big]g'(\tau-t) d\tau \nonumber\\
+\epsilon(X_i,t)-\epsilon(X_j,t)-\epsilon(Y_i,t)+\epsilon(Y_j,t) \, .
\end{align}
Using this differential signal cancels the contribution of common frequency fluctuations of the interrogation laser, the only  differential contribution coming from horizontal movement of the beam-splitter that creates a frequency noise $\delta\nu_{Bs}$ in the Y-arm of the detector:
\begin{equation}
\frac{\delta\nu_{Bs}(\tau)}{\nu}=\frac{\delta x'_{Bs_X}}{c}-\frac{\delta x'_{Bs_Y}}{c} \, ,
\end{equation}
where $\delta x_{Bs_X}$ and $\delta x_{Bs_Y}$ are the variation of position of the beam-splitter along $X$ and $Y$ direction.
Considering that the detection noise, the end mirror and the beam-splitter displacements are uncorrelated, and supposing $S_{\delta x_{M_X}}$=$S_{\delta x_{M_Y}}$=$S_{\delta x_{Bs_X}}$=$S_{\delta x_{Bs_Y}}$; we can write the power spectral density of  the average signal $H_N$ as:
\begin{equation}\label{eq:PSDnetwork}
S_{H_N}(\omega)=(2nk_l)^2 \Big[\Big(S_h(\omega)+\frac{4\omega^{2}}{c^{2}}S_{\delta x_{M_X}}(\omega)\Big)L^2+S_{NN}(\omega)\Big]|\omega G(\omega)|^2+\frac{4S_{\epsilon}(\omega)}{N}\, .
\end{equation}
In this last equation,  $S_{NN}$ is the PSD of the differential displacement introduced by the Newtonian Noise on the test masses of the network $NN(t)$, defined by:
\begin{equation}
NN(t)=\frac{1}{N}\sum^{N}_{i=1}\Big[\delta x_{at}(X_j,t)-\delta x_{at}(X_i,t)-\delta x_{at}(Y_j,t)+\delta x_{at}(Y_i,t)\Big]\, ,
\end{equation}
Following the method discussed in the previous section, we obtain the strain sensitivity of the detector exploiting the average signal as:
\begin{equation}\label{eq:SHnetwork}
S_h(\omega)=\frac{4\omega^{2}}{c^{2}}S_{\delta x_{M_X}}(\omega)+\frac{S_{NN}(\omega)}{L^2}+\frac{4S_{\epsilon}(\omega)}{N(2nk_l)^2L^2|\omega G(\omega)|^2}\, .
\end{equation}
In comparison with the result obtained for a single gradiometer, we observe that this configuration enables to mitigate the influence of the frequency noise of the interrogation laser, while preserving sensitivity to GWs with + polarization. Evenmore, considering the average signal also enables to partially mitigate the influence of gravity gradient noise exploiting the space-time correlation properties of its different sources. This process will be detailed later in Sec.~\ref{sec_NN_reduction}: in brief, assuming that the main sources of gravity gradient noise comes from isotropic density fluctuations of the medium surrounding the detector linked to seismic activity and atmospheric pressure variations, the averaging and correlation of the gradiometric phase from all participating gradiometers in the two arms enables to significantly reduce the unwanted signal from the gravity gradient noise~\cite{Chaibi2016}, related to the term $S_{NN}$ in Eq.~\ref{eq:SHnetwork} . 
Indeed, in units of strain$/\sqrt{\rm{Hz}}$, this technique can reduce the contribution from GGN by a factor $1/\sqrt{N}$ in comparison with the one of a single gradiometer, and can perform even better than $1/\sqrt{N}$ if the appropriate considerations are taken for optimizing the position of the gradiometers and the detector site has adequate properties. For what concerns direct effect of seismic noise, related to the term $S_{\delta x_{M_X}}$ in Eq.~\ref{eq:SHnetwork}, this configuration has a similar sensitivity to the one of a single gradiometer. Using a dedicated low frequency seismic attenuation system for the mirrors of the detector will be necessary to reduce its effects. We evaluate in Sec. \ref{sec_seis} the necessary high quality isolation and suspension system, which adopts and pushes forward key concepts devised for GW detection at low frequency based on optical interferometry. 

After mitigation of the different noise sources, the sensitivity of the detector is related to detection noise which is the last term in Eq.~\ref{eq:SHnetwork}.  This term is strongly related to the atomic species used in the atom interferometer as well as to the transition and techniques used for detection. The ELGAR detector can be run with different atom sources - see section~\ref{sec_Atom_source} for an overview of considered atomic species. Considering the use of $^{87}$Rb atoms launched onto a ballistic trajectory at a flux of $10^{\text{12}}$~atoms/s, an atom shot noise limited detection, a number of photon transferred during the interrogation of $2n=$1000, and an integration time of $4T=800$~ms, this sets a detection noise limited strain sensitivity of about $4.1 \times 10^{-21}/\sqrt{\rm{Hz}}$ at 1.7 Hz for a single gradiometer of the network. Considering the detection noise of the $2N$ gradiometers is independent, the shot noise limited sensitivity of the whole detector goes as $1/\sqrt{2N}$ and improves to about $3.3 \times 10^{-22}/\sqrt{\rm{Hz}}$ at 1.7~Hz.

In this section, we considered the main noise sources listed in Eq.~\ref{eq:Sh} which are relevant for the functioning principle and geometry of the detector. In Sec.~\ref{sec_other_noise_couplings} we will consider exhaustively other relevant backgrounds which impinge on the instrument's sensitivity, defining for each one the specific differential phase noise contribution, and outlining how we envision to contain the related effect below the targeted detector performance. Specifically, we will study the coupling to the instrument signal of parameters associated to the atomic sources (e.g. position, velocity, temperature and momentum spread of each atomic cloud), the manipulation beams (e.g. their pointing jitter, and relative alignment), and the environment fluctuations (e.g. gravity and its gradient, the magnetic and electric field, the blackbody radiation).

%
\section{Detector site}\label{sec_Detector_site}
Here, we discuss the list of requirements for the installation site of ELGAR, which are crucial to the performance and sensitivity of the detector. 
Following this, we discuss the properties of candidate sites in France and Italy which could host ELGAR: In France, the Laboratoire Souterrain \`{a} Bas Bruit (LSBB), an underground low-noise laboratory located in Rustrel, east of Avignon, which hosts the MIGA prototype antenna \cite{Canuel2018}; in Italy, two candidate sites on Sardinia contained within former mining concessions. 
\subsection{Site requirements}\label{sec_site_requirements}
Previous metrological studies carried out for both AI and GW detection have shown that a candidate site for ELGAR should account for stringent environmental requirements which otherwise could impact its functionality.
This includes controlling or monitoring signals that spoil GW detection and those that affect the individual atom interferometers.
Practical aspects are also important and must be examined, as well as feasibility of installation. 
The detector requires an analysis of environmental impact and that it be far from present and future anthropogenic disturbances, environmental pollution originating from human activity.
The construction must consider preexisting infrastructure, additions to infrastructure, impact on the environment and the local community, as well as identify suitable infrastructure for users. 
Finally, a cost model for each candidate site is required to compare the differing cost between suitable sites.
\par Regarding metrological aspects, seismic noise has proven to be a major concern for both AI and GW detection.
On an AI, vibrations during the launch and preparation of an atomic sample translate into fluctuation of the atomic readout but the most important impact is to imprint spurious atomic phase into the AI from movement of the retroreflection mirror.
A gradiometric configuration reduces the sensitivity of the detector to vibrations, see section~\ref{sec_Detector_configuration}, but given ELGAR's projected strain sensitivity, vibrations and rotations must be carefully managed.
It is necessary to counter the noise in rotation and in displacement to measure GWs.
Another metrological aspect with impact on AI are time-varying stray magnetic fields and field gradients and the difference in these fields between each AI. 
All atomic species, regardless of their nuclear spin, have some dependency on magnetic fields in all AI techniques. 
Magnetic fields and magnetic field gradients changing in time present a technical noise that can only be met with magnetic shielding or active compensation.
A candidate site requires mapping of magnetic fields and field gradients to mitigate this technical noise concern.
Related to seismic noise, local gravity gradient noise is a important  source of technical noise for GW detection in the ELGAR observation band, see sections~\ref{sec_NN_reduction}.
This noise can be separated into a seismic component and an atmospheric component, which places importance on local geological features and the local climate of the candidate site. 
Finally, we must consider the feasibility of installing the detector at a candidate site; construction of such a detector requires significant anthropogenic disturbances, the need to ensure future ecological protection, the availability or construction of a host facility, and local infrastructure for staff and installation of the detector.
It is crucial to examine the impact of the construction and operation of the detector to the candidate site and surrounding community.
\par A candidate detector site requires characterization and monitoring of the local seismic activity.
Correlation of this seismic noise across the detector site poses another technical noise problem - this movement of mass around the detector, amplified or attenuated by the local soil/rock composition, is seismic contribution to gravity gradient noise along one arm.
In these ways, a site that has low seismic noise properties year round is advantageous; continuous monitoring by a network of seismic detectors is required.
The data provided by a network is required to model the seismic noise and design seismic isolation units for the detector - the noise can be either up-converted or down-converted in frequency to remove it from within the observation band, see section~\ref{sec_seis}.
Further compounding the problem, broad contrasts in seismic impedance, e.g. ore deposits or geological levels, at a site complicate seismic analysis via reflection, conversion, or diffraction of surface waves - geological study of a site is required to mitigate such phenomena. 
In addition to concerns about seismic properties at the candidate site, anthropogenic noise is endemic throughout the ELGAR observation band - sites with nearby industrial and/or human activity must be ruled out due to vibrations. 
As a starting point, we consider sites appropriate for an underground detector as it provides a well controlled environment for large scale atom interferometry \cite{Canuel2018} in that the magnitude of seismic noise and GGN is reduced throughout the ELGAR observation band.
\par Seismic noise considerations are related to the geology and surrounding environment of the candidate site.
Soil and rock composition of the detector site is critical for both seismic GGN contributions, but also for magnetic noise; mechanical properties like the density, homogeneity, and other geological aspects have an impact on infrastructure works, changing the feasibility and cost. 
Geological features such as nearby flowing water are of interest to geophysicists and hydrologists, which enhances a candidate site's potential.
For example, while water flow near to the detector could constitute a large GGN background, it is a potential candidate for a major hydrogeological studies~\cite{Henry, Geiger2015, karsth20} like ground water transfer in the critical zone and the layout of underground water resources.
The technical noise added from such a background can be mitigated with a network of seismometers, tiltmeters, flowmeters, and gravimeters and the models created through study of these hydrogeological phenomena. 
\par The atmospheric contribution to gravity gradient noise places restrictions on the detector location.
Any atmospheric gravity gradient noise contribution is attenuated in an underground detector; monitoring, modeling, and characterization are still required.
Atmospheric temperature, propagating infrasound fields ~\cite{Fiorucci,Junca19}, barometric pressure readings, and wind direction/magnitude all must be monitored to account for their effect within the detector's observation band. 
A candidate site should not be at the confluence of multiple weather systems, to avoid large barometric pressure fluctuations.
In addition, we seek to mitigate the risk posed by storms, floods, and excessive humidity to an underground detector by analyzing climate data. 
\par Feasibility of construction and operation of a detector must be considered in choosing a candidate site.
Rock with strong impedance contrasts, like a high heterogeneity with composite rock of varying mechanical resistances are more difficult to bore through and could prove problematic for reasons of AI metrology and GW detection. 
Rural regions where candidate sites are located may lack the infrastructure required to host and support a detector; it could prove costly to transport the tunnel boring equipment along safe roads.
Removing the material from boring requires road access that local infrastructure may not be well equipped for; this waste material must be carefully managed and stored appropriately considering its possible environmental impact. 
This strain on local communities in remote locations must be studied for each candidate site.
A study on the impact of construction at candidate site must be performed to ensure minimum disruption to the ecosystem, safety of ground water, the power grid, and the local population. 
\par An output of candidate site surveys would be to approximate installation costs through a cost model at various locations for an ELGAR detector.
These costs include labor, power requirements, travel, geological studies, legal and environmental administration, as well as maintenance for existing facilities and the construction of new ones to help facilitate a large detector and it's community - interactions with existing initiatives could reduce costs significantly. 
Practical considerations necessitate that the detector not be too far from local facilities for staff, like schools and residences, as well as reasonable access to the facility via road, rail, and access to the detector via access shafts or tunnels. 
In addition, there are water, power, and sewage requirements for a site, as well as environmental restrictions and managing the risk of developing anthropogenic noise in the future.
The site would need to be located on protected land or land that could be declared as such to mitigate the risk of nearby human development that could impede or jeopardize the detector's operation. 
All these requirements, in addition to construction and maintenance, constitute a major disruption to any local community and considerations must be made about how the ELGAR detector affects the detector site. 
A full survey of candidate sites could be conducted, like what was accomplished for the Einstein Telescope~\cite{ET-design}.
\subsection{Candidate Sites in France and Italy}
In this white paper, we include two candidate sites that are presently under study. The first site considered is the Laboratoire Souterrain \`{a} Bas Bruit (LSBB), located in the hamlet of Rustrel, in southern France; it is the location of the MIGA experiment. 
The second set of sites are located on the Italian island of Sardinia and are all former mining concessions considered for both GW detection and gravity gradiometry with AI.
\subsubsection{The LSBB facility.}
\begin{figure}[htp!]
\centering
\includegraphics[width=1\linewidth]{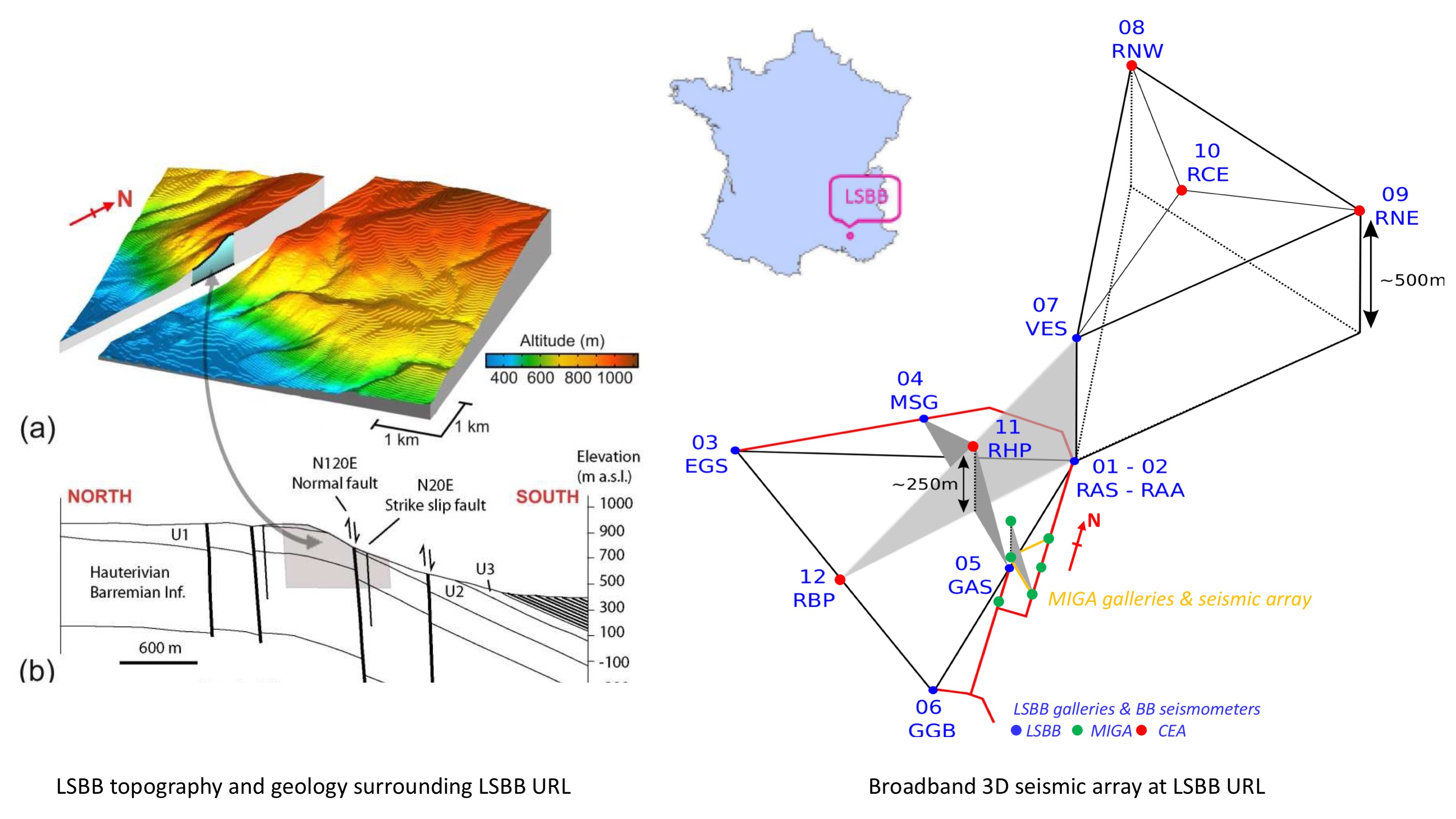}
\caption{The LSBB facility. (A) The laboratory is underneath a carbonate massif in southeastern France. (B) The underground facility is shown in thin red lines with a schematic representation of the permanent 3D broadband seismic array \cite{labonne_seismic_2016, labonne_seismic_2016-1} - installed at the surface and in the galleries. Inset; the LSBB overlooks the hamlet of Rustrel, 60 km east from the high-speed train station in Avignon and 100 km north of Marseille, with the Fontaine-de-Vaucluse watershed, part of the Regional Natural Park of Luberon - this region is sparsely inhabited. Modified from~\cite{maufroy_travel_2014}, with permission. 
\label{fig:LSBB1}}
\end{figure}
The Laboratoire Souterrain \`{a} Bas Bruit (LSBB), a Low Background Noise Underground Research Laboratory located in Rustrel, near the city of Apt in Vaucluse, France, and about 100 km from the international airport in Marseille, is a European-scale interdisciplinary laboratory for science and technology created in 1997.
The laboratory was formed after the decommissioning of a launch control facility for the French strategic defense initiative (direct translation: nuclear dissuasion strategy) during the Cold War. 
The LSBB is now a ground and underground based scientific infrastructure~\cite{Gaffet2010} characterized by an ultra-low noise environment, both seismic and electromagnetic, as a result of the distance of the site from anthropogenic disturbances and its location under a large massif. 
The LSBB fosters multidisciplinary interactions and interdisciplinary approaches, pursuing both fundamental and applied research and has broad scientific and industrial expertise.
For these reasons, this site was chosen for the location of the MIGA project \cite{Canuel2018}, which aims to study gravity gradient noise and test advanced detector geometries for its mitigation~\cite{Chaibi2016}.
New galleries were blasted specifically for the MIGA project, giving LSBB and the MIGA consortium expertise in the administration of large infrastructure works.
Further infrastructure work at this facility would benefit from this acquired experience.
\par The LSBB facility is in the Fontaine-de-Vaucluse watershed, the fifth largest karst aquifers on Earth, covering 1115 km$^{2}$ in surface area. 
LSBB has 4.2 km long horizontal drifts at depths below ground ranging from 0 m to 518 m; its orientation lies primarily north to south and north-east to south-west. 
Carbonate rock surrounding the facility result in a thermal blanketing effect - a passive temperature stability better than 0.1~$\degree$C is observed. 
Internal air pressure and circulation are controlled via a series of air locks (SAS) that protect underground areas against anthropogenic disturbances.
The whole facility is connected to power lines, high-speed telecommunications (RENATER), and GPS time. 
The facility's location in the karst system has generated interest from hydrogeologists and geophysicists~\cite{carriere_role_2016, gaillardet_ozcar:_2018, jourde_sno_2018, cappa_stabilization_2019, barbel-perineau_karst_2019} studying such watershed systems; this has resulted in a series of sensors installed at the facility to help measure the tilting of rock mass \cite{lazaro_roche_design_2019, hivert_muography_2017, lesparre_new_2017} and the influence of gravity gradient variations - the local environment around LSBB is completely monitored, from wind speed, temperature, humidity, local gas composition (CO, O$_{2}$, CO$_{2}$, Rn) and microbarometric variations to local seismic and hydrogeological activities, as well as gravimetric perturbations measurements via superconducting magnetometers and gravimeters \cite{pozzo_di_borgo_minimal_2012, collot_operation_2013, henry_simultaneous_2016, Rosat2018}.
\par
\begin{figure}[htp!]
\centering
\includegraphics[width=.9\linewidth]{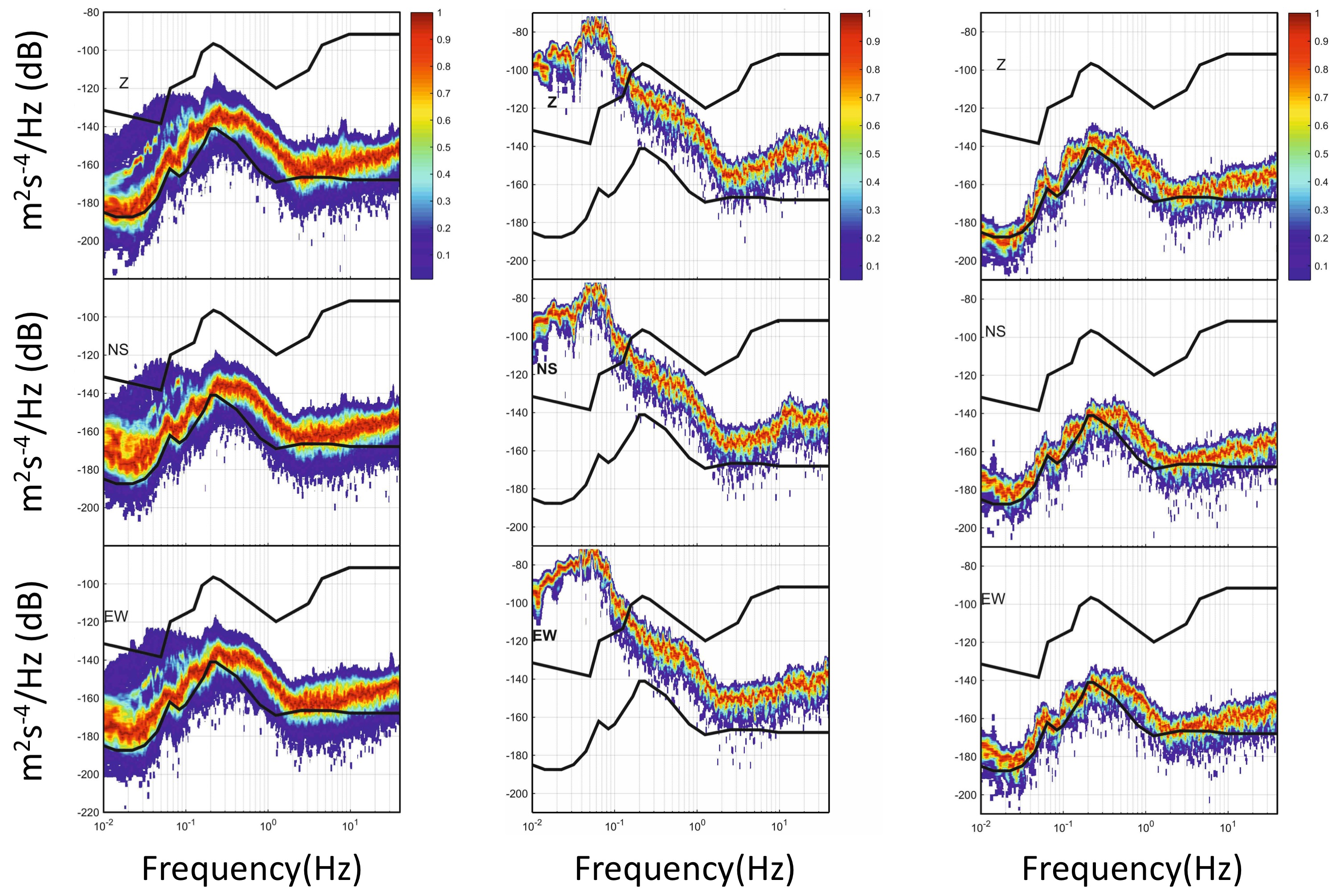}
\caption{Probability density functions samples of ambient seismic noise (color scale) of RUSF.01 broadband station for year 2011, including quiet days and days with earthquakes (left column); 6 hours of ground motion which include the Tohoku-Oki earthquake of March 11, 2011 (center column); a quiet day (June 26th, 2011) without seismic events (right column). The seismic noise PSD for three components (top row, Z; middle row, N-S; bottom row, E-W) is compared to the Peterson’s high and low noise models (black lines).
\label{fig:LSBB3}}
\end{figure}

 A network of 17 broadband seismometers monitors the seismic ground motion at the facility.
A sample of seismic noise data recorded at the station RUSF.01 (the underground seismometer named RAS in Fig.~\ref{fig:LSBB1}) is depicted in Fig.~\ref{fig:LSBB3}. 
This figure shows the probability density function (PDF) compared to Peterson's models for this site.
These data show three orthogonal components for three separate measurement intervals: the first column includes, roughly, the whole year of 2011, including the Mw 9.1 Tohoku-Oki mega-thrust earthquake in Japan from March 11$^{\rm{th}}$ of that year, the second column is a six hour interval during the earthquake, and the third column is a 24 hour period with no seismic activity. 
These nine plots show low seismic noise, baring large transient signals from earthquakes. 
The highest probability of noise occurrence, over a long time interval, is close to Peterson's low-noise model for all three components over the entire frequency band considered. 
Below 200 mHz, the PDF at one year spreads between the high and low noise models.
This low frequency band includes worldwide seismic activity and, more specifically, all surface waves that superficial earthquakes produce; considering a quiet day, the PDF can dip significantly below the low noise model with high probability.
The seismic noise spectra recorded are close to worldwide minima.
\par The low-noise seismic properties of LSBB have been confirmed using superconducting gravimeters (model iOSG from GWR Instruments Inc.~\cite{Rosat2018}).
This device was installed at the underground site in 2015 in order to define a gravimetric baseline dedicated and complementary to the MIGA experiment. 
These instruments installed at LSBB have a demonstrated stability of 1.8 $\text{nm} / s^{2} / \text{Hz}^{1/2}$ at 1 mHz, a noise performance among the best in a worldwide network of superconducting gravimeters \cite{farah_underground_2014, Rosat2018}.
\par A measurement campaign near to the location of the MIGA project found that the observed magnetic field in the frequency band 1 mHz to 400 Hz is much lower than the expected background~\cite{Canuel2018, magstuff}.
Magnetic field fluctuations around 2 pT/$\sqrt{\text{Hz}}$ at 1 Hz were observed in the tunnels near to the MIGA installation - throughout the measurement galleries in the facility, fluctuations ranged from 10 to 0.08 pT/$\sqrt{\text{Hz}}$, depending upon the shielding in each measurement hall. 
\par LSBB is protected against anthropogenic disturbances within the Regional Natural Park of Luberon, which is lightly industrialized.
The location of LSBB beneath a massif nets a unique sheltering effect with respect to electromagnetic noise~\cite{waysand_first_2000}.
There is a two-kilometer exclusion zone near the facility, reducing magnetic interference from high-voltage power lines and railways. 
The facility is 500 m below carbonate rocks loaded with water - this gives a high frequency cut-off around 200 Hz for incoming electromagnetic waves. 
\par The LSBB has established maintenance facilities, running water, power, sewage, nearby villages and towns with residences, schools, restaurants and shops, as well as the reputation of an established pan-European large scale research facility.
The facility's low-noise characteristics as well as the suitability of the surrounding area has already been extensively studied for the MIGA project, among other European projects.
\subsubsection{Sardinia facility.}
\par
\begin{figure}[htp!]
\centering
\includegraphics[width=.9\linewidth]{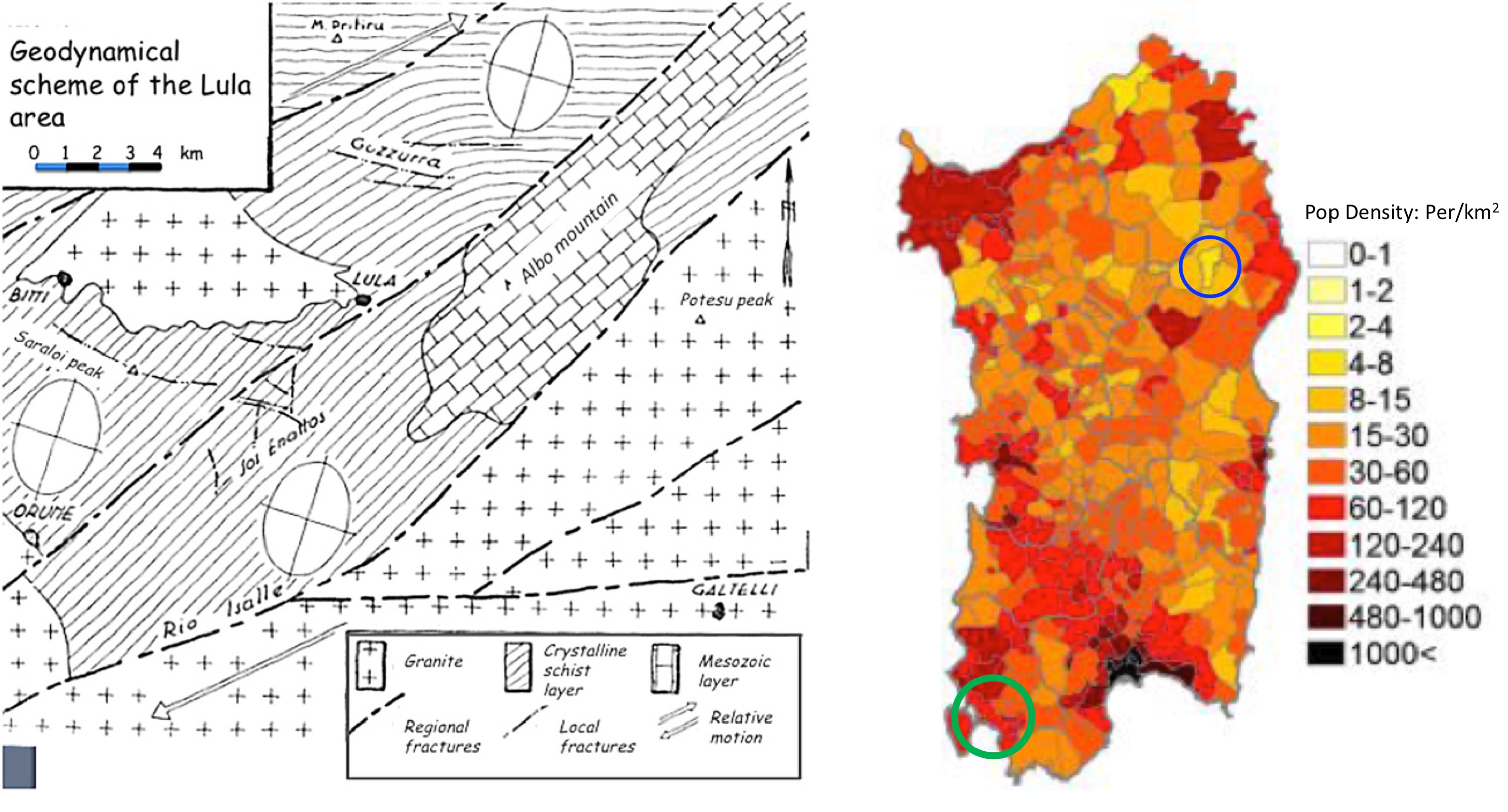}
\caption{The Sos Enattos site on Sardinia. Left: Geodynamical scheme of the Lula area, around Sos Enattos on Sardinia.
Right: distribution of population density in Sardinia. 
The blue circle indicates the location of the Sos Enattos site and the green circle indicates the location of the Seruci site.     
\label{fig:Sard1}}
\end{figure}

 The island of Sardinia in Italy offers several candidate sites which could host a detector of the scale of ELGAR. 
It contains areas with low anthropogenic disturbances, given a population density on the island that is among the lowest in Europe. 
The geological structure of the island is ancient; due to this, micro-seismic activity is among the lowest on Earth. 
The continential landmass, the Sardinia-Corsica block, is isolated and partially removed from the Alps block.
Addition, this block is located on the European tectonic plate and far from any fault lines.
From a feasibility standpoint, the island offers several interesting locations - former mine shafts have appealing characteristics.
Two potential candidate sites are under study; the Sos Enattos site near Lula, in the north-east, and the Seruci/Nuraxi Figus mining sites in the southern area of Sulcis, see Fig.~\ref{fig:Sard1}.
\par
\begin{figure}[htp!]
\centering
\includegraphics[width=1\linewidth]{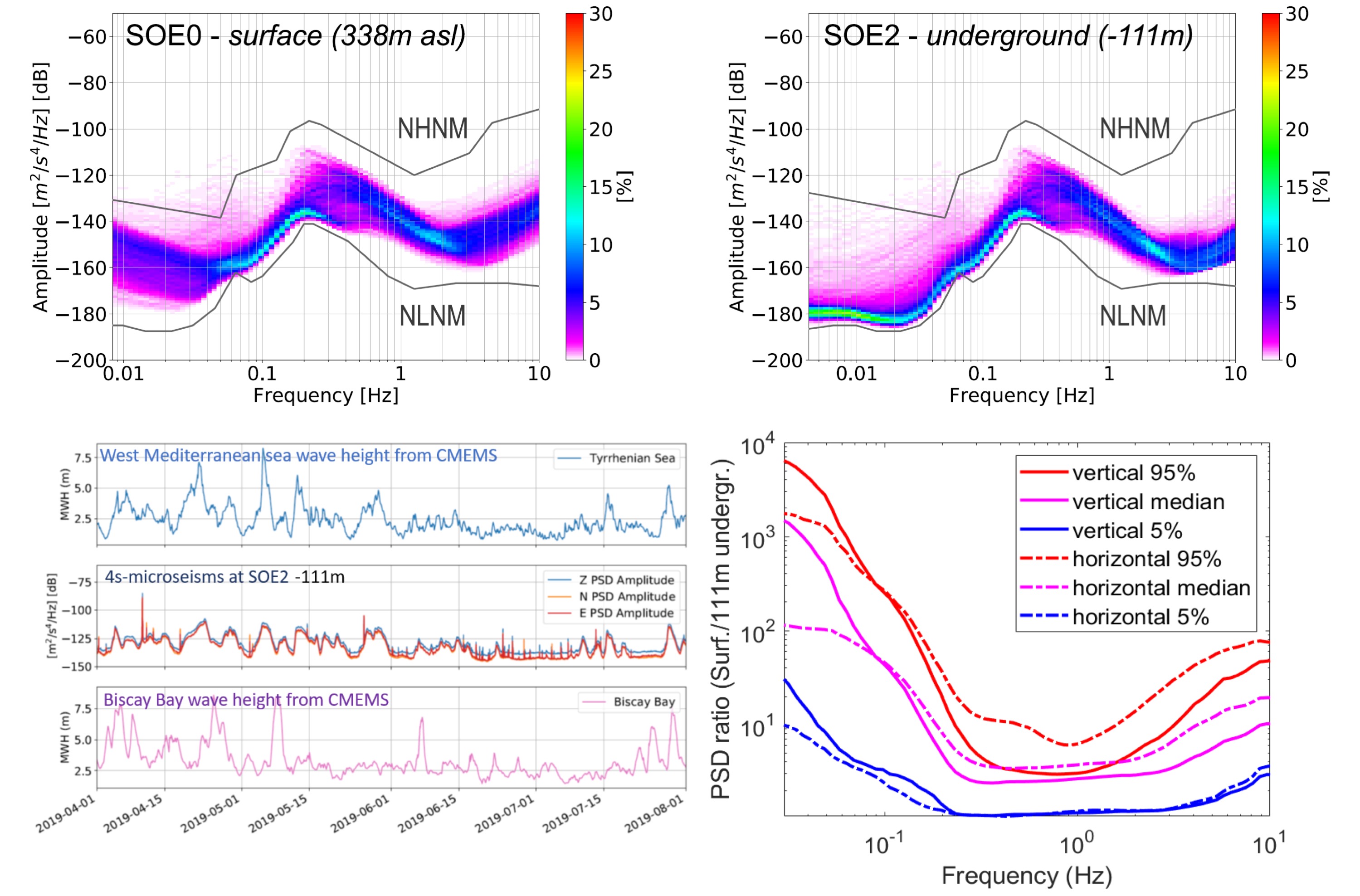}
\caption{Seismic noise at the Sos Enattos site. 
Figures courtesy of \cite{bignat19}.
Top: Power spectrum percentiles of seismic noise (vertical component), for the period April-August 2019. 
Surface measurements (338 m above sea level) on the left, underground measurement below 111 m of rock on the right. 
The continuous lines represents Peterson's models. 
Bottom left: Sea wave height from CMEMS and microseismic amplitude at Sos Enattos. 
Bottom Right: Ratio between surface and underground seismic noise. 
\label{fig:Sard3}}
\end{figure}

 The center and eastern sections of the island are dominated by carbonate rock.
The first site consider is located there, the Sos Enattos mine, and is located 40 km north by northwest from the city of Nuoro and 5 km from the village of Lula. 
The mine is composed of sphalerite ([Ze,Fe]S) and galena (PbS) rock.
The mine focused on the excavation of lead and zinc deposits for 130 years - well maintained underground caverns have left the site usable since 1864. 
This site is particularly favorable due to ancient geology; a carbonate platform covers a base of Hercynian granites. 
The layers have few fractures and are more likely to bend than create breaks. 
Despite these favorable mining conditions, the population density of the Albo mountain area is a factor seven lower than the mean for the island of Sardina. 
The site is managed by the I.G.E.A. S.p.A. company, which has entertained the Einstein telescope collaboration's interest in this site for their detector.
A series of seismic, acoustic, and magnetic measurement campaigns are presently underway to fully characterize this facility \cite{bignat19}.
Fig.~\ref{fig:Sard3} shows the seismic power spectral density (PSD) at the Sos Enattos site. 
Seismic noise is close to the Peterson's New Low-Noise Model (NLNM). 
Underground measuring stations at depths of 84 and 111 m show a large attenuation of anthropic noise above few Hertz, and of slow thermal and pressure fluctuations below 80 mHz. 
Seismic noise below 1 Hz is dominated by microseismic peaks from sea waves. 
Correlation of the seismic PSD with wave height from Copernicus Marine Environment Monitoring Service (CMEMS) in the western Mediterranean sea and in Biscay bay shows that that the dominant contribution comes from the closer Mediterranean sea, in particular for a period of 4.5 s \cite{bignat19}. 
Population density in this region is among the lowest in Europe, so the anthropic noise background, which is usually dominant above 1 Hz, is low even at the Earth's surface, and is additionally attenuated underground as shown in Fig.~\ref{fig:Sard3}.
\par On the southwestern portion of Sardinia, the carbonate rock gives way to coal bearing substrate. 
Here, the Monte Sinni mining concession C233, extracts coal; it covers an area of 59.4 km$^{2}$, falling entirely within the municipalities of Gonnesa, Carbonia, and Portoscuso in the province of Carbonia-Iglesia, South Western Sardinia. 
The company Carbosulcis S.p.A., owner of the mining concession, resumed mining coal in 1976, which had been interrupted by ENEL a few years prior. 
Until recently, mining and industrial activities were concentrated at the two work sites of Seruci and Nuraxi Figus. 
After the centralization of the services in Nuraxi Figus, the mining activities of Seruci were halted.
\par The mining site includes an underground area for the primary phases of the coal extraction cycle, comprising excavation and preparation of tunnels, cultivation of coal, and transportation of crude coal to the surface.
The coal mine extends underground through a network of tunnels whose total length is about 30 km. 
At present, the structure of the mine is developed to a depth between 350 and 500 m below the surface (between 200 m and 400 m above sea level).
The connection between the surface and underground is maintained through four main shafts (two in Seruci and two in Nuraxi Figus) and a winze with inclined ramps. 
The ventilation system is operated by two aspiration fans. 
The connection between the shafts by structural galleries.
The dimensions of the main galleries allow for the transport and installation of mining and excavation equipment. 
The shafts and the main galleries have lighting, electricity, water, and compressed air.
\par
\begin{figure}[htp!]
\centering
\includegraphics[width=0.9 \linewidth]{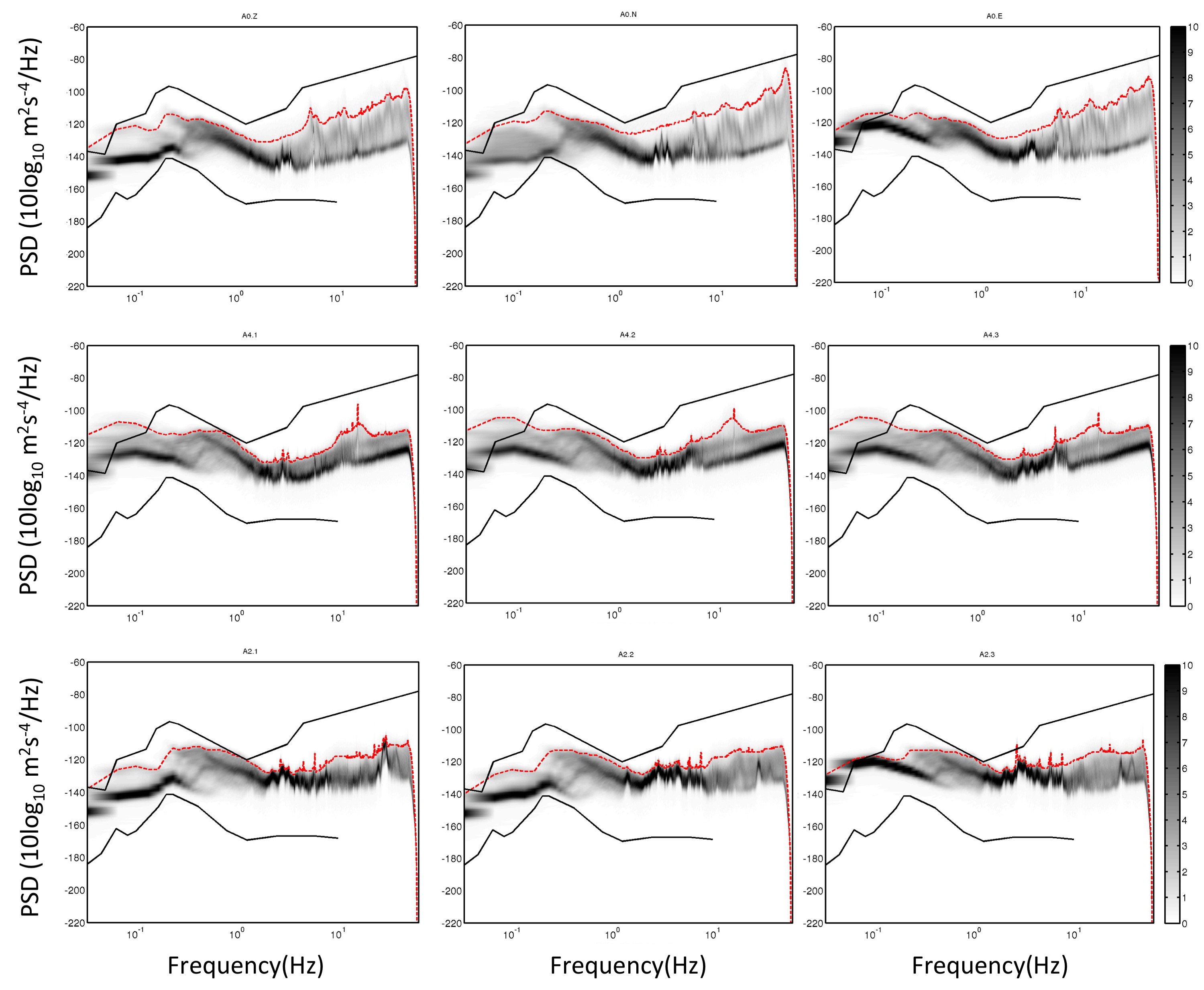}
\caption{Probability density function of the spectral powers for the three stations near the Seruci site; figure courtesy of INGV.
The distributions refer to 30 days of continuous ground motion recording. From left to right, the graphs refer to the Z, N-S and E-W components. 
From top to bottom, the plots refer to the A0, A4, and A2 stations respectively. 
The red line indicates the 95th-percentile of the spectral power. 
The probability density is indicated in gray scale. 
The black lines indicate Peterson's upper and lower models of Earth's seismic noise.
\label{fig:Sard2}}
\end{figure}

 Forced ventilation guaranteed everywhere in the mine for safety reasons; the air speed through the gallery section, depending if it is a primary, secondary or service gallery, is between 0,8 and 1,6 m/s. 
The typical section of a primary gallery is up to 24 m$^{2}$ (7.2 m wide and 4.9 m high), decreasing to 18 m$^{2}$ for a secondary gallery. 
In working areas, there can be special electrical devices or instruments.
It is possible to smooth and level the gallery pavement, or cover the ceiling by sputtering concrete. 
The roof supports are made with bolts and iron arches,
Transport of people and materials through the mine is available dedicated underground vehicles. 
Special teams are trained and ready for emergency and support. 
There is a system of continuous underground environmental monitoring in place, which consists of analyzers in fixed locations near air inlets of the reflux wells, in the secondary reflux, and in the active cultivation yards - all places where harmful gases may develop. 
Presently, the control station is outfit to monitor the following gases: CH4, CO, O2, CO2, NOX, and the following parameters: ambient temperature, relative humidity, and, air velocity.
We show data from continuous seismic monitoring of the site to illustrate the location's suitability. 
The data in Fig.~\ref{fig:Sard2}, power spectral densities, were obtained from the average of the Fourier transforms calculated on 10 consecutive windows of length 30s, with 50$\%$ overlap. 
The estimates were then repeated on successive intervals along 30 days of continuous recording, leading to a total of over 16,000 independent spectral estimates (523 PSD / day x 30 days). 
The results, represented by the probability densities (PDF) of the spectral power versus frequency, are compared with Peterson's Earth seismic models NLNM and NHNM of Low- and High-Noise. 
Before PSDs computation, the time series were corrected for the instrumental transfer function, regularizing the deconvolution by band-pass filtering in the 0.1-50 Hz interval. 
Out of this interval the spectra powers must be taken with caution.
The data show that site displays quiet seismic properties. 
\par The mine maintains significant infrastructure and is now transitioning into scientific exploration.
The Seruci 1 shaft is presently being adapted for a 350 m veritcal cryogenic column, required for the $^{40}$Ar distiller for the ARIA project \cite{Aria}. 
This requires an upgraded bearing framework that will be able to hold more than one column - for this reason, it is under consideration as a potential distributed vertical arm of the ELGAR detector at reduced cost. 
Other main shafts are also available and will be compared for available column depth, environmental noise, and crossing with horizontal galleries.
\section{Atom optics}\label{sec_Atom_optics}
As outlined in previous sections, two atom interferometers in a gradiometric configuration separated by a baseline $L$, and coherently manipulated by the same light fields with a wavenumber $k_{\mathrm{eff}}$ can be utilized as a differential phasemeter~\cite{dimopoulos_atomic_2008,Canuel2018}.
An incident gravitational wave with amplitude $h$ and frequency $\omega$ modulates the baseline, leading to a differential phase shift between the two atom interferometers.
Single-loop (3 pulses)~\cite{dimopoulos_atomic_2008,Canuel2018,Hogan2016PRA}, double-loop (4 pulses)~\cite{Hogan2011}, and triple-loop (5 pulses)~\cite{Hogan2011} geometries in gradiometric configuration were proposed for vertically~\cite{dimopoulos_atomic_2008} or horizontally~\cite{Canuel2018} oriented gravitational wave detectors on ground and in space.
The beam splitters, mirrors, and recombiners therein forming the interferometer typically consist of composite pulse and / or higher-order processes based on Bragg- / Bloch transitions to enhance the wavenumber $k_{\mathrm{eff}}=2 n k_{\mathrm{l}}$ corresponding to the relative momentum between the two trajectories in the atom interferometer, $n$ two-photon transitions, and a (single-photon) wavenumber $k_{\mathrm{l}}$ of the driving light field.
Single photon processes were also proposed to relax requirements on laser frequency noise~\cite{Graham2013PRL}.
All these geometries share similarities as free falling atoms, linear scaling of the phase shift in the effective wavenumber, the distance between the two atom interferometers, and a frequency response dependent on the pulse separation time $T$, leading to a general form of the phase shift $\Delta\phi\sim h k_{\mathrm{eff}} L f(\omega T)$.
\begin{figure}
\centering
\includegraphics[width=0.48\linewidth]{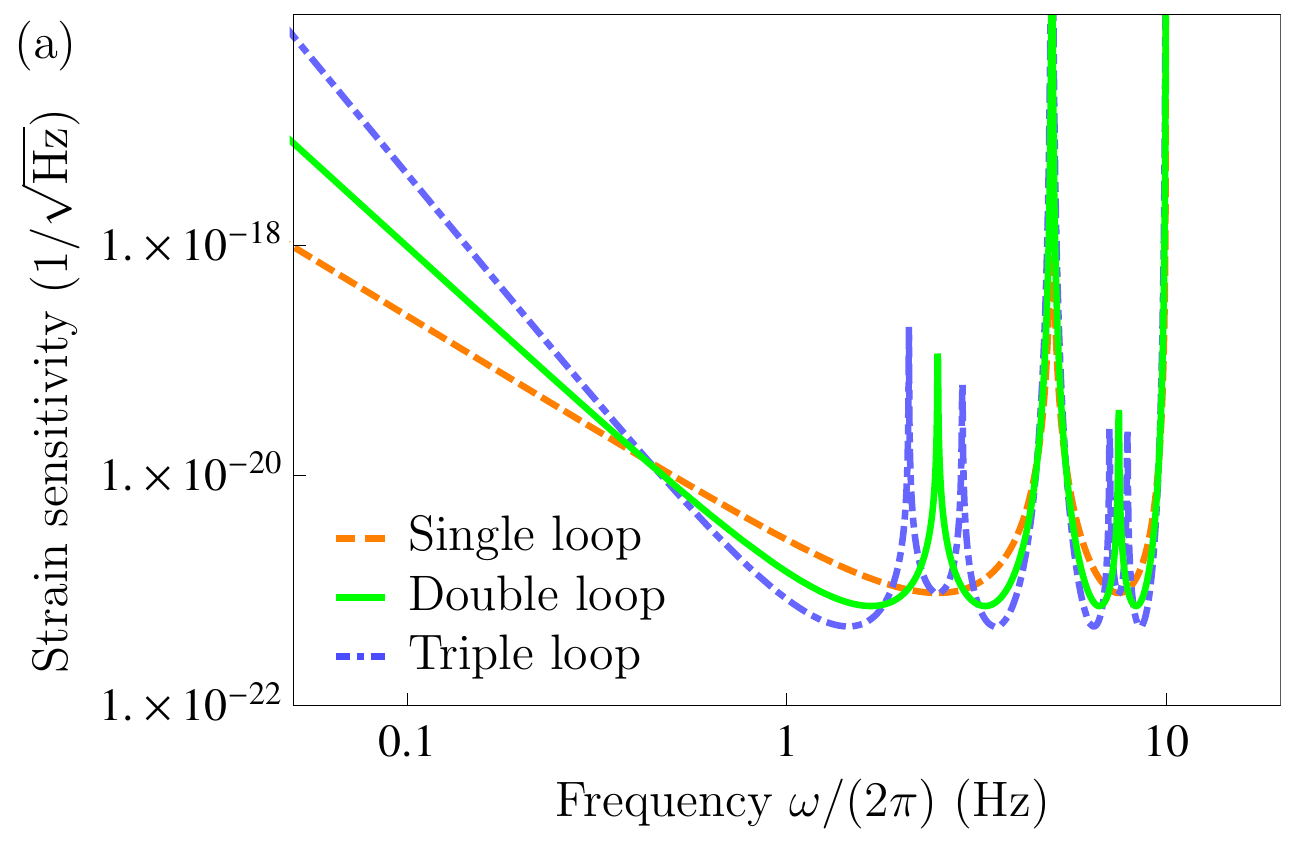}
\includegraphics[width=0.48\linewidth]{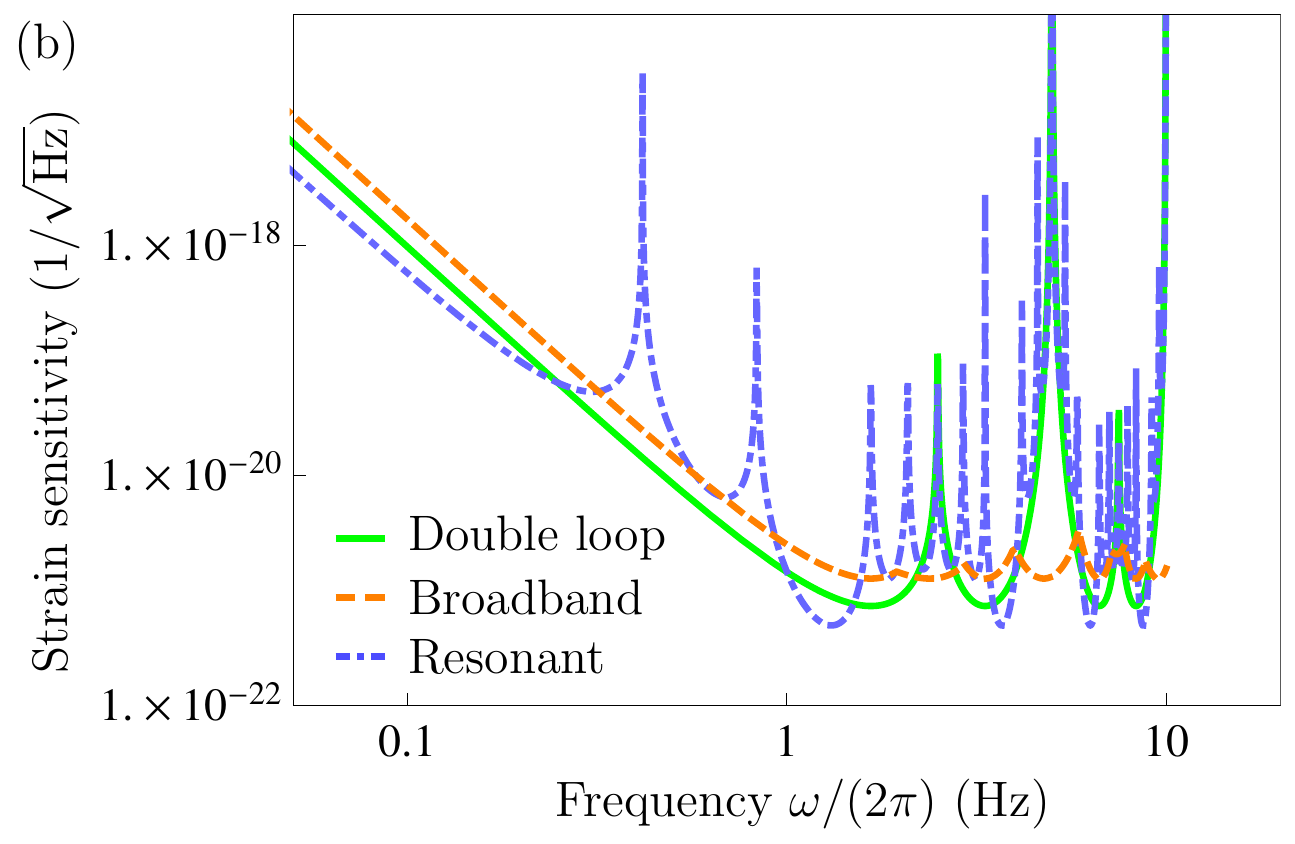}
\caption{Shot-noise limited strain sensitivities for different geometries. 
(a) Comparison of single-, double-, and triple-loop interferometers.
The graph assumes an effective wavenumber $k_{\mathrm{eff}}=1000\cdot2\pi/(780\,\mathrm{nm})$, a pulse separation time $T=200\,\mathrm{ms}$, a baseline $L=16.3\,\mathrm{km}$, a phase noise of $1\,\mu\mathrm{rad}/\sqrt{\mathrm{Hz}}$, and 80 gradiometers per arm, as well as sky and polarization averaging~\cite{Moore2015CQG}.
(b) Comparison of a double-loop interferometer as in (a) to a double-loop configuration in broadband mode and in resonant mode with a tripled number of loops.
The broadband mode uses interleaved interferometers with the three different pulse separation times $0.7\,T$, $0.9\,T$, and $T$. 
}
\label{fig:atom-optics-strain-sensitivities}
\end{figure} Differences are that multiple loops change the response $f(\omega T)$ at low frequencies (Fig.~\ref{fig:atom-optics-strain-sensitivities} (a)), lead to a resonant enhancement~\cite{Graham2016PRD} (Fig.~\ref{fig:atom-optics-strain-sensitivities} (b)), and can suppress spurious phase terms~\cite{Hogan2011}.  
Depending on the implementation, one laser link for a vertical~\cite{dimopoulos_atomic_2008} or more for a horizontal setup~\cite{Canuel2018} are required for the geometries.

\subsection{Overview of beam splitters}
Being the cornerstone of atom interferometry, the atom optics essentially does two tasks: imprinting the phase of light field into atoms and manipulating the quantum states of the atoms.
Currently there are three popular choices of atom optics based on stimulated Raman transition~\cite{Kasevich1991}, Bragg diffraction~\cite{Giltner1995}, and Bloch oscillations~\cite{Cadoret2008}.
Recently, there are new development using single-photon transitions in demonstration experiments~\cite{Hu2017} including large momentum transfer~\cite{Rudolph2019}.
The Raman atom optics employs the stimulated Raman transition between the long-lived ground hyperfine states. Two counter propagating light beams give the atoms a momentum transfer \(k_{\mathrm{eff}}\approx 2 k_{\mathrm{l}}\).
Both Bragg diffraction and Bloch oscillations utilize the ac Stark shift induced by the optical lattice to interact the atoms.
In the Bragg case, the atoms are coherently scattered by the spatially periodic potential created by the ac Stark shift and gain the momentum transfer from the lattice.
In the Bloch case, the atoms are put into a moving lattice by slightly detuning the frequencies of two light fields.
Under certain conditions, the atoms will be coherently accelerated by the moving lattice and at the end of each oscillation period the atoms will receive the momentum transfer \(2 \hbar k_{\mathrm{l}}\).
Comparing with Raman atom optics, which affect both the internal and external quantum states, the Bragg and Bloch atom optics will only interact with the external quantum states and therefore the systematic effects caused by the environment can be greatly suppressed.
However, the requirements of the optical power for consecutive Bragg pulse and Bloch lattices are increased to suppress spontaneous emission by having the individual laser frequencies far-off resonant.
The power requirement for higher order Bragg diffraction is even more stringent~\cite{Szigeti2012NJP,Muller2008}.   
All these atom optics can be combined with other techniques including a rapid adiabatic passage to improve the efficiency and reduce the contrast loss~\cite{Weitz1994,Peik1997,Kovachy2012}.
One major research direction for atom optics is the large-momentum-transfer optics~(LMT) because the sensitivity of an atom interferometer is typically proportional to the transferred momentum.
Several LMT schemes including consecutive pulses sequence~\cite{Jaffe2018,McGuirk2000}, higher-order atom optics~\cite{Muller2008}, Bloch oscillations~\cite{Cadoret2008}, the combination of schemes~\cite{Gebbe2019arxiv,Muller2009PRL}, and dual-lattice beam splitters~\cite{Pagel2017arXiv} have been demonstrated.
So far the highest LMT separating the trajectories in an atom interferometer utilises the combination of double Bragg diffraction~\cite{ahlers2016} and a twin lattice driving Bloch oscillations for symmetric beam splitting to reach an effective wave number of \(k_{\mathrm{eff}}=408\ k_{\mathrm{l}}\)~\cite{Gebbe2019arxiv}, where $k_{\mathrm{l}}=2\pi/(780\,\mathrm{nm})$ is the wave number for a single photon.
By mitigating the technical limitations due to spontaneous emission and intensity inhomogeneities across the atom-light interaction zone, effective wave numbers of $k_{\mathrm{eff}}=1000\,k_{\mathrm{l}}$ appear to be feasible for ELGAR.

\subsection{Suppression of spurious phase terms}
In a single-loop geometry, rotations and gravity gradients give rise to phase terms $2\,\mathbf{k}_{\mathrm{eff}} \cdot (\mathbf{v}_0 \times \boldsymbol{\Omega}) T^2$ and $\mathbf{k}_{\mathrm{eff}}^\text{T}\,\Gamma{T^2}(\mathbf{x}_0 + \mathbf{v}_0 T)$ with couplings to the initial central position $\mathbf{x}_0$ and central velocity $\mathbf{v}_0$~\cite{Bongs2006,Hogan2009}. 
Here $\boldsymbol{\Omega}$ denotes the angular velocity and $\Gamma$ is the gravity gradient tensor, which corresponds to minus the Hessian of the gravitational potential $U(\mathbf{x})$ with components $\Gamma_{ij} = - \partial^2 U / \partial x^i \partial x^j$.
Moreover, we have employed a convenient vector-matrix notation where boldface characters correspond to vector quantities and the superscript T denotes the matrix transpose.
If the gravity gradient is known, the corresponding phase term can be suppressed in a gradiometric configuration by adjusting the wave number of the mirror pulse to $\mathbf{k}_{\mathrm{eff}}+\Delta{\mathbf{k}_{\mathrm{eff}}}$ with $\Delta{\mathbf{k}_{\mathrm{eff}}}=(\Gamma{T^2}/2)\,\mathbf{k}_{\mathrm{eff}}$~\cite{Roura2017PRL}.
The Sagnac term $2\,\mathbf{k}_{\mathrm{eff}} \cdot (\mathbf{v}_0 \times \boldsymbol{\Omega}) T^2$, however, remains and implies stringent requirements on the residual expansion rate which is connected to the mean velocity~\cite{Schubert2019arxiv}.
This challenge is mitigated by multi-loop geometries.
A double-loop geometry retains sensitivity to DC rotations, but decouples the leading order term from the initial velocity~\cite{Hogan2011,Canuel2006,Dubetsky2006PRA}.
Contributions from gravity gradients can again be suppressed by adjusting the wave number of the two mirror pulses~\cite{Roura2017PRL}.
Further details are provided in sec.~\ref{sec_other_noise_couplings}.
To first order, the triple-loop scheme suppresses both the coupling of rotations and gravity gradients to the initial conditions and shows no sensitivity to DC accelerations and rotations for adjusted wave numbers of the three mirror pulses~\cite{Hogan2011}, but requires an additional atom-light interaction zone as compared to the other two geometries if no techniques for suspension or relaunch are employed.

\subsection{Broadband and resonant detection modes}\label{sec_operating_mode}
Intrinsically, the transfer function of an atom interferometer features peaks at frequencies corresponding to multiples of the pulse separation time~\cite{Cheinet08}, as depicted in Fig.~\ref{fig:atom-optics-strain-sensitivities}~(a).
For the detection of gravitational waves, this can be an undesired feature if the frequency of the gravitational wave is not known, and the pulse separation time cannot be adjusted accordingly.
The solution is to drive $m$ interleaved interferometers~\cite{Savoie2018,Biedermann2013PRL} with different pulse separation times, so that the peaks are slightly shifted between subsequent cycles, effectively leading to a broadband detector with a flat response over a chosen frequency range~\cite{Hogan2016PRA} at the cost of a factor of $\sim\sqrt{m}$ in the strain sensitivity (Fig.~\ref{fig:atom-optics-strain-sensitivities}~(b)).
Provided, that the signal of a gravitational wave is identified, the pulse separation time can be adjusted to its frequency to maximise the response.
In addition, the geometry can be extended by adding additional pulses before the final recombination pulse, so that the total number of loops gets multiplied by an integer factor $b$ Fig.~\ref{fig:atom-optics-strain-sensitivities}~(b).
This leads to an amplification of the response by the multiplication factor $b$ and corresponds to a resonant detection mode~\cite{Graham2016PRD}.
Depending on the implementation, e.g. the size of the vacuum vessel or the beam diameter in a horizontal detector, the tuneability of the pulse separation time $T$ will be limited.
The resonant detection is also affected by the total free-fall time that the vacuum vessel can support, may need relaunching for a ground detector, requires highly efficient beam splitters due to the added pulses, and has implications for the detection due to the extended time of flight.
For pulse separation times $T=200\,\mathrm{ms}$ and $b=3$, the total time of flight would be $2.4\,\mathrm{s}$, constraining residual expansion rates to $\sim100\,\mu$m/s, which requires a well-collimated atomic ensemble~\cite{Rudolph2016Diss,Kovachy2015PRL,Muntinga2013PRL}.

\subsection{Vertical and horizontal arms}
In a vertical arm~\cite{Coleman2018arXiv,dimopoulos_atomic_2008}, the free fall of the atoms is aligned with with the axis of the laser link for coherent manipulation.
This implies a tuneability in the pulse separation time $T$ enabling the broadband detection mode, adjusting $T$ to the frequency of interest, and the possibility of resonant detection within the limit of the area in which the atoms can efficiently be manipulated.
An interleaved mode requires a labelling of the concurrent interferometer, e.g. by different Doppler shifts~\cite{Hogan2016PRA,Hogan2011}.
The accessibility of deep boreholes may limit the maximum baseline $L$, in the case of Ref.~\cite{Coleman2018arXiv} reported to 1\,km.
If beam splitting techniques other than single-photon transitions~\cite{Graham2013PRL} are implemented, this implies the requirement of an additional horizontal arm to suppress laser frequency noise.
For horizontal arms, baselines of several kilometres as in LIGO~\cite{Abbott2016} or VIRGO~\cite{Acernese2014} and possibly more~\cite{Chaibi2016} appear feasible.
This relaxes the requirements on the beam splitting order and the intrinsic phase noise of the interferometer.
In the horizontal configuration, the atoms travel orthogonal to the beam splitters, constraining the minimum beam size for efficient manipulation, and defining the pulse separation $T$ if more than one atom-light interaction zone is required.
Depending on the chosen geometry, this may limit the possibilities of a broadband or resonant detection mode.
An advantage of spatially separated atom-light interactions zones which can address atoms only at specific times of flight is the easier accommodation of interleaved operation~\cite{Savoie2018}.

\subsection{Double-loop geometry}\label{sec_double_loop}
The double-loop geometry~\cite{Marzlin1996PRA} consists of an initial beamsplitting pulse, two mirror pulses which invert the momenta, and finally a recombination pulse.
\begin{figure}
\centering
\includegraphics[width=0.48\linewidth]{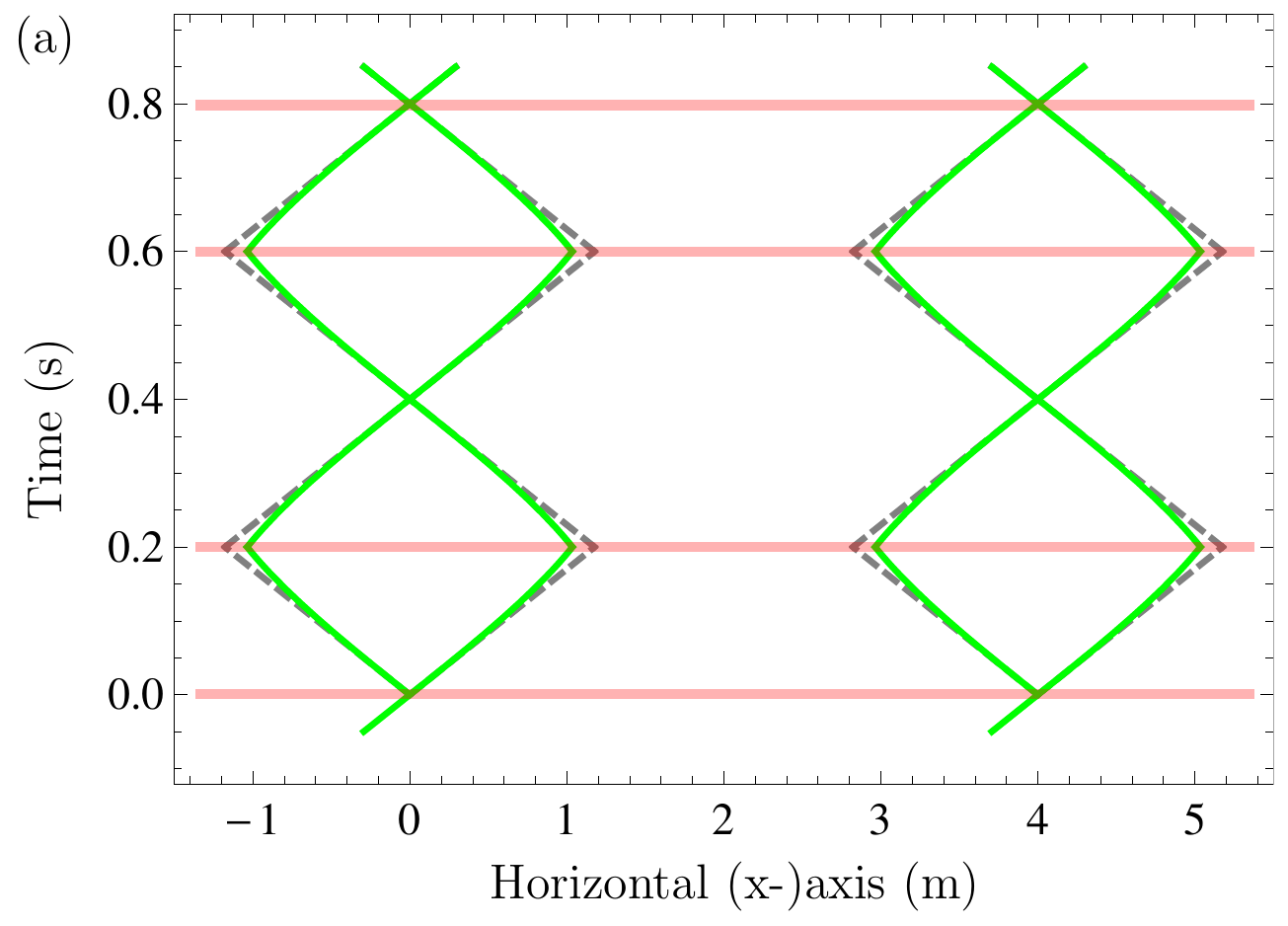}
\includegraphics[width=0.48\linewidth]{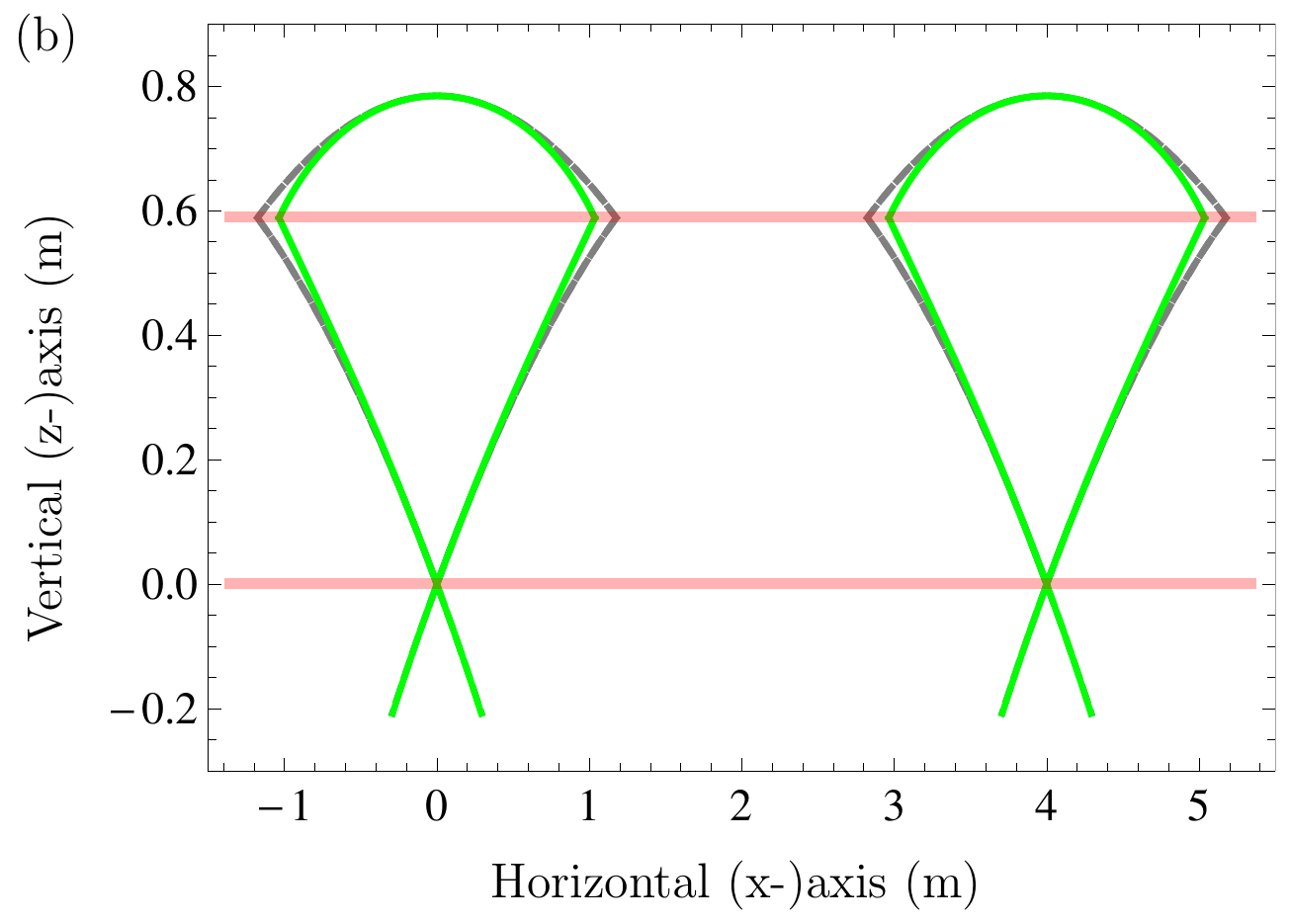}
\caption{Sketch of the double-loop geometry. Two interferometers (gray dashed / green lines) are manipulated by the same beam splitters (red lines).
Gray dashed lines correspond to the central trajectories in absence of gravity gradients, and green lines to the central trajectories with gravity gradients and the compensation technique.
(a) Time evolution of the trajectories in horizontal axis.
(b) Trajectories in vertical and horizontal axis. Herein, the input port and one output port overlap (see also Fig.~\ref{fig:expsch}).
For clarity, the impact of the gravity gradient is exaggerated, and the baseline between the two interferometers is shortened.
}
\label{fig:atom-optics-double-loop-geometry}
\end{figure} Herein, the pulse spacing is $T$ -- $2T$ -- $T$, leading to a total interferometer time of $4T$.
A sketch of the geometry is shown in Fig.~\ref{fig:atom-optics-double-loop-geometry}.
Neglecting other terms, the differential phase between two double-loop interferometers induced by an impinging gravitational wave is $8k_{\mathrm{eff}}hL\sin^2(\omega{T}/2)\sin(\omega{T})$~\cite{Hogan2011}. 
As discussed previously, the double-loop geometry suppresses spurious phase shifts, especially when adjusting the wavenumber of the two central mirror pulses to cancel effects of gravity gradients~\cite{Roura2017PRL}.
To implement the double-loop geometry~\cite{Savoie2018,Hogan2011,Canuel2006,Dubetsky2006PRA}, the atoms are initially launched upwards and subsequently interact with the two spatially separated beamsplitting zones as depicted in Fig.~\ref{fig:atom-optics-double-loop-geometry} and Fig.~\ref{fig:expsch}.
At the lower zone, the intial beam splitter and the final recombiner are applied, whereas the two mirror pulses are flashed onto the atoms in the upper zone.
The distance between these two zones defines the free fall time $T$.
Depending on the size of the interaction zones, $T$ may be tuneable within a limited range.
The double-loop geometry is also compatible with the implementation of a single-loop sequence.
Fig.~\ref{fig:atom-optics-strain-sensitivities} compares the strain sensitivities of these schemes and depicts the broadband and resonant detection modes for the case of a double-loop interferometer.

\subsection{Folded triple-loop geometry}
The folded triple-loop geometry is an alternative which features symmetric beam splitters~\cite{Gebbe2019arxiv,Abend2017Diss,ahlers2016} in the horizontal direction and relaunches~\cite{Abend2016PRL} at the intersections of the trajectories so that only a single laser link is required~\cite{Schubert2019arxiv}.
This enables a scalability in $T$, consequently a broadband detection mode~\cite{Hogan2016PRA}, and a resonant detection mode~\cite{Graham2016PRD} by adding additional relaunches and beam splitting pulses.
In addition, the triple-loop geometry is robust against fluctuations of the mean position and velocity of the wave packet entering the interferometer.
The scheme requires a pointing stability of the relaunch vectors at the level of $\sim$\,prad, comparable to the requirement on the initial launch vector in a single-loop geometry~\cite{Schubert2019arxiv}.
Omitting other terms, the differential phase shift between two triple-loop interferometers in a gradiometric configuration caused by a gravitational wave is $8k_{\mathrm{eff}}hL\sin^4(\omega{T}/2)\left[ (7+8\cos(\omega{T}))/2 \right]$~\cite{Hogan2011}.
A shot-noise limited strain sensitivity curve is shown in Fig.~\ref{fig:atom-optics-strain-sensitivities}~a) (dash-dotted blue line).
As an option, this geometry can be implemented at a later stage for broadband and resonant detection modes.

\subsection{Beam-splitter performance estimation}
Common to all configurations described here is the need for large momentum transfer of around $k_{\mathrm{eff}}=1000\,k_{\mathrm{l}}$ while the atoms are moving perpendicular to the light beams at velocities in the order of a few m/s. Supposing atom-light interaction in the order of several $\mu$s per transferred photon recoil implies centimeter beam waists to cover the motion of the atoms. In the following, we estimate the power and waist of the beam required to realize 1000\,$k_{\mathrm{l}}$ momentum transfer for transverse atomic motion of $v_a = 4$~m/s, with numerical models both for sequential first-order Bragg diffraction and for accelerated optical lattices driving Bloch oscillations. 

\begin{figure}
\centering
\includegraphics[width=1\linewidth]{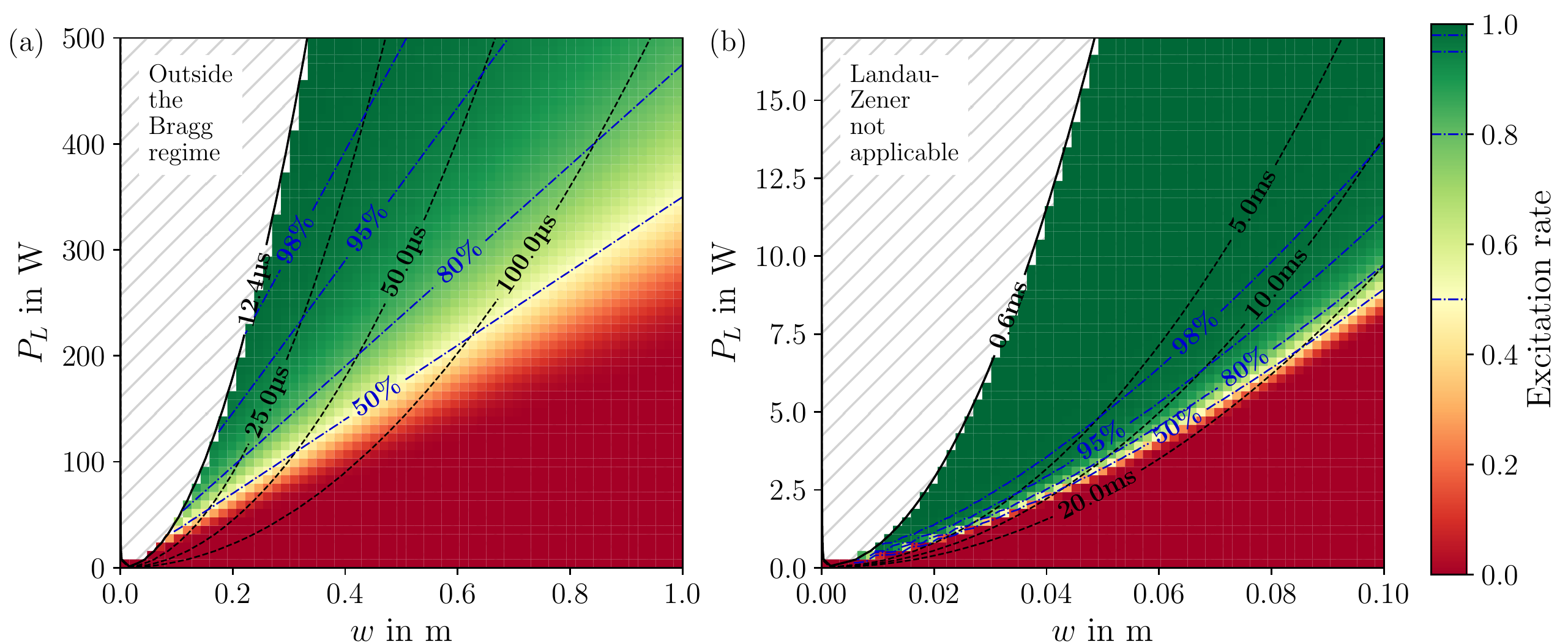}
\caption{Estimated efficiency of a $1000\,k_{\mathrm{l}}$ momentum transfer, assuming a transverse atom velocity of 4~m/s, as a function of laser waist $w$ and power $P_L$. The dashed blue lines denote parameters resulting in equal probabilities, which value is written on the line. \textbf{a)} 500 subsequent first-order Bragg pulses. The dashed lines denote the pulse duration of a $\pi$-pulse, which is determined by the peak Rabi frequency $\Omega_0=\Omega_0(P_L,w)$. The gray area excludes the parameters outside the deep Bragg regime. \textbf{b)} Accelerated Bloch lattice. The dashed lines denote the full duration of the acceleration sequence. The gray area excludes parameters for which the Landau-Zener model is not valid.}
\label{fig:bs_eff}
\end{figure}

To this end, we first consider an effective two-level system in the deep Bragg regime, which is the case when Rabi frequencies are small compared to the recoil frequency~\cite{Mueller2008}. The efficiency  $P_i$ of a single Bragg pulse is determined by the atoms' position at time $t_i$ within the transverse intensity profile of the beam. Therefore, we evaluate the integral~\cite{CheinetPhD}
\begin{equation}\label{eq:p_i}
P_{i} = \int dr \frac{1}{\sqrt{2\pi}\sigma_r}e^{-\frac{(r-r_0(t_i))^2}{2\sigma_r^2}} p(r)
\end{equation}
to weigh the spatially dependent excitation rate
\begin{equation}
p(r) = \sin^2\left(\frac 1 2 \int_0^\infty \Omega(t,r)~dt\right)
\end{equation}
with the density distribution of an atomic cloud of width $\sigma_r$, which is centered around $r_0(t_i)=r_0(0)+v_a t_i$. $\Omega(t,r)$ is the Rabi frequency given by the transverse intensity distribution of the light beam and the temporal shape of the light pulse~\cite{szigetiPhD}. Here, we consider Gaussian distributions in time as well as in position. The total efficiency of a $1000\,k_{\mathrm{l}}$ beam splitter is calculated by multiplying the individual single-pulse excitation rates. 
Evidently,  the pulses, for which the atomic density distribution is centered around the wing of the beam, are less efficient than the ones where the cloud passes its center. As a consequence, the beam waist needs to be sufficiently large to accommodate the atoms at all times, with appropriately enhanced power. Fig. \ref{fig:bs_eff} a) illustrates the required parameters range. Typically, efficient Bragg pulses are realized with pulses of about $100~\mu$s duration, such that the atoms are expected to travel about 20~cm perpendicular to the beam. Indeed, in order to provide a sufficient light intensity over this distance, the required beam waist is in the order of several 10~cm, which in turn calls for a lot of power. 

Moreover, we similarly study the performance of a beam splitter based on Bloch oscillations in a moving optical lattice with an adapted Landau-Zener model. To account for the inhomogeneities given by the beam divergence and atomic motion, we decompose the coherent lattice acceleration into single Bloch oscillations, of which the efficiency is quantified by 
\begin{equation}
\eta_{LZ}(r) = 1-\exp\left(-\frac{\pi \Omega_{bg}^2(r)}{2\alpha}\right). 
\end{equation}
The band gap $\Omega_{bg}(r)$ is numerically determined through the energy difference of the lowest Bloch bands at the boundary of the Brillouin-zone~\cite{Abend2017Diss}, and is a function of beam intensity. The lattice acceleration time $\tau_{acc}$, which enters the lattice chirp rate $\alpha = 1000\,k_{\mathrm{l}}/\tau_{acc}$, is chosen such that $1000\,k_{\mathrm{l}}$  are transferred with 99\% probability at the center of the beam. 
By replacing $p(r)$ with $\eta_{LZ}(r)$ in Eq.~\ref{eq:p_i}, we can convolute the time-varying spatial distribution of the atoms with the inhomogeneous excitation rate. Fig.~\ref{fig:bs_eff} b) displays the results of this study. Notably, accelerated Bloch lattices transfer momentum in a shorter time than Bragg pulses, namely in the order of a few $10~\mu$s per photon recoil. As a consequence, the total momentum transfer can be achieved faster, such that the transverse distance travelled by the atoms is around one order of magnitude smaller as compared to Bragg. For similar intensities $I\sim P/w^2$, the required power is hence relaxed by two orders of magnitude. Comparing the requirements in power and waist of the two operation modes a) and b) of Fig.~\ref{fig:bs_eff}, one concludes that achieving the needed beam splitting is more realistic when Bloch lattices are implemented.  

\subsection{Conclusion on geometry and beam-splitting type}
Motivated by the available sites for ELGAR, the possibility to exploit baselines of several kilometres, and the differential suppression of laser frequency noise between two arms, the horizontal configuration is favoured.
To date, the highest momentum transfer between the two trajectories of an interferometer was realised by the combination of Bragg transitions and Bloch oscillations, and an effective wave number of $k_{\mathrm{eff}}=1000\,k_{\mathrm{l}}$ can be expected.
A double-loop interferometer with a gravity gradient compensation scheme offers robustness against spurious phase terms.
Therefore, the baseline design foresees a double-loop interferometer with a gravity gradient compensation scheme, an effective wavenumber of $k_{\mathrm{eff}}=1000\,k_{\mathrm{l}}$, a baseline $L=16.3\,\mathrm{km}$, and a pulse separation time $T=200\,\mathrm{ms}$ for an intrinsic phase noise of $10^{-6}$\,rad/$\sqrt{\mathrm{Hz}}$.
This design keeps the possibility open to implement other geometries such as the single-loop or folded triple-loop interferometer.
A vertical arm can be added in addition, with a trade-off in sensitivity due to a shorter baseline.
The latter is expected to prevent the full exploitation of the rejection scheme~\cite{Chaibi2016} for Newtonian noise.

\section{Atom source}\label{sec_Atom_source}
The design of the atomic source is essential for the successful operation of the  interferometer as it determines the sensor sensitivity as well as the susceptibility to environmental effects. 
The intrinsic noise of a two-mode sensor with uncorrelated input states is characterized by the standard quantum limit (SQL) $\delta \phi_{\rm{SQL}} \sim 1/\sqrt{n_{\rm{meas}}~N}$, scaling with the number of atoms $N$ and interferometric measurements $n_{\rm{meas}}$. Therefore, the generation of large ensembles of atoms at a fast rate is desirable to attain sufficient stability level in a shortest possible integration time. Matching the target sensitivity of $\SI{1}{\mu rad/\sqrt{Hz}}$ for a single atom interferometer operating at SQL, requires the flux of $10^{12}$ atoms per second.

In order to mitigate various systematic and statistical contributions to the measurement uncertainty, it is necessary to engineer the external (spatial and momentum distribution) as well as the internal (e.g. magnetic) states of the atoms. The interferometry scheme under consideration requires the uncertainties in the center-of-mass position and velocity of the ensemble to be kept below $\SI{400}{nm}$ and $\SI{3}{nm/s}$, respectively, given by coupling of wave front imperfections, beam misalignment and pointing jitter etc. to the phase-space properties of the atoms (see section \ref{sec_other_noise_couplings}). Therefore, the careful design of the atomic source lies at heart of both experimental and theoretical efforts as will be outlined in the following.

\subsection{Species choice}
Cold-atom experiments typically employ alkaline species (Li, Na, K, Rb, Cs) as they offer multiple pathways to ultra-cold temperatures and quantum degeneracy.
Being the workhorse solution for atom interferometry for a wide range of applications, Rb is the prime choice for the gravitational wave antenna under consideration and is the work hypothesis for the entire paper. State-of-the-art atom optics techniques, as outlined in the previous section, have been primarily explored with Rb, which provides comfortable wavelengths in the optical range. Moreover, the lowest effective expansion energies in the order of a few tens of pK, expressed in temperature equivalents, have been demonstrated with Rb~\cite{Kovachy2015PRL,Rudolph2016Diss}. 
Another promising candidate is Cs, as employed for example in experiments for measurements of the hyperfine structure constant~\cite{Parker2018}, where efficient large momentum transfer was demonstrated.

Alkaline-earth and alkaline-earth-like atoms such as Yb, Sr and Ca are routinely utilized in optical atomic clocks due to the very narrow linewidth of their intercombination transition. 
Additionally, their immunity to quadratic Zeeman shifts is a distinct asset and renders them auspicious for high-precision experiments.
Indeed, these species are well-suited~\cite{Loriani2019} for space-borne gradiometers with large baselines based on single-photon transitions, which mitigate laser phase noise to serve as gravitational wave antennas~\cite{Graham2013PRL, Hogan2016PRA}. In these configurations, the baselines in the order of $10^9$~m alleviate the need for ambitious atom optics, requiring a few photon recoils per beam splitter only.
The distinct advantage of these schemes over two-photon excitation mechanisms is the inherent mitigation of laser phase noise, which is common-mode in a gradiometric setup. Schemes based on Raman- or Bragg-diffraction are, even in a gradiometric setup, prone to laser phase noise due to the finite speed of light, which prohibits simultaneous interaction of the two interferometers with the light over large baselines~\cite{Graham2013PRL}.
Recently, large momentum transfer through consecutive application of single-photon $\pi$-pulses was realized~\cite{Rudolph2019} in $^{88}$Sr on the weakly-allowed $^1$S$_0$--$^3$P$_1$ transition. However, the extension to higher momentum transfer is limited by the finite life time of the excited state, such that working on the narrower $^1$S$_0$--$^3$P$_0$ line is required. In bosons, this forbidden transition is enabled by application of a static magnetic field~\cite{Taichenachev2006}. However, the need to reach reasonable Rabi frequencies and to mitigate the second-order Zeeman shift put strict requirements on the magnitude and stability of the magnetic field~\cite{Hu2019}. In the fermionic isotopes, the intercombination line is allowed due to the nuclear spin and they seem to be a more promising solution in the long term despite their low isotopic abundance. Overall, in terms of heritage, the prime candidates for for ambitious experiments based on alkaline-earth species are Sr and Yb, both of which are being explored in large fountain interferometers~\cite{Hartwig2015,Rudolph2019}.

\subsection{Atomic source preparation}
Atomic sources for alkaline atoms are routinely operated in a double MOT configuration.
While a 2D+MOT can provide a high flux of atoms loaded from background vapor, either fed by dispensers or ovens, the 3D MOT is separated via a differential pumping stage. 
This has the advantage of maintaining good ultra-high vacuum in the main experimental chamber. 
Current 2D+MOT setups achieve loading rates in the order of $10^{11}$ atoms per second~\cite{Rudolph_2015}.
In the case of alkaline-earth (like) elements the 2D+MOT may be exchanged with a Zeeman slower, while the source has the same functionality and can achieve a comparable flux.
In the 3D MOT, atoms are first cooled down to the Doppler temperature, followed by sub-Doppler cooling in a molasses configuration reaching almost recoil temperature. Cooling protocols based on gray molasses are inherently faster, and maintain a higher atomic density~\cite{Rosi2018,Naik2019}. Alkaline-earth (like) atoms can be directly cooled on a narrow line achieving much lower Doppler temperatures. 
The laser-cooled sample is then loaded either in a magnetic or optical trapping potential, where the sample is evaporated to quantum degeneracy. 
The fastest evaporation rates are so far achieved either using atom chips~\cite{Rudolph_2015} or painted optical potentials \cite{Roy_2006,Condon2019} reaching condensation in $\simeq$1~s or less.
There are also recent examples of experiments which reach quantum degeneracy with direct laser-cooling, for example, by Raman sideband cooling in an optical lattice~\cite{Urvoy_2019}, but these do not reach so far performances comparable to conventional methods.
After reaching quantum degeneracy the samples expansion rate can be further reduced by delta-kick collimation.
Residual expansion rates in the pK regime have been previously achieved~\cite{Rudolph2016Diss,Kovachy2015PRL,Muntinga2013PRL}.
If the chosen isotope provides a magnetic sub-structure, the atoms may be transferred into the non-magnetic sub-state by pulsed or chirped RF fields prior to further manipulation.

From the region where the atomic sample has been prepared, it has to be transported to the interferometry region.
Atomic transport via shifting or kicking in trapping potentials suffers of several limitations due to required adiabaticity and delicate experimental implementation~\cite{Amri_2019}.
Therefore, the atomic transport to reach large enough distances away from the atom chip via an optical lattice is the preferred option, although requiring additional optical access. 
Bloch oscillations in an optical lattice~\cite{Peik1997} can be driven with very high efficiency and due to the discrete momentum transfer, the transport can be controlled very.

\subsection{Atomic Flux}
The high degree of control over the atomic ensemble required to mitigate stochastic and systematic effects ultimately suggests the use of quantum degenerate sources. 
However, the flux of condensed atoms is arguably one of the most challenging aspects of the present proposal. In order to reach target performance of $\SI{1}{\mu rad/\sqrt{Hz}}$ phase sensitivity, a flux of $10^{12}$ atoms per second is required in shot-noise limited operation. This exceeds state-of-the-art flux \cite{Rudolph2016Diss} by six orders of magnitude; however, there are several promising pathways for improvement, and their combination is expedient to reach that goal. 
The natural effort is a higher atom number per experimental shot. Multiple techniques aim at increasing significantly the atom number before evaporation, such as improved mode-matching through time-averaged potentials. At the same time, in the present scheme, the preparation and interrogation zones in the interferometer are spatially separated, allowing for the concurrent operation of multiple interferometers~\cite{Savoie2018}. This might be accompagnied by multiple source chambers and a system, which transfers the atomic sources to the interferometric region, for example based on moving optical lattices~\cite{Dahan1996}. Instead of a mere increase of the atom number, the phase sensitivity can  be also increased by employing entangled atomic ensembles (section \ref{subsec:squeezing}). Eventually, the realisation of an atom source needs a careful trade-off between sensor performance requirements and technical possibilities in terms of flux. Ultra-cold sources close to degeneracy might suffice to comply with the requirements and naturally offer a higher atom number and shorter preparation time as investigated in~\cite{Loriani2019}.

\subsection{Atomic sources for entanglement-enhanced interferometry}
\label{subsec:squeezing}
The requirements on the atomic flux can be reduced by employing atomic sources that can surpass the SQL. 
Sensitivities beyond the SQL can only be reached with nonclassical input states.
Such entangled atomic source can, to date, be generated by two main methods: Atom-light interaction (quantum non-demolition measurements or cavity feedback) or atomic collisions~\cite{Pezze2018}.
The best results so far were obtained with cavity feedback~\cite{Hosten2016, Cox2016} and allowed for a demonstrated enhancement of 18.5 dB, which is equivalent to a 70-fold increase of the atom number.
These techniques were so far only demonstrated with thermal ensembles, the best results with Bose-Einstein condensates were obtained with atomic collisions~\cite{Hamley2012} with a metrological gain of 8.3 dB.
A technologically interesting concept, that is also followed in laser interferometry, is the application of squeezed vacuum~\cite{Kruse2016}, where only the formerly empty input port is squeezed.
The development of a reliable, high-flux, quantum degenerate atomic source with a metrological gain of 20~dB for a 100-fold reduction of the required atom number presents an active research goal.

\subsection{Source engineering and transport} 

The operation of the atom interferometer requires the transport of the atoms from the cooling and evaporation chamber to a science or interferometry chamber where several interleaved interferometers are occurring. In order not to affect the cycling time, and hence the performance of the interferometer, this transport has to be time efficient. To this end shortcut-to-adiabaticity protocols, as proposed in~\cite{Corgier2018} and implemented in~\cite{Rudolph2016Diss}, will be engineered. These fast transports induce very low residual dipole oscillations in the final traps, with a typical amplitude of the order of 1\,$\mu$m. If a quantum control is needed i.e. to ensure the gases are in the ground state of the target traps, optimal control solutions are available and proved to be equally fast~\cite{amri2018} for relevant degenerate atomic systems. 

Moreover, the atomic ensembles need to be slowly expanding with effective temperatures down to 10--100\,pK. This is realized by implementing the so-called delta-kick collimation technique. The atomic gases are freely expanding for a few tens of ms before being subject to an optical or magnetic trap flashed for a few ms that collimates the ensemble~\cite{Chu1986,Amman1997}. This drastically reduced the expansion rate of the cloud to the desired level in recent experiments~\cite{Muntinga2013PRL,Kovachy2015PRL,Rudolph2016Diss}.

Moving the atomic ensembles into the interferometry region might require the displacement over rather long distances (few tens of cm) in a short time. This is possible using an accelerated optical lattice driving Bloch oscillations. In a few ms, velocities corresponding to 200 recoils~\cite{Cadoret2008,Gebbe2019arxiv} could be imparted. Thanks to the atomic cloud collimation step, it would be possible to restrict the Bloch beam waists to 1--2 mm keeping the power usage at a reasonable level. In combination with the Bloch beams, symmetric schemes, that reduce the impact of spatially dependent systematics, will be sought for. They would rely on double-Raman-~\cite{Leveque2009} or double-Bragg-~\cite{ahlers2016} diffraction beams. 

\section{Seismic Isolation}
\label{sec_seis}
\bigskip

GW detectors, either using laser interferometry or matter-wave interferometry, share the same basic principle: the distance between two “free-fall” inertial test masses is precisely measured with a highly stable laser in order to detect tiny modulations that can be attributed to a GW. Noise due to seismic vibrations or to fluctuating gravitational forces can induce motions of the test masses, which limit the sensitivity of ground based detectors. For laser interferometry, the “free-fall” inertial test masses are mirrors suspended from high performance vibration isolation systems presently allowing for the detection at frequencies 10 Hz and above. 

In the ELGAR concept a matter-wave interferometer relies on atoms that are naturally in free fall and represent ideal test masses that are naturally isolated from seismic vibrations. Using as test masses free falling atom sources instead of suspended mirrors enables overcoming most of the technical limitations presented by optical GW detectors at low frequency. 
See Eq.~\ref{eq:SHnetwork} for the strain sensitivity of the ELGAR detector.
An examination of Eq.~\ref{eq:SHnetwork} displays an important difference between the AI GW detector sensitivity and that of laser interferometric GW detectors. The GW strain noise, $S^{1/2}_h(\omega)$ is related to the mirror and beam-splitter displacement noise, $S^{1/2}_{\delta x_{M_X}}$ by a factor of $4 \pi f_{GW}/c$. For laser interferometric GW detectors the displacement noise is divided by the arm length (3 km for Virgo and 4 km for LIGO) to produce the noise on the strain signal, so a factor of about $3 \times 10^{-4}$ m$^{-1}$. For the AI GW detectors the factor of $4 \pi f_{GW}/c$, and a frequency of $f_{GW} = 1$ Hz, provides a numerical factor of $\sim 2 \times 10^{-8}$ m$^{-1}$. This is a definite advantage in terms of seismic isolation. 
Still, it will be important to have adequate vibration isolation for the components of ELGAR. 
Indeed, the retro-reflection mirror
acts as an inertial reference for the atom interferometers in the detector. Although the differential signal between two atom interferometers, which contains the signal of the GW, already implies a rejection of mirror movement, a spurious coupling due to imperfections will remain.
Therefore, a sophisticated suspension system will need to be realized.


\subsection{Status of Seismic Isolation Presently for Gravitational-Wave Detectors and Test Systems}


The requirements for the isolation of the Advanced LIGO~\cite{TheLIGOScientific:2014jea} test masses (consisting of the input and end mirrors of the Fabry-Perot cavities in each arm of the interferometer) are such that the longitudinal noise must not exceed $1 \times 10^{-19}$ m Hz$^{-1/2}$ at 10 Hz. The Advanced LIGO test masses are suspended from a quadruple pendulum suspension with three stages of cantilevered blade springs~\cite{Aston_2012}. This configuration isolates the test mass in six degrees-of-freedom for frequencies of 1 Hz and above. In addition, there is an active damping system for suspension rigid body modes below 9 Hz, as acted up from the top stage of the suspension system. This is accomplished with a magnetic actuation system for the upper three levels of the suspension system. At the lowest level, an electrostatic actuation system is used. Much noise can be generated from the friction at the clamping points of the suspension~\cite{doi:10.1063/1.1148692,CAGNOLI1999230}. The use of silica wires has been determined to be the optimal way to minimize these sources of noise; these wires are attached to the mirror by welding or the use of silica bonding. This will replicate the contact between materials at the molecular level~\cite{doi:10.1063/1.1150040,PhysRevLett.85.2442,Grote_2008}. This is called a monolithic suspension~\cite{TheVirgo:2014hva}. The Advanced LIGO test masses are suspended via such a monolithic fused silica assembly. The suspension system is mounted upon seismically isolated optics tables. For the Advanced LIGO beamsplitter, it is a triple pendulum isolation. The requirement for the isolation of the beamsplitter is such that the longitudinal noise must not exceed $6 \times 10^{-18}$ m Hz$^{-1/2}$ at 10 Hz. A comprehensive description of the Advanced LIGO seismic isolation system can be found in~\cite{TheLIGOScientific:2014jea}.

Advanced Virgo has similar seismic isolation goals to that of Advanced LIGO~\cite{TheVirgo:2014hva}. The test masses (mirrors) are suspended in a monolithic suspension configuration. The mirrors are suspended via four wires from the {\it marionette}. The marionette is suspended by a single steel the so called super attenuator.
The super attenuator is a combination (both active and passive) seismic isolation system. The super attenuator for Advanced Virgo reduces the seismic noise by more than 10 orders of magnitude in six degrees of freedom above a few Hz. 
The super attenuator for Virgo consists of three principal systems: an inverted pendulum, a chain of seismic filters, and the mirror suspension.
The inverted pendulum~\cite{doi:10.1063/1.1149783} is made up of three 6-m long aluminum monolithic legs, each of which is connected to ground via flexible joints and supports an inter-connecting structure on the very top~\cite{TheVirgo:2014hva}. 
The chain of seismic filters consists of an 8 m long set of five cylindrical passive filters (each of which will reduce the seismic noise by 40 dB in both the horizontal and vertical degrees of freedom from a few Hz). Magnetic actuators are used to adjust the final alignment of the test masses~\cite{TheVirgo:2014hva}.
A comprehensive summary of the Virgo super attenuator and its performance characteristics are presented in~\cite{ACERNESE2010182}.

Research is already underway to extend the GW detection technology past that of Advanced LIGO and Advanced Virgo. For example, the Einstein Telescope (ET) is proposed to be a third generation ground-based gravitational wave detector with a sensitivity 10 times better than that of Advanced LIGO and Advanced Virgo~\cite{Punturo2010,Hild_2011,ET-design}. The current plan for ET is to have an equilateral triangle configuration, with arms lengths of 10 km. There will be 3 low-frequency interferometers, and three high frequency interferometers~\cite{Punturo2010,Hild_2011,ET-design}. The observational band for ET will start at a few Hz. The design of ET is such that seismic, gravity gradient, and suspension thermal noise will be the limitation at the very low frequencies of 1 - 10 Hz~\cite{Punturo2010}.
The upper part of the ET suspension for the mirrors will the super attenuator, similar to that of Advanced Virgo. This will provide the needed seismic and acoustic isolation. In order to obtain a sufficiently low cut-off frequency the height of the individual pendulum stages of the super attenuator will be 2 m per stage. The proposed modified super attenuator will consist of of six pendulum stages, each stage providing horizontal and vertical isolation). The total height of the super attenuator will be 17 m~\cite{Hild_2011}. The last suspension stage, or payload, will be crucial for setting the thermal noise performance and the mirror control. The payload is the system that couples the test mass to the super attenuator; it compensates the residual seismic noise and steers the mirror through internal forces exerted from the last super attenuator element. The payload is similar to that of Advanced Virgo, namely a marionette, a recoil mass, and the test mass (mirror). The marionette is the first stage used to control the mirror position by means of coil-magnet actuators acting between the last stage upper part of the suspension and the marionette arms, while the recoil mass is used to protect and to steer the mirror. On the mechanical structure electro-magnetic actuators (coils) act on the mirror~\cite{ET-design}. The goal of the ET seismic isolation system will be to have a strain noise sensitivity of $5 \times 10^{-21}$ Hz$^{-1/2}$ at 1 Hz, and $1 \times 10^{-24}$ Hz$^{-1/2}$ at 6 Hz~\cite{Hild_2011}.  

Another important method for observing low-frequency GWs is the Torsion-Bar Antenna (TOBA), which will attempt to observe GWs in the 1 mHz to 10 Hz band~\cite{PhysRevLett.105.161101,Shimoda2019}. The target GW strain sensitivity for TOBA is $10^{-19}$ Hz$^{-1/2}$ at 0.1 Hz. Because a mechanical oscillator behaves as a freely falling mass above its resonant frequency, the low resonant frequency of the torsion pendulum could provide for the observation of GWs, and in the case of TOBA, this would correspond to observational frequencies of a few tens of mHz. The TOBA design is to have two orthogonal bars suspended horizontally by wires. The tidal force from a GW would produce differential rotations on the bars, thereby providing the means to observe GWs through the measurement of the rotation of the bars and observed via interferometric sensors~\cite{Shimoda2019}. Because of the low-frequency attempt to measure GWs, a sophisticated seismic isolation system would be needed. The current suspension and isolation system for TOBA consists of the torsion pendulum suspended from an intermediate mass, and then the whole system is connected to an active vibration isolation table. The goal is to produce a near-term demonstration of a sensitivity of $10^{-15}$ Hz$^{-1/2}$ at 0.1 Hz. See~\cite{Shimoda2019} for more explicit details on the seismic isolation goals for TOBA.

\subsection{Seismic Isolation for Atomic Interferometer Systems}
Vibration isolation systems are extremely important for experiments involving atom gravimeters~\cite{doi:10.1063/1.4919292} and atom interferometers~\cite{doi:10.1063/1.4895911}. A recent study displayed the results of a three-dimensional active vibration isolator (essentially an isolated table) applicable for atom gravimeters. The results were the suppression of vertical vibration noise by a factor of 50, and a factor of 5 for horizontal noise in the 0.1 Hz and 2 Hz band~\cite{doi:10.1063/1.4919292}. 
Experiments like these use a both passive isolation and vibration correction methods~\cite{LeGouet2008}. Another study demonstrated a programmable broadband low frequency active vibration isolation system for atom interferometry~\cite{doi:10.1063/1.4895911}. They started with a passive isolation platform where the vertical resonance frequency was to 0.8 Hz. A digital control system was then activated to provide feedback, and in doing so, the intrinsic resonance frequency of the system was reduced to 0.015 Hz. It was then demonstrated that the use of this active isolation improved the vertical vibration in the 0.01 to 10 Hz band by up to a factor of 500 over the passive isolation. In the end this experiment had a vibrational noise spectrum density as low as $6 \time 10^{-10} g$ Hz$^{-1/2}$in the 0.01-10 Hz band.

\begin{figure}[ht]
\centering
\includegraphics[width=.8\linewidth]{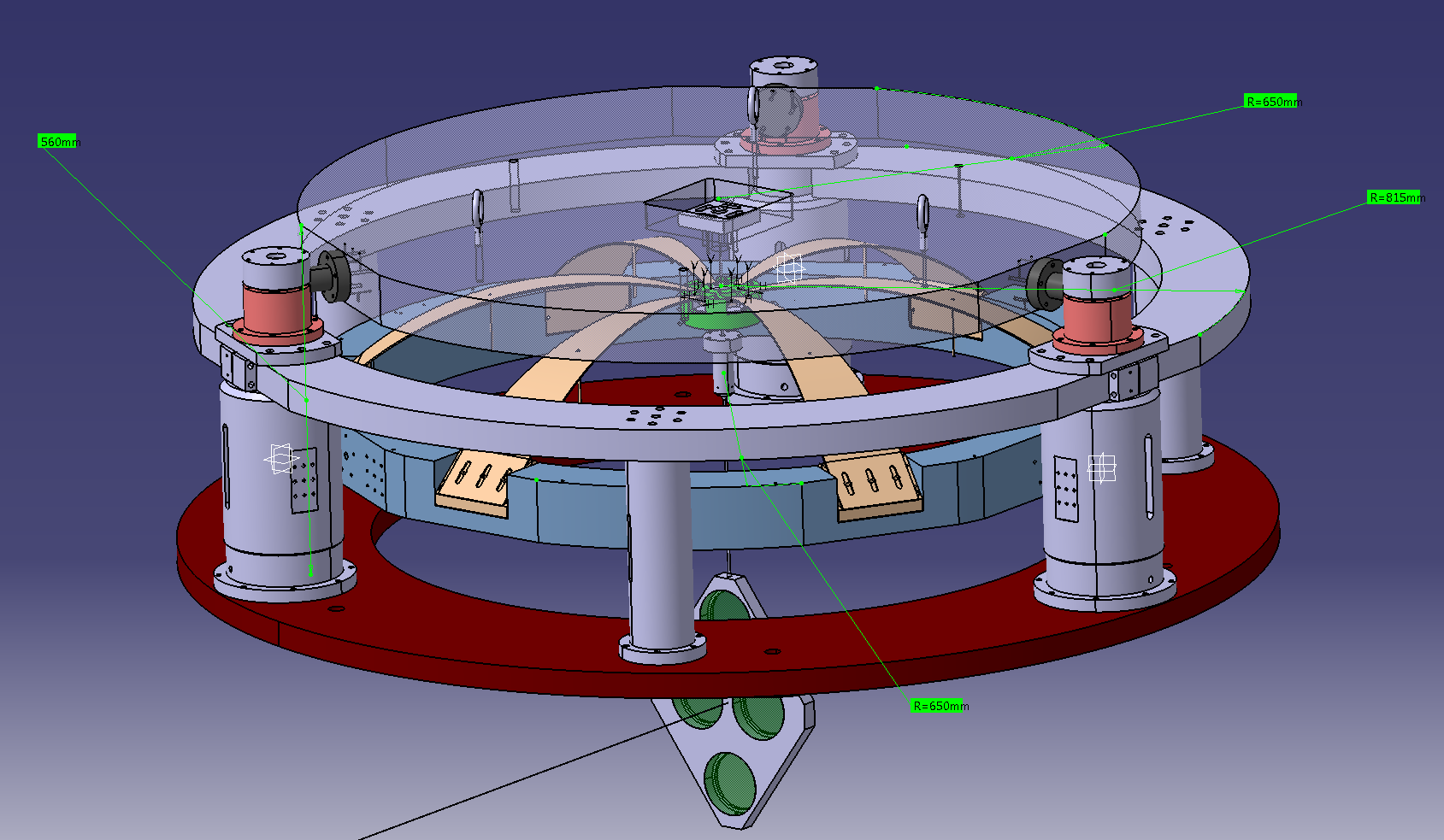}
\caption{A drawing of the MIGA seismic isolation system for its beamsplitter and mirrors. This system has been constructed and its performance is presently being measured. The system consists of two 20 cm pendula that provide isolation for horizontal noise, and pre-constrained blades for vertical isolation.}
\label{fig:MIGA_seismic}
\end{figure}

There is already a very active research program for seismic isolation for atom interferometric GW detectors. As part of the research for the matter-wave laser interferometer gravitation antenna (MIGA) there has been the design and construction of an isolation system for the MIGA mirrors and beamsplitter~\cite{MIGO-Canuel-2014}. The prototype is a similar to Virgo's super attenuator isolation system~\cite{ACERNESE2010182}. The MIGA prototype isolation system has two 20 cm pendula that provide isolation for horizontal noise, and pre-constrained blades for vertical isolation. A drawing of this system can be seen in Fig.~\ref{fig:MIGA_seismic}. The characteristics of this system are presently being measured and its actual performance will be published in the near future.

\subsection{Requirements for the suspension of ELGAR's optics}
The motion of the retroreflecting mirrors and the beamsplitter will contribute to the noise of the detector, and could be a limitation on its sensitivity.
As can be seen from Eq.~\ref{eq:SHnetwork}, the spectral density of the optics motion, $S_{\delta x_{M_X}}$, will be the limity on the sensitivity if
\begin{equation}\label{eq:Sh_mirror}
S_{h}=\frac{4\omega^{2}}{c^{2}}S_{\delta x_{M_X}}(\omega).
\end{equation}
Consider a target sensitivity of $3.3 \times 10^{-22}$ Hz$^{-1/2}$ at 1 Hz.
The requirement for on the noise for the motion of the retroreflecting mirror is be to less than $8 \times 10^{-15}$m Hz$^{-1/2}$ at 1 Hz.
Similarly, for a target of a strain sensitivity of $2 \times 10^{-19}$ Hz$^{-1/2}$ at 0.1 Hz, the limit on the retroreflecting mirror noise would be $5 \times 10^{-11}$m Hz$^{-1/2}$.
To reach such performances, the ELGAR detector will need to take advantage of innovative and effective seismic isolation techniques. Possible scenarios are discussed below.


\subsection{Possible Suspension Designs}
The requirements for the development of adequate vibration isolation system will be challenging, but not impossible. This will be an active and important research area for the types of AI GW detectors discussed in this paper. One could naively imagine, for example, a double suspension~\cite{doi:10.1063/1.1142030} with a resonance frequency at 10 mHz. Such a system would have a transfer function for seismic noise of $10^{-4}$ at 0.1 Hz,  $10^{-8}$ at 1 Hz, and  $2 \times 10^{-12}$ at 10 Hz. However a simple pendulum with a resonant frequency of 10 mHz would require a length of 25 m, which becomes both impractical and expensive. Other solutions must clearly be found.

\begin{figure}[ht]
\centering
\includegraphics[width=.8\linewidth]{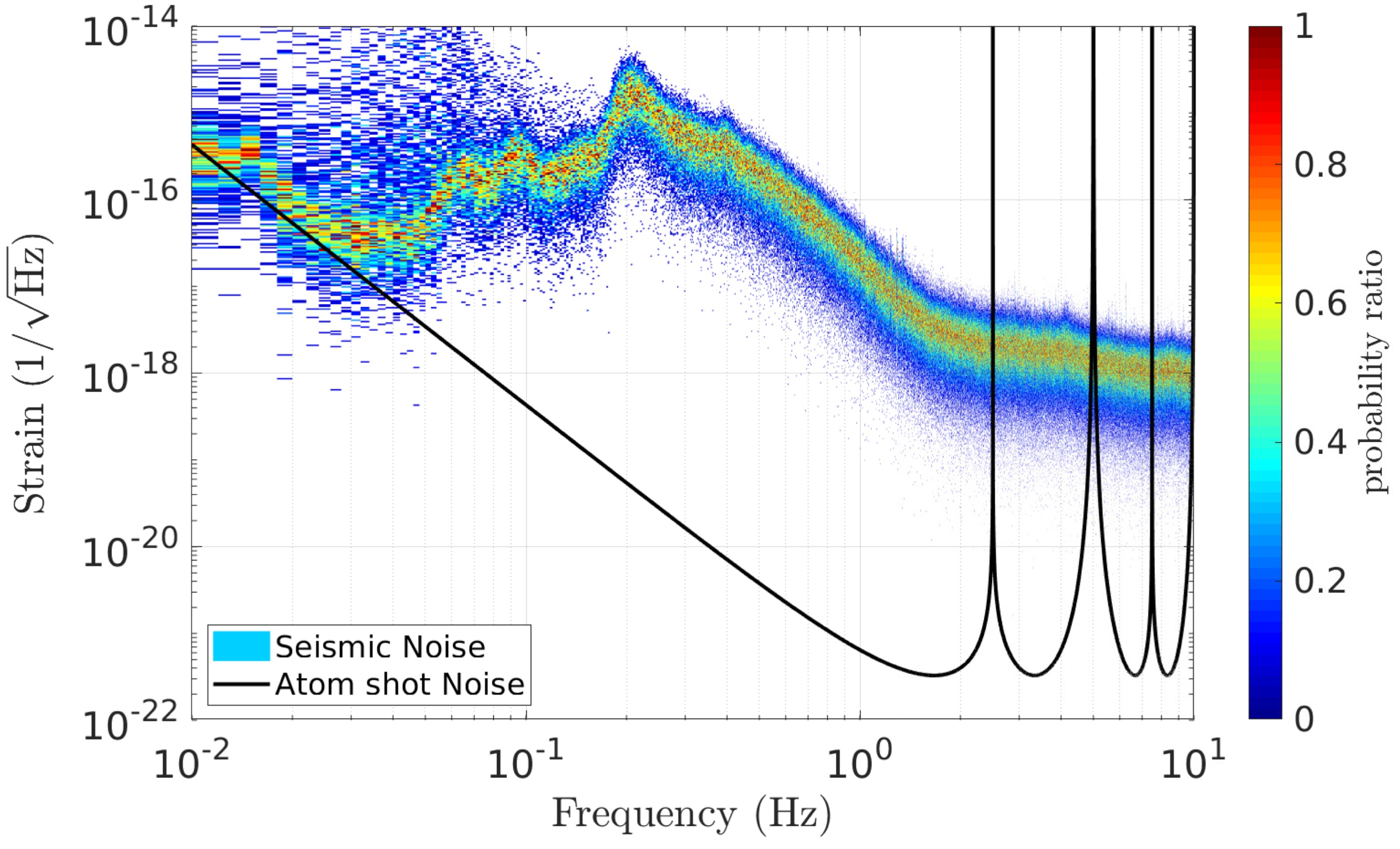}
\caption{An example of input seismic noise (as measure in the galleries of the Laboratoire Souterrain \`a Bas Bruit, Rustrel, France) is displayed in colors. The black curve shows the estimated atom shot noise.}
\label{fig:seismic}
\end{figure}

As an example of the isolation requirements derived by considering the desired sensitivity and the seismic noise of a quiet location, see Fig.~\ref{fig:seismic}. The seismic noise was measured in the galleries of the Laboratoire Souterrain \`a Bas Bruit, Rustrel, France. Also presented is the target sensitivity of ELGAR. The isolation transfer function required is approximately $1.4 \times 10^{-3}$ at 0.1 Hz, $4 \times 10^{-5}$ at 1 Hz, and $3 \times 10^{-4}$ at 10 Hz. 

One promising strategy is to adapt the design developed for the Australian International Gravitational Observatory (AIGO)~\cite{AIGO_2009,doi:10.1063/1.3250841}. 
The AIGO Compact vibration isolation and suspension system demonstrated two stages of horizontal pre-isolation and a single stage of vertical pre-isolation with resonant frequencies around 100~mHz. 
The demonstrated system contained multiple pre-isolation stages, self-damping pendulums, Euler springs~\cite{Winterflood_2002}, and niobium ribbons suspension. There were three pre-isolation stages consisting of an inverse pendulum for horizontal pre-isolation, a LaCoste linkage for vertical pre-isolation~\cite{274fe9face5f4a5fbcd42fb43eb7db16}, a second stage of horizontal pre-isolation based on a Roberts linkage~\cite{doi:10.1063/1.1583862}. At the end of the suspension system there was a a control mass which interacted with the test mass via niobium flexures~\cite{AIGO_2009,doi:10.1063/1.3250841}. This demonstration from AIGO was not quite at the lower frequencies needed by ELGAR, but the methods presented there could possibly be further extended toward lower frequencies. For example, an Euler-LaCoste linkage has been demonstrated for vertical isolation with a resonance frequency of 0.15 Hz~\cite{274fe9face5f4a5fbcd42fb43eb7db16}; the same group has also constructed a coil spring based LaCoste stage for vertical isolation with a resonant frequency of 0.05 Hz, and indications that it could be further improved~\cite{doi:10.1063/1.3250841}. The same group also reports a resonant frequency of 0.0006 Hz for a Scott-Russell horizontal linkage~\cite{doi:10.1063/1.3250841,WINTERFLOOD1996141}. The pursuit of such low frequency isolation will always face difficulties due to losses and anelasticity~\cite{doi:10.1063/1.3250841,doi:10.1063/1.1144774}.

The lowering of the resonance frequency of the seismic isolation system cannot just be an upgrade of the super attenuators used in optical GW detectors~\cite{accadia:in2p3-00603336}; a new design is required. This is why this vibration isolation project for ELGAR should be dedicated to creating an innovative suspension design for integration into a GW detector. The Roberts linkage (converting a rotational motion to approximate straight-line motion) for horizontal isolation and the Euler-Lacoste linkage (using pre-compressed springs) for vertical isolation were previously proposed and their performance (as noted above) has been subsequently demonstrated~\cite{doi:10.1063/1.1583862,274fe9face5f4a5fbcd42fb43eb7db16}. A possible strategy could be the optimization of these designs to make them robust by reducing the number of degrees of freedom. For ELGAR the goal is to produce an upgraded innovative design to push the vibration isolation system to an even lower resonance frequency. Whereas the Roberts linkage can have, by its design, as low frequency a mechanical adjustment as can be accurate, the Euler-Lacoste linkage is based on a low resonance frequency mechanical spring. A new design can then be proposed and its feasibility fully studied. It is through this effort that the vibration isolation around 1 Hz can be such that
the sensitivity of the GW detector can be achieved. A vertical antenna configuration, where the laser propagates between vertically displaced atomic probes, would require a different geometry for the retro-reflecting mirror suspension. 

In addition, a trade-off between active / passive systems, post-correction techniques, and their combinations must be studied.
Lessons can be learned form the Advanced LIGO seismic isolation system~\cite{Matichard:2015eva}. The optics for Advanced LIGO hang from suspension systems which are mounted on a pair of tables that are actively isolated. Attached to the ground is a hydraulic platform, or a Hydraulic External Pre-Isolators (HEPI). On top of the HEPI is an Internal Seismic Isolation within what is called the Basic Symmetric Chamber (BSC-ISC). Finally, a quadruple pendulum suspension supporting the test mass is attached to the BSC-ISC. The HEPI platform is used to align and position the entire system, while the BSC-ISC uses active isolation down to 0.05 Hz~\cite{Matichard:2015eva}. This combination system of active isolation and passive suspension provides an example of what ELGAR might be able to use, although research and development work will be necessary to decrease the operating frequency band.

\subsection{Control Strategy}
The control strategy for maintaining the position and operation of the optical components, and providing seismic isolation, will be necessary and inherently complicated. For the mirrors and beamsplitter it will ultimately be necessary to use sensing and feedback systems to control all degrees of freedom. It is likely that actuation forces will be necessary at all stages of the isolation system, from pre-isolation tables to the final suspension stages holding the mirror or beamsplitter. It will be necessary to maintain a locked laser and optical system while maintaining the alignment of the various components, accounting for drifts, and damping excitations. Pushing these requirements to the low frequency band for ELGAR will require research and development work.

The control system must slow-down and align the optical components, putting their operational dynamics within the gain range of the servo systems and the error signals generated by wavefront sensing for angles,  and locking for longitudinal position. This should be done in such a way that only internal forces are used.

There are different options available for applying actuation forces within the suspension system. For gravitational wave detectors there is typically the use of magnetic or electrostatic actuation. Magnetic actuation is typically done with coil and permanent magnetic combinations; for example, note how they are used in Advanced Virgo~\cite{Acernese2014}. A small adjustment, done quickly, of the position of the optical component can be made; one must account for the mechanical transfer function of the system. When forces are applied to the suspension system the energy associated with the actuation will be primarily absorbed at mechanical resonant frequencies; this creates potential difficulties. To apply actuation on a mirror will require the use of a reaction mass. Similar issues were considered for the proposed seismic isolation system of the Einstein Telescope~\cite{Punturo2010}. There is a down-side to magnetic actuation, namely the susceptibility to noise coupling from magnetic fields~\cite{Cirone_2019}. This can be especially problematic with respect to the globally coherent Schumann resonances, which can be a source of coherent noise between gravitational wave detectors at different locations around the world~\cite{PhysRevD.87.123009,PhysRevD.97.102007}.

Electrostatic actuation is another possibility for use in the control of the mirrors and beamsplitter for ELGAR. This is what is used for actuation of the Advanced LIGO test masses~\cite{TheLIGOScientific:2014jea,Carbone_2012}. There are benefits to using electrostatic actuation. No magnets need to be attached to the mirrors, and in that way, this reduces thermal noise by not deleteriously affecting the quality factor of the mirror. There is also no susceptibility to magnetic fields with the electrostatic actuators. A down side would be when the mirror acquires stray charge, which is known to happen~\cite{Martynov:2016fzi}, then additional noise can be induced, although modulating the sign of driving actuation field has been show potentially alleviate this problem~\cite{Acernese_2008}.


A control scheme for the suspension system must mitigate mechanical resonances if the system is to be brought into lock. The attenuation of resonances by the control system will be necessary, but these techniques have already been demonstrated in other systems~\cite{doi:10.1063/1.3250841}. However, when this is done the energy must be extracted from the system. Techniques to do this have used magnetic induction and dissipation (resistivity)~\cite{BRAGINSKY1996164,AIGO_2009,doi:10.1063/1.3250841}. It is known that these suspension control systems can add noise to GW detectors. Advanced LIGO and Advanced Virgo measure this control noise, and subtract it when producing their calibrated strain data~\cite{Meadors_2014,Davis_2019,Acernese:2018bfl}.

In summary, ELGAR will require a sophisticated control systems to align and lock its retro-reflecting mirrors and the beamsplitter. Only in this way can the detector be properly operated and achieve its desired sensitivity. One can imagine a system of pre-isolation tables, similar to what is used by Advanced LIGO~\cite{Matichard:2015eva}. The suspension system will reside on a pre-isolation table, and the optical component (mirror or beamsplitter) will be suspended and have seismic noise filtered to an appropriate level (as discussed above). Finally, the optical component will be controlled via magnetic or electrostatic actuation. Such actuation and control is necessary to keep the detector locked. 
Clearly the ultimate seismic isolation system for ELGAR will be complex, and to arrive at its construction and installation will take much research. Such a low-frequency isolation system would likely have additional applications outside of GW detection~\cite{Harms2019}.



\externaldocument{seismic_isolation}

\section{Newtonian-noise reduction}\label{sec_NN_reduction}

As we saw in Sec.~\ref{sec_Detector_configuration}, local gravity perturbations can create a spurious atom phase variation on the signal of an atom gradiometer, also known as Newtonian noise~(NN), which can limit the sensitivity of a GW detector~\cite{Saulson}. Indeed, in the gradiometric configuration considered in Sec.~\ref{sec_Detector_configuration}, the freely falling atoms play the role of reference masses corresponding to the suspended cavity mirrors in an optical GW detector: the equivalence principle implies that atom gradiometers will be equally sensitive to NN as an optical GW detector.

Newtonian noise has long been identified as important issue for optical GW detectors and has been thoroughly studied since their first generation \cite{Beccaria1998, Hughes1998, Creighton2008, Harms2019}. Up to now, Newtonian noise has not been revealed as a limiting effect, given the sensitivity and the bandwidth of state-of-the-art detectors, which currently operate above 10~Hz~\cite{PhysRevD.86.102001}. However, for instruments planned to operate below this frequency, studies have shown that NN will start to be limiting \cite{Beker2012,Fiorucci2018}. The operating window of the ELGAR project (0.1~Hz -- 10~Hz) makes NN a very important topic to address. We present here a preliminary study based on existing work applied to the ELGAR configuration.

\subsection{Sources of Newtonian noise}
Two main categories of sources can been defined: moving masses close to the test masses and density variations of the surrounding medium of the detector \cite{Harms2019}. In the first category fall all perturbations linked to human activity such as movement of vehicles or vibration from close-by apparatus, but also different geophysical processes such as for example hydrological effects accounting for water transfers. 
Human activity can be constrained, and defining a no disturbance perimeter for the antenna will be a necessity~\cite{Thorne1999} to limit its impact on the antenna. For what concerns geophysical effects, their impact for a given potential candidate site must be carefully evaluated: in particular, hydrological effects can be characterized using superconductive gravimeters, which enables the development of precise models of the induced gravity variations~\cite{Rosat2018}. In the second category stands the variation of density of the medium surrounding the antenna which could arise from local seismic activity and perturbations of the atmospheric pressure, contributions that were previously identified as the dominant effects for GW detectors. In the following, we briefly introduce the models used to calculate the gravity perturbation and the strain limitation that both effects will induce on the antenna.

The ground is in continuous motion, driven by the deformation of the Earth criss-crossed by seismic waves propagating within the upper mantle. Beside inducing a direct displacement noise felt by objects mechanically linked to the ground, which may directly impact the sensitivity of an antenna (see discussion in Sec.~\ref{sec_seis}), seismic activity also creates a density variation of the surrounding medium that modifies the local gravitational field leading to a spurious displacement of the detector test masses. In the frequency window of interest of ELGAR (0.1~Hz -- 10~Hz) lies the oceanic microseismic peak, which can show strong seasonal and regional variations. Supposing that within this window, the vertical motion of the ground can be modelled as an incoherent sum of Rayleigh waves~\cite{Beccaria1998, Hughes1998}, one can derive the expression of the seismic NN introduced into a single gradiometer of the detector. The PSD of the difference of the local gravity field induced between the points $X_i$ and $X_j$ near the surface where the gradiometer test masses are located can be written~\cite{Junca19}:
\begin{equation}\label{eq:SDaSeismic}
S^R_{\Delta a_x}(L,\omega)=\kappa_R^2(k_R)S_{\xi_z}(\omega) (1-J_0(k_R L)+J_2(k_R L))\, ,
\end{equation}
where $\kappa_R(k_R)$ is a function of the mechanical properties of the ground and the frequency of the seismic wave,
$J_n$ the $n^{\text{th}}$ Bessel function of the first kind, $S_{\xi_z}(\omega)$ the power spectral density of the vertical displacement of the ground.

From the same mechanism, fluctuations of atmospheric pressure can create density variations within the atmosphere that may also impact ELGAR's measurements.
Considering that pressure variation inside the atmosphere are adiabatic and described by an isotropic superposition of acoustic plane waves, the PSD of the difference of the local gravity field induced on the test masses of a single gradiometer can be written~\cite{Junca19}:
%
%
\begin{equation}\label{eq:SDaAtmo}
S^I_{\Delta a_x}(L,\omega)=\kappa_I^2(k_{I})S_{\delta p}(\omega) \big\langle 2(1-\cos(k_I L\sin\theta\cos\phi)) e^{-2hk_I\sin\theta}\sin^2\theta\cos^2\phi \big\rangle_{\theta,\phi}\, ,
\end{equation}
where: $\kappa_I(k_{I})=\frac{4\pi G\rho_0}{\gamma p_0 k_{I}}$, 
%
$S_{\delta p}(\omega)$ is the power spectral density of the pressure noise considered constant around the whole detector and $\langle.\rangle_{\theta,\phi}$ defines an averaging over all the propagating directions $(\theta,\phi)$ of the sound waves.
 \begin{figure}[t!]
  \centering
    \includegraphics[width=0.8\textwidth]{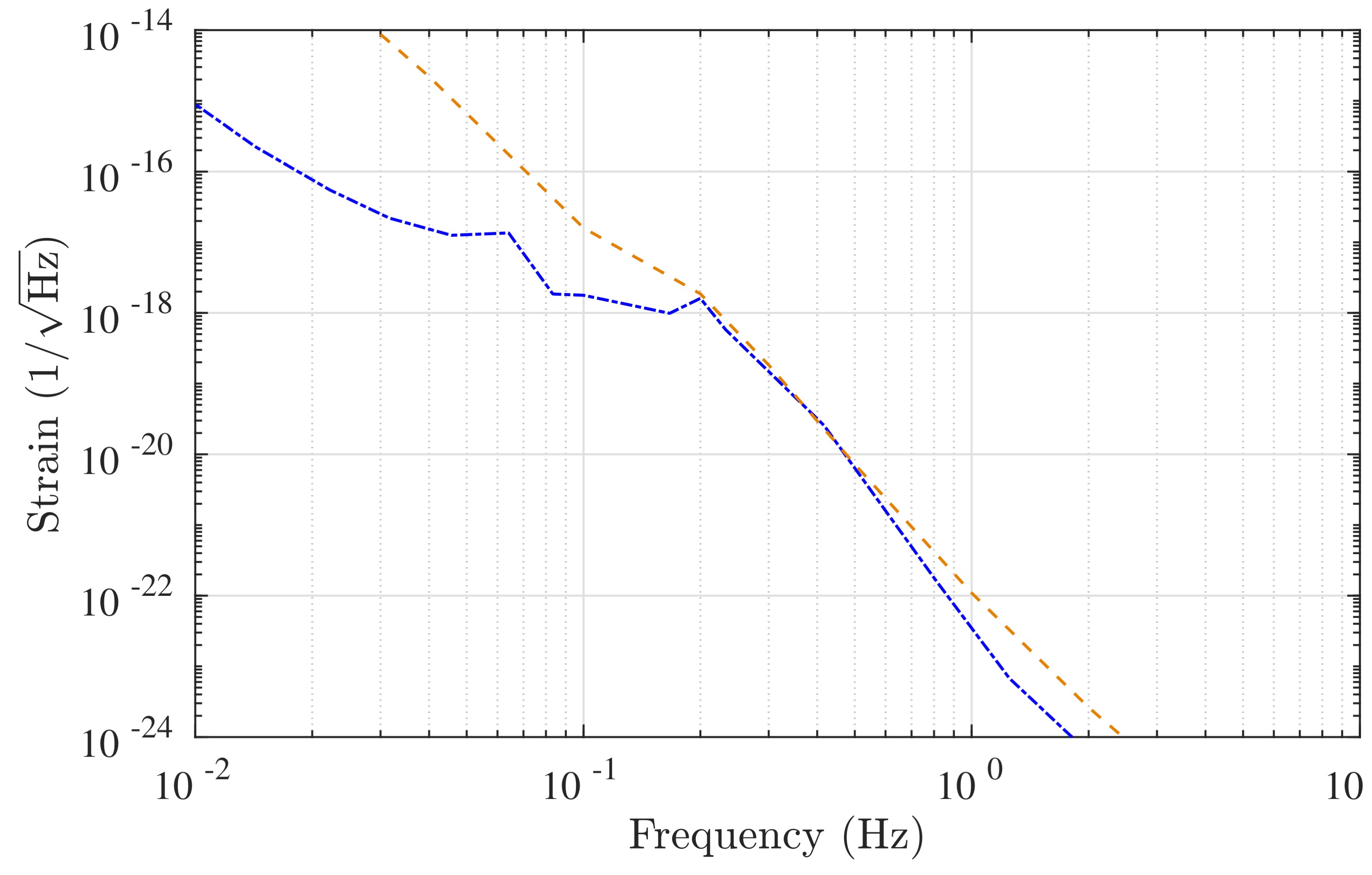}
    \caption{Comparison of NN from seismic and atmospheric sources. 
    Dot-dash line (blue) represents seismic NN using Peterson's low-noise model (NLNM) \cite{Peterson1993} for the vertical displacement noise, dash line (orange) represents atmospheric NN using mid-Bowman model \cite{Bowman2005} for the pressure noise.}
    \label{fig:NN1gradio}
\end{figure}

Fig.~\ref{fig:NN1gradio} shows the strain limitation induced by seismic and acoustic NN on a single gradiometer of baseline $L=16.3$~km. The strain projections are derived from Eq.~(\ref{eq:Sh}), using the NN acceleration PSDs of Eq.~(\ref{eq:SDaSeismic}) and Eq.~(\ref{eq:SDaAtmo}), and the relation: 
\begin{equation}\label{eq:Sdeltax}
S_{NN_1}(\omega)=\frac{S_{\Delta a_x}^{R,I}(\omega)}{\omega^4} \, ,
\end{equation}
We observe that both contributions induce a strong limitation for the detection of GWs in the band $0.1$ -- $1$\,Hz where it stands well above the level of $10^{-21}/\sqrt{Hz}$. In the following section, we show how the geometry of the ELGAR detector can help reducing the NN impact by averaging over an array of single gradiometers.

As a final note, it should be expected that other ambient fields produce significant NN including seismic body waves \cite{CoEA2018b} and advected, atmospheric temperature fields \cite{Fiorucci2018}. While seismic body waves are typically weaker than Rayleigh waves and therefore produce a minor contribution to NN, they can interfere significantly with any of the mitigation methods discussed in the following. Atmospheric temperature fields can be considered stationary perturbations, but they convert into fluctuations at a fixed point when this field is transported by wind, which leads to NN. 

\subsection{Mitigation of Newtonian noise using an atom-interferometer array}
\begin{figure}[htbp]
  \centering
	\includegraphics[width=0.8\linewidth]{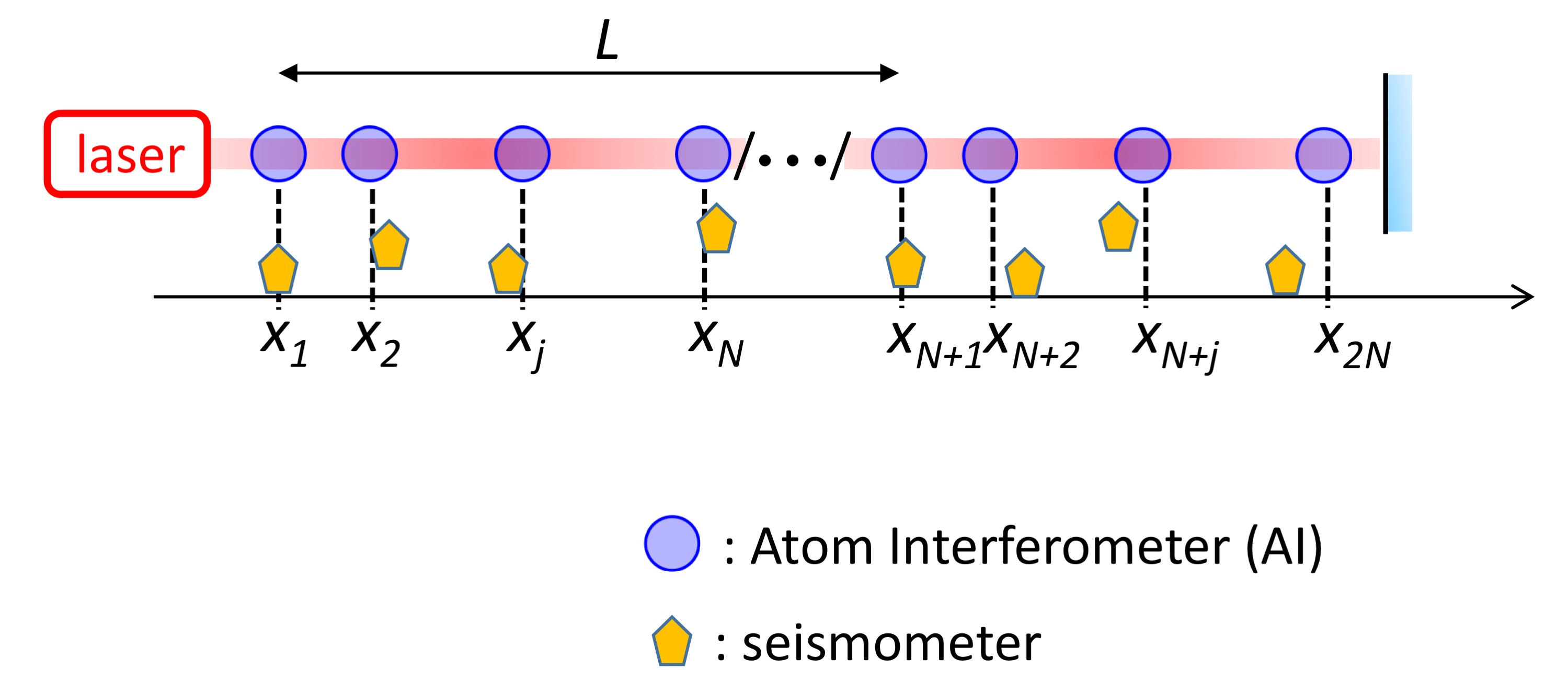}
  \caption{Arm geometry of the ELGAR detector. Each arm composed by N gradiometers regularly spaced over the total detector baseline.}
  \label{AI_network}
\end{figure}

As shown in Sec.~\ref{sec_Detector_configuration}, we propose for ELGAR a geometry using a 2D array of atom interferometers where each arm of the detector is composed by $N$ gradiometers regularly spaced over the total baseline. We now explain how the average signal of the different gradiometers enables to reduce the influence of Newtonian noise. For sake of clarity, we will limit this discussion to the averaging of a single arm of the detector, shown in see Fig.~\ref{AI_network}; the two-arm case being mostly similar.
In the following, we assume that the fields are stationary, homogeneous, and leading to Gaussian fluctuations. This means that we can describe the relevant properties of the field and its impact on the GW detector in terms of two-point correlations. Stationarity and homogeneity of the field $\zeta(\vec x,t)$ generating NN imply
\begin{equation}
\label{Correlation}
\mathcal{C}\left(\tau\right) = \left\langle\zeta\left(t\right)\zeta\left(t+\tau\right)\right\rangle\hspace{0.5cm}\textrm{and}\hspace{0.5cm}\Gamma\left(\vec{\xi}\,\right) = \left\langle\zeta\left(\vec{X}\right)\zeta\left(\vec{X}+\vec{\xi}\,\right)\right\rangle
\end{equation}
Averaging the signals from different gradiometers suppresses the contribution of NN at the same time maintaining the GW signal, because GWs and NN have different spatial signatures: the GW is a plane wave and its effect on a gradiometer is uniform on the earth scale whereas NN has characteristic scale of the order of the wavelength $\lambda$ of the considered source (seismic waves of different types, infrasound). 

The network configuration of $N$ aligned AIs, shown in Fig.~\ref{AI_network}, offers the possibility to repeat the GW measurement with different NN contributions without adding any significant noise in the considered frequency range. The average signal:
\begin{equation}
H_N^X\left(t\right)=\frac{1}{N}\sum^{N}_{i=1} \psi(X_i,X_{N+i},t)=2n k_l\int^{\infty}_{-\infty}\left(\frac{Lh(\tau)}{2}+NN^X(\tau)\right)g'(\tau-t) d\tau
\end{equation}
where:
\begin{equation}
NN^{X}(\tau)=\frac{1}{N}\sum^{N}_{i=1}
\Big[\delta x_{at}(X_j,\tau)-\delta x_{at}(X_i,\tau)\Big]
\end{equation}
represents an approximation of the GW signal $Lh/2$. The residual NN on this measurement procedure is given by:
\begin{equation}
\sqrt{S_{H_N^X}\left(\omega\right)}=2n k_l|\omega G(\omega)|\sqrt{S_{NN^X}\left(\omega\right)}
\end{equation}
According to Monte-Carlo theory~\cite{Caflisch1998}, we have:
\begin{equation}
\sqrt{S_{NN^X}\left(\omega\right)}=\frac{K\left(\omega\right)}{N^{1/2}}\sqrt{S_{NN_1}\left(\omega\right)}
\end{equation}where $K\left(\omega\right)$ is the Monte-Carlo variance reduction factor that drives the NN reduction gain $K\left(\omega\right)/N^{1/2}$ of the network with respect to the one of a single gradiometer. This factor, depending on the Fourier frequency through the residual correlation of NN between the considered AIs locations, ranges from 0 to $N^{1/2}$ :         
\begin{itemize}
\item when successive AIs are uncorrelated: $K = 1$, the NN is reduced by a gain $1/N^{1/2}$
\item when successive AIs are correlated: $K>1$, the gain diminishes till the minimal value of 1, reached when correlations are maximum: $K = N^{1/2}$. 
\item when successive AIs are anti-correlated, $K<1$, the NN is reduced by more than $1/N^{1/2}$, which is the best configuration.
\end{itemize}
The Newtonian force is a short range interaction. Hence, the spatial correlation of gravitational acceleration in a given direction simply reproduces spatial correlation of its source. NN is related to the fluctuations of mass density surrounding the AIs and if the field of density fluctuations is homogeneous, and isotropic, correlation between two points only depends on the separating distance $d$. In homogeneous media, this correlation roughly evolves as follows~\cite{Mykkeltveit1983} : (i) $\Gamma\left(d<\lambda/2\right)>0$, (ii) $\Gamma\left(\lambda/2<d<\lambda\right)<0$ and (iii) $\Gamma\left(d>\lambda\right)\simeq 0$ and so does NN. Since $\lambda$ depends on $\omega$, one needs to optimize the distribution of AI along the laser optical axis in order to reduce the $K$ factor on the frequency band of interest. While highest anti-correlations observed in natural seismic fields impose a limit $K(\omega)\gtrsim 0.5$, interesting mitigation factors of NN can be reached using adequate configuration by noise averaging as shown in Ref.~\cite{Chaibi2016}. 

The case of an inhomogeneous medium is much more difficult, because the spatial correlation function depends on the considered points~\cite{Braun2008}. Hence, NN correlation between AIs is \textit{a priori} unknown and can only be accurately determined once the detector is settled and running. As for Wiener filters applied for NN mitigation~\cite{Harms2019}, estimation of correlation functions requires a wide distribution of sensors which is difficult to conduct for an underground detector at sub-Hz frequency. Therefore, the distribution of AIs does not offer any degree of freedom to compensate the inhomogeneity of the spatial correlation function. For example, positive correlation between AIs might dominate and the factor $K$ remains above 1 for the whole frequency band of interest. Intuitively, one can think of applying weights when combining gradiometer signals,
\begin{equation}
\label{average_generalized}
H_N^X\left(t\right) = \sum^{N}_{i=1}\alpha_i\psi(X_i,X_{N+i},t)\hspace{0.5cm}\textrm{with}\hspace{0.5cm}\sum^{N}_{i=1}\alpha_i=1,
\end{equation}
in order to promote the effect of anti-correlations. The factors $\alpha_i$ are determined using Wiener filters except that the corresponding optimization procedure requires Lagrange multiplier in order to fulfill the condition $\sum \alpha_i = 1$. Further simulations are required to prove the efficiency of this procedure and determine its limitation.

\subsection{Coherent cancellation of Newtonian noise}
It was proposed that NN can be suppressed by coherent cancellation using environmental sensors like seismometers and microphones~\cite{HuTh1998}. One possibility is to calculate a Wiener filter based on correlations between environmental sensors and ELGAR data~\cite{Cel2000}. The Wiener filter is a linear filter that takes data from all environmental sensors as input, and it outputs an estimate of the ELGAR instrument noise that is correlated with the environmental data. This output is subtracted from ELGAR data ideally providing a time series without significant NN. However, there are practical limitations of the cancellation performance including numerical and statistical limits, incomplete information content in environmental data, and limited sensitivity of the environmental sensors~\cite{BaHa2019}. It might be possible to overcome some of these limitations with more sophisticated filtering techniques, e.g., based on Kalman filters or machine-learning algorithms, which can deal in a straight-forward way with non-stationary data or reducing statistical errors. 

Newtonian-noise cancellation may already help with geophysical observations~\cite{BaCr1999,JuEA2018a}, and a few orders of magnitude NN reduction is required to sufficiently reduce NN for GW observations~\cite{HaEA2013,FiEA2018}. This challenge can only be partially ameliorated when combined with other mitigation approaches. Newtonian-noise cancellation is therefore an essential technology for the realization of ELGAR as GW detector. 

The cancellation of NN below 10\,Hz from seismic fields was recently investigated for the Einstein Telescope (ET)~\cite{BaHa2019}, which is the European concept of a large-scale underground laser-interferometric GW detector, and for a sub-Hertz GW detector concept~\cite{HaPa2015}. One important aspect of these studies is that they focus on different types of seismic waves. The most relevant types of waves are the body waves (shear and compressional), which can propagate along all directions inside Earth, and Rayleigh waves, which are evanescent surface waves~\cite{AkRi2009}. While Rayleigh waves typically dominate surface displacement since they are produced efficiently by surface sources including human activities and weather phenomena, body waves can still make important contributions to the seismic displacement at underground sites~\cite{OlEA2015} and even to surface displacement~\cite{CoEA2018b}. A NN cancellation system needs to address contributions from all relevant wave types.

One of the major challenges of NN cancellation in ELGAR is the configuration of the environmental monitoring system. This comprises the choice of sensors and the array configuration. Seismic sensors can be simple broadband seismometers common in earthquake monitoring, but it can also be advantageous to consider other types of sensors such as tiltmeters~\cite{HaVe2016}. The optimal choice of sensors is still an open problem for ELGAR, but it seems possible in principle to provide sufficient reduction of seismic NN albeit at substantial cost for the installation of (likely) hundreds of sensors at the surface and inside deep boreholes (compare with~\cite{BaHa2019}). 

Far more challenging is the monitoring of atmospheric temperature, humidity, and pressure fields, which all give rise to NN~\cite{Cre2008,Har2015}. Microphones can in principle be used to monitor the pressure fluctuations, but data quality is severely degraded by wind noise, i.e., local pressure fluctuations from wind-driven turbulence around the microphones~\cite{WaHe2009}. Microphones have the additional disadvantage that they cannot be easily deployed at greater heights to form 3D arrays. It seems that the only reliable future option of atmospheric monitoring is through LIDAR; a laser-based observation of atmospheric disturbances. Current systems have enough sensitivity to monitor in a scanning fashion temperature fields~\cite{HaEA2015b}, humidity fields~\cite{SpEA2016}, and velocity fields~\cite{CLN2004}. However, the smallest perturbations of atmospheric density associated with pressure fluctuations, e.g., sound waves, cannot be resolved with LIDAR yet. It is expected though that acoustic NN needs to be substantially reduced by NN cancellation for ELGAR~\cite{FiEA2018}. This means that technological developments of LIDAR are still required, or other options for atmospheric sound monitoring are being developed that do not suffer from significant wind noise.

\subsection{Compatibility of mitigation methods}

Mitigation of NN on a single reference mass using an array of sensors can be very effective: each sensor measures its surrounding density fluctuation whose contribution to the mass gravitational motion is estimated using a Wiener filter~\cite{Cel2000}. The efficiency grows rapidly with the number of sensors $N_s$ and is above all limited by the sensor noise, ability of an array configuration to extract information about the field, and statistical estimation errors of the Wiener filter~\cite{Harms2019}. The diameter and density of the array determine the frequency band where noise suppression is effective.

Having an underground detector offers the advantages to significantly reduce atmospheric NN. However, NN from seismic body waves remains important at any depth, and also Rayleigh-wave NN is only suppressed to some extent underground. In this case, a three dimensional distribution of the sensors is required~\cite{BaHa2019}, which is technically difficult to achieve at low frequency ($<1\,\textrm{Hz}$) with a necessary array diameter of about $0.3\lambda$, i.e., more than a kilometer.

Nevertheless, a Wiener filter based on sensor arrays can be combined with the AIs averaging procedure. This process is shown in Fig.~\ref{AI_network} and requires the following steps:
\begin{itemize}
\item distribution of the sensors along the optical axis.
\item estimation of NN gravitational effect for each AI using a Wiener filter based on surrounding sensors.
\item use of an averaging procedure on the Wiener filter residuals.               
\end{itemize}
In the general case, the detector output is inferred from Eq.~\eqref{average_generalized}:
\begin{equation}
\label{Combination_Wiener_average}
H_N^X= \sum^{N}_{i=1}\alpha_i\left(\psi(X_i,X_{N+i})-\sum^{N_s}_{j=1}\sum^{3}_{k=1}\beta_{i,j,k}\,\left[\eta_{j+N_s,k}-\eta_{j,k}\right]\right)\hspace{0.5cm}\textrm{with}\hspace{0.5cm}\sum^{N}_{i=1}\alpha_i=1
\end{equation}where $\eta_{j,k}$ designates the $j$th sensor output in the direction $k$. $\beta_{i,j,k}$ are Wiener filter coefficients optimized to reduce the NN residual. 

The last equation can be approached in a simple case : (i) $N=N_s$ and to each AI corresponds a single sensor; (ii) all AIs are uncorrelated and so are the sensors: then $\alpha_i=1/N$; (iii) the media is homogeneous and all sensors are placed at the same position with respect to the AIs; (iv) only one sensor output is relevant, for example $k=1$. Hence, all AIs are equivalent and so are the sensors, i.e.  and $\beta_{i,j,k} = \delta_{i,j}\delta_{k,1} \beta$. Eq.~\eqref{Combination_Wiener_average} becomes:
\begin{equation}
\label{Combination_Wiener_average_uncorrelated}
H_N^X= \frac{1}{N}\sum^{N}_{i=1}\left(\psi(X_i,X_{N+i})-\beta\left[\eta_{i+N,1}-\eta_{i,1}\right]\right)
\end{equation}
In this case, the optimization procedure to reduce the NN contribution, i.e. $S_{NN}^X\left(\omega\right)$ at a given frequency is straightforward and gives :
\begin{equation}
\beta\left(\omega\right) = 2n k_l|\omega G(\omega)|\frac{\left\langle\delta x_{at}(X_i)\times\eta_{i,1} \right\rangle_\omega}{S_\eta\left(\omega\right)}\hspace{0.5cm}
\end{equation}for which
\begin{equation}
S_{NN}^X\left(\omega\right)=\frac{S_{NN1}\left(\omega\right)}{N}\left(1-\frac{\left\langle \delta x_{at}(X_i)\times\eta_{i,1} \right\rangle_\omega^2}{S_{\eta}\left(\omega\right)S_{\delta x_{at}(X_i)}\left(\omega\right)}\right)
\end{equation}
Hence, Eq.~\eqref{Combination_Wiener_average_uncorrelated} represents an observable, which combines the average procedure and the Wiener filter at the same time with a combined mitigation efficiency. For a correlation coefficient $\left\langle \delta x_{at}(X_i)\times\eta_{i,1} \right\rangle_\omega/\sqrt{S_{\eta}\left(\omega\right)S_{\delta x_{at}(X_i)}\left(\omega\right)}\simeq 0.5$, the NN is only reduced by $13\%$. This factor should be improved when several sensors are correlated with a single AI, also their distribution geometry is not optimal. The general case described by Eq.~\eqref{Combination_Wiener_average} gives a mitigation of the form:
\begin{equation}
\sqrt{S_{NN}^X\left(\omega\right)}=K\left(\omega\right)\sqrt{R\left(\omega\right)}\frac{\sqrt{S_{NN1}\left(\omega\right)}}{N^{1/2}}
\end{equation}
where $R\left(\omega\right)$ represents the residual of the Wiener filter and $K\left(\omega\right)$ represents the Monte-Carlo variance reduction factor. Sensor positions, their number $N_s$, $\alpha$ and $\beta$ factors are to be optimized in order to reduce the product $K\left(\omega\right)\sqrt{R\left(\omega\right)}$. A complete set of simulations are required to confirm this combination of NN mitigation methods and gives a realistic order of magnitude for its efficiency.

\subsection{Site evaluation criteria with respect to NN}

Based on the analysis described above, its appears that the following properties of the site are crucial to ensure the efficiency of NN mitigation :
\begin{itemize}
\item Reducing the correlation length $\lambda$ offers the possibility to increase the number $N$ of AIs for a given antenna length. For an 
underground detector for which air contribution to the NN is naturally reduced, this concerns a low sound propagation velocity. In practice, this is the case for media of low density which in this same time has high seismic noise. Seismic NN being given by the product of the density by the seismic noise, it clearly appears that site choice must balance between the sound velocity and the seismic noise.
\item Inhomogeneities within the site medium increases the probability to have pure positive correlations between AIs and $K<1$ might not be reached for any frequency $f$. Therefore, homogeneity must also be considered as a crucial factor for the detector site choice.
\end{itemize} 

\section{Noise couplings}
\label{sec_other_noise_couplings}

In section \ref{sec_Detector_configuration} we have outlined the connection between the strain sensitivity floor level for the whole detector and sensitivity limit of a single gradiometer, which can be recast as follows:
\begin{equation}\label{sensitivity_link}
\sqrt{S_h(\omega)}=\frac{1}{\sqrt{2N}}\times\frac{1}{nk_{l}L|\omega G(\omega)|}\times\sqrt{2S_{\epsilon}(\omega)},
\end{equation}
The target floor sensitivity level of $3.3\times 10^{-22}/\sqrt{\rm Hz}$ for the array of $2N$ uncorrelated gradiometers corresponds to the strain sensitivity for a single gradiometer of $4.1\times 10^{-21}/\sqrt{\rm Hz}$ which is reached  at the peak frequency of 1.7~Hz. This, in turn, sets the differential atom phase sensitivity limit of $\sqrt{2S_{\epsilon}^{\rm{lim}}}= \sqrt{2}\times 10^{-6}~\rm{rad}/\sqrt{\rm{Hz}}$, with $\sqrt{S_{\epsilon}^{\rm{lim}}}$ being the atom shot noise limited phase sensitivity of a single interferometer. In the following, we thus restrict the analysis to the case of a single  interferometer. We set a conservative requirement for each of the spurious  phase noise contributions to not exceed the level of $0.1~\mu\rm{rad}/\sqrt{\rm{Hz}}$ at 1.7~Hz for a single atom interferometer.

\subsection{Couplings of rotations, gravity gradients, beam misalignment and beam pointing jitter to the initial position and velocity and to the gravitational acceleration.}

The phase-shift contributions due to rotations and gravity gradients in light-pulse atom interferometry have been calculated in Refs.~\cite{Audretsch1994,Borde2001,Antoine2003,Bongs2006,Hogan2009,Roura2014,Kleinert2015,Bertoldi2019}.
Here we will make use of the convenient theoretical tools developed in Refs.~\cite{Roura2014,Roura2018b} to derive the relevant terms for the interferometry scheme introduced in Sec.~\ref{sec_double_loop}.
We will first focus on those contributions that depend on initial central position and velocity of the atomic cloud and then consider the couplings to the gravitational acceleration $g$.
Cross-couplings between the different contributions considered here are typically further suppressed and will not be explicitly discussed.

\subsubsection{Rotations.}\label{sec_coupling_rotation}

The proposed atom-interferometer geometry can successfully suppress phase-shift contributions of order $\big( \Omega\, T \big)$ that depend on the initial velocity of the atomic cloud. However, some dependence on the initial conditions remains at higher orders in $\big( \Omega\, T \big)$. The corresponding contributions to atomic phase shift are given by
\begin{align}
\Delta \phi_\Omega &= -\,6\, \mathbf{k}_\text{eff}^\perp \cdot \mathbf{v}_0\, T \, \big( \Omega\, T \big)^2
+ 4 \Big( \big( \bold{\Omega}\, T \big) \times \mathbf{k}_\text{eff} \Big) \cdot \mathbf{v}_0\, T \, \big( \Omega\, T \big)^2 \nonumber \\
&\quad + 2 \Big( \big( \bold{\Omega}\, T \big) \times \,\mathbf{k}_\text{eff} \Big) \cdot \mathbf{x}_0 \, \big( \Omega\, T \big)^2
\,+\, O \Big( \big( \Omega\, T \big)^4 \Big) ,
\end{align}
where $\mathbf{k}_\text{eff}^\perp \,=\, - \, \hat{\bold{\Omega}} \times \big( \hat{\bold{\Omega}} \times \mathbf{k}_\text{eff} \big)$ is the component of $\mathbf{k}_\text{eff}$ perpendicular to the direction of Earth's angular velocity, characterized by the unit vector $\hat{\bold{\Omega}}$. The dominant contribution comes from the first term, which gives the maximum allowed amount of initial-velocity jitter. For the interferometer configuration under  consideration, 
and given Earth's angular velocity $\Omega \approx 73 \, \mu\text{rad/s}$, an initial-velocity stability of $\sqrt{S_{v_0}} \lesssim 0.05 \, \mu \text{m/s} / \sqrt{\text{Hz}}$ is required for each velocity component in order to reach the targeted sensitivity.

\subsubsection{Relative beam misalignment.}\label{sec_coupling_beam_misalignment}

A small misalignment between the two interferometer beams also leads to the following velocity-dependent phase-shift contribution \cite{Savoie2018,Altorio2019}:
\begin{align}
\Delta \phi_\theta \, &= \, 4\, k_\text{eff}\,\Delta\hat{\mathbf{n}} \cdot \mathbf{v}_0\, T
= \, 4\, k_\text{eff}\,\big( \sin \delta\theta \ v_0^{y'} T + (\cos \delta\theta -1)  \ v_0^{x}\, T \big)
\nonumber \\
&\approx 4\, k_\text{eff}\,\big( \delta\theta_y\, v_0^{y} + \delta\theta_z\, v_0^{z} \big) \, T ,
\label{eq:misalignment}
\end{align}
where $v_0^{y'}$ in the second equality is the initial velocity component along the $\Delta\hat{\mathbf{n}}$ direction projected onto the $y-z$ plane
and in the last equality we have kept just the lowest-order terms corresponding to $\Delta\hat{\mathbf{n}} \approx  \big( 0, \delta\theta_y, \delta\theta_z \big)$.
Assuming the stability requirement for the initial velocity obtained in the previous paragraph,
the maximum allowed relative beam misalignment is  $\delta\theta \lesssim 0.32 \, \text{nrad}$.

\subsubsection{Beam pointing jitter.}\label{sec_coupling_beam_jitter}
Small fluctuations in the direction $\hat{\mathbf{n}}_j$ of each laser pulse (with $j=1,\ldots,4$) can be characterized by $\delta\hat{\mathbf{n}}_j \approx \big( 0,\boldsymbol{\xi}^\perp_j \big) = \big( 0,\xi^y_j,\xi^z_j \big)$, where $\boldsymbol{\xi}^\perp_j$ are stochastic processes. For \emph{low-frequency} pointing jitter with $\omega\, T \ll 1$ the directions of the two pulses associated with each beam are correlated, so that $\boldsymbol{\xi}^\perp_1 = \boldsymbol{\xi}^\perp_4$ and  $\boldsymbol{\xi}^\perp_2 = \boldsymbol{\xi}^\perp_3$. This reduces then to shot-to-shot fluctuations of the relative beam misalignment considered in Sec.~\ref{sec_coupling_beam_misalignment}, but with $\Delta\hat{\mathbf{n}} \approx \boldsymbol{\xi}^\perp_2 - \boldsymbol{\xi}^\perp_1$, and leads to the following coupling between $\boldsymbol{\xi}^\perp_j$ and the mean initial velocity $\langle \mathbf{v}_0 \rangle$ that would result from averaging over many shots:
\begin{equation}
\Delta \phi_\xi \,\approx\, 4\, k_\text{eff}\, \big( \boldsymbol{\xi}^\perp_2 - \boldsymbol{\xi}^\perp_1 \big) \cdot \langle \mathbf{v}_0 \rangle\, T
\label{eq:pointing_jitter1} .
\end{equation}

On the other hand, for \emph{higher-frequency} pointing jitter with $\omega\, T \gtrsim 1$ the above correlations no longer hold and one gets the following coupling with the mean initial position and velocity:
\begin{equation}
\Delta \phi_\xi \,\approx\, \sum_{j=1}^{4} \varepsilon_j \, k_\text{eff}\, \boldsymbol{\xi}^\perp_j \cdot
\Big(  \langle \mathbf{x}_0 \rangle + \langle \mathbf{v}_0 \rangle\, t_j \Big)
\label{eq:pointing_jitter2} ,
\end{equation}
where $t_j$ is the time at which the $j$-th pulse is applied and we have introduced $\varepsilon_1 = -\varepsilon_4 = 1$ and $-\varepsilon_2 = \varepsilon_3 = 2$. From Eq.~(\ref{eq:pointing_jitter2}) one can derive bounds on the amplitude of the power spectral density for beam pointing jitter depending on the values of the mean initial position and velocity. However, since it is important to take into account that beam pointing jitter also couples to the local gravitational acceleration and to consider their combined effect, we defer this discussion to Sec.~\ref{sec_coupling_g}, where the stability requirements for a differential measurement compatible with the targeted sensitivity will be obtained.

\subsubsection{Gravity gradients.}\label{sec_coupling_grav_grad}
The phase-shift for the double-loop interferometer geometry proposed in Sec.~\ref{sec_double_loop} is free from any coupling of the gravity gradient to the initial position, but not to the initial velocity, which contributes as follows:
\begin{align}
\Delta \phi_\Gamma = - 2\, \mathbf{k}_\text{eff}^\text{T}\, \big( \Gamma\, T^2 \big) \, \mathbf{v}_0 T
\,+\, O \Big( \big( \Gamma\, T^2 \big)^2 \Big), 
\end{align}
where the transposed vector $\mathbf{k}_\text{eff}^\text{T}$ is introduced to underline the vector-matrix notation conventionally employed in the literature. Given Earth's gravity gradient in the horizontal direction, $\Gamma_{xx} = \Gamma_{yy} \approx 1.5 \times 10^{-6}\, \text{s}^{-2}$, this would lead to more stringent (by several orders of magnitude) requirements on the initial velocity, than those obtained above as a consequence of Earth's rotation.

Fortunately, in order to substantially relax these requirements, one can make use of the compensation technique proposed in Ref.~\cite{Roura2017PRL} and involving in this case a single-photon frequency change $\Delta\nu \sim 12\, \text{MHz}$ for the two intermediate pulses. Indeed, taking the stability requirement for the initial-velocity jitter that follows from the effect of rotation, one finds that the gravity gradient needs to be compensated at level of $1 \%$. Compensation at this level has already been demonstrated experimentally in both gradiometry measurements \cite{d_Amico2017} and tests of the universality of free fall \cite{Overstreet2018}.

It is, however, important that the gravity gradient is the same (within that level of accuracy) for every atom interferometer on the baseline so that the sensitivity to their independent initial velocities can be simultaneously minimized \cite{Roura2018a}. If necessary, this can be achieved by making small adjustments in the local mass distribution. Furthermore, in the whole process (including the determination of the appropriate value of $\Delta\nu$) the compensation technique itself can be employed for self-calibration \cite{Overstreet2018}.

\subsubsection{Couplings to the local gravitational acceleration $g$.}\label{sec_coupling_g}

Time variations of the local gravitational acceleration $\mathbf{g}$ at the positions of each interferometer also couple to rotation, gravity gradient, relative beam misalignment and beam pointing jitter.
The corresponding phase-shift contributions are given by
\begin{align}
\Delta \phi_g \,&=\, 2 \Big( \big( \bold{\Omega}\, T \big) \times \,\mathbf{k}_\text{eff} \Big) \cdot \mathbf{g} \, T^2
\,+\, 8\, k_\text{eff}\,\Delta\hat{\mathbf{n}} \cdot \mathbf{g}\, T^2 
+ \frac{1}{2} \sum_{j=1}^{4} \varepsilon_j \, k_\text{eff}\, \boldsymbol{\xi}^\perp_j \cdot \mathbf{g}\, t_j^2
\nonumber \\
& \quad\quad 
- 4\, \mathbf{k}_\text{eff}^\text{T}\, \big( \Gamma\, T^2 \big) \, \mathbf{g}\, T^2
+\, O \Big( \big( \Omega\, T \big)^2 \Big) \,+\, O \Big( \big( \Gamma\, T^2 \big)^2 \Big)
\label{eq:grav_coupling}.
\end{align}
Here we have neglected any time dependence of $\mathbf{g}$ within a single shot for simplicity. This can be taken into account, but simply gives rise to lengthier expressions without substantially 
altering the main conclusions. Below we discuss the stability requirements implied by each term.

(\emph{i}) Shot-to-shot variations of $\mathbf{g}$ with characteristic frequencies comparable to the GW frequencies of interest can give rise to spurious signals. However, such variations are directly related to the Newtonian noise discussed in Sec.~\ref{sec_NN_reduction}. Thus, any methods employed to mitigate the effects of Newtonian noise would also help in this case.
Moreover, assuming that Earth's \emph{angular velocity} is sufficiently stable, the first term is suppressed by a factor $\big( \Omega\, T \big) \sim 10^{-5}$ compared to standard Newtonian noise.

(\emph{ii}) On the other hand, shot-to-shot variations of the \emph{relative beam misalignment} also couple to the static part of $\mathbf{g}$.
Taking into account that a similar coupling to the mean initial velocity $\langle \mathbf{v}_0 \rangle$ appears in Eq.~(\ref{eq:pointing_jitter1}), both contributions can be combined to give $4\, k_\text{eff}\,\Delta\hat{\mathbf{n}} \cdot \!\big( 2\, \mathbf{g}\, T + \langle \mathbf{v}_0 \rangle \big)\, T$
with $\Delta\hat{\mathbf{n}} \approx \boldsymbol{\xi}^\perp_2 - \boldsymbol{\xi}^\perp_1$.
Therefore, since it is dominated by the directions orthogonal to the interferometer beam, the net contribution can be minimized by choosing $\langle \mathbf{v}_0^\perp \rangle = - 2\, \mathbf{g}^\perp T$ for those directions.
In practice this optimal value of the initial velocity can be determined by varying $\Delta\hat{\mathbf{n}}$ on purpose by a substantial amount and finding the value of $\langle \mathbf{v}_0^\perp \rangle$ that minimizes the dependence on $\Delta\hat{\mathbf{n}}$ \cite{Savoie2018,Altorio2019}.
If one could achieve this at the p.p.m.\ level, so that $\big| 2\, \mathbf{g}^\perp T + \langle \mathbf{v}_0^\perp \rangle \big| \lesssim 10^{-6} g\, T$,
a beam-alignment stability of $\sqrt{S_{\delta\theta}} \lesssim 4 \times 10^{-12} \, \text{rad} / \sqrt{\text{Hz}}$ would be required in order to reach the target sensitivity.

(\emph{iii}) In contrast, for higher-frequency \emph{beam pointing jitter} (with $\omega\, T \gtrsim 1$) the cancellation between the contributions from Eqs.~(\ref{eq:pointing_jitter1}) and (\ref{eq:grav_coupling}) no longer holds because the coefficients will be different in general and one needs to consider both contributions separately.
As far as the contribution involving the static part of $\mathbf{g}$ is concerned, the relevant quantity for the differential measurement of two interferometers $A$ and $B$ is proportional to $k_\text{eff}\,\,\boldsymbol{\xi}^\perp_j  \!\cdot \!\big( \mathbf{g}_A - \mathbf{g}_B \big) \, T^2$. Hence, assuming $\big| \mathbf{g}^\perp_A - \mathbf{g}^\perp_B \big| \lesssim 10^{-5} g$, which is compatible with a 30-km total baseline%
\footnote{The difference due to Earth's curvature and projected onto the transverse plane will typically be of this order for baselines of $30\,\text{km}$.}, one concludes that a beam-alignment stability of $\sqrt{S_{\boldsymbol{\xi}^\perp_j}} \lesssim 3 \times 10^{-12} \, \text{rad} / \sqrt{\text{Hz}}$ is required at those frequencies. Similar conclusions are reached for the contribution involving the mean initial velocity if one assumes $\big| \langle \mathbf{v}_0^\perp \rangle_A - \langle \mathbf{v}_0^\perp \rangle_B \big| \lesssim 10^{-5} g\, T$. This could be accomplished by linking $\langle \mathbf{v}_0^\perp \rangle$ with $\mathbf{g}^\perp T$ independently for each interferometer as described in the previous paragraph.

In addition, there is a third contribution involving the mean initial position transverse to the interferometry beam. If one wants to have comparable requirements on the beam pointing stability, the condition $\big| \langle \mathbf{x}_0^\perp \rangle_A - \langle \mathbf{x}_0^\perp \rangle_B \big| \lesssim 10^{-5} g\, T^2$ needs to be fulfilled. In this case, however, one will need to vary the relative alignment of the two laser baselines for different laser pulses within a single shot 
in order to calibrate that.

(\emph{iv}) Finally, for the static part of the \emph{gravity gradient} similar considerations to those made above for Earth's angular velocity apply. In this case there is an additional suppression factor $\big( \Gamma\, T^2 \big) \sim 6 \times 10^{-8}$ with respect to standard Newtonian noise. Furthermore, although time variations of $\Gamma$ can couple to the static gravitational acceleration $\mathbf{g}$, preliminary estimates suggest that this contribution is smaller than the regular Newtonian noise and one expects that it can be tackled through the same means as the latter.

\subsection{Magnetic fields, electric fields, blackbody radiation}\label{sec_mag_elect_temp}
In this section we estimate the limiting noise levels on various background fields (magnetic field, black body radiation and electric field) that produce a local acceleration noise $a_x$ along the direction of the Bragg beams at the position of the atoms in each AI forming the array of the antenna. An acceleration noise of PSD $S_{a_{x}}(\omega)$ translates into an interferometer phase noise given by:
\begin{equation}\label{eq:anoise}
    \sqrt{S_{\phi}(\omega)}=2nk_{l}\frac{|\omega G(\omega)|}{\omega^2}\sqrt{S_{a_{x}}(\omega)}\,
\end{equation}
We assume uncorrelated acceleration noise between the atom interferometers forming the antenna.
In this section, we estimate a requirement on the acceleration noise level  at the peak frequency of 1.7~Hz in order to ensure an  associated phase noise contribution  below $0.1~\mu\rm{rad}/\sqrt{Hz}$. When going to lower frequencies, the contribution to phase noise of a white acceleration noise reduces due to the factor $H_a(\omega)=\frac{|\omega G(\omega)|}{\omega^2}$ that diminishes when going to lower frequencies. 

\subsubsection{Second order Zeeman force.}
We consider rubidium atoms prepared in the first order Zeeman-insensitive ground state ($F=1$,  $m_{F} = 0$). In the presence of a fluctuating, non-homogeneous linearized magnetic field $B(x,t)=B_{0}(t) + \beta x$, the second order Zeeman shift induces an acceleration $ a_{x}(t)=\frac{2\alpha B_{0}(t) \beta }{m_{\rm{at}}} $ which results in a sensitivity to magnetic field fluctuations :
\begin{equation}
    S_{a_{x}}(\omega) \simeq\left(\frac{2\alpha \beta }{m_{\rm{at}}}\right)^{2}S_{B_{0}}(\omega)\, ,
\end{equation}
where $\alpha \simeq \frac{h}{2} \cdot 575$ Hz$/$G$^{2}$ is half the second order Zeeman shift on the rubidium clock transition and $m_{\rm{at}}$ is the atomic mass. For a magnetic field gradient of $\beta = 1$ $\mu$G/cm, the magnetic field stability requirement is then $\sqrt{S_{B}(\omega)}\le 1$ $ \mu$G/$\sqrt{\rm{Hz}}$ at $1.7$ Hz.

\subsubsection{Blackbody radiation force.}

Blackbody radiation emitted from the walls of the vacuum chamber induces an AC Stark shift on the atomic levels. If we consider a thin horizontal cylindrical chamber surrounding the interferometers, a temperature gradient along the horizontal axis will induce a gradient of potential energy and thus an acceleration for the atoms of \cite{haslinger_BBR_2018}: 
\begin{equation}
    a_{x}=\frac{\partial}{\partial x} \frac{2 \alpha_{\rm{at}} \sigma T^{4}(x,t)}{m_{\rm{at}} c \epsilon _{0}}\, ,
\end{equation}
where $\alpha_{\rm{at}}=h \times 0.0794$ Hz$/$(V$/$cm)$^2$ is the atom's static polarizability in the ground state, $\sigma$ is the Stefan-Boltzmann constant, T is the temperature of the walls and $\epsilon_{0}$ is the vacuum permittivity. If we consider a small linear temperature gradient $\tau$ and small temperature fluctuations $\delta T$ so that $T(x,t) = T_{0} + \delta T(t)+\tau x$, we can write at first order:
\begin{equation}
    S_{a_{x}}(\omega) \simeq \left(\frac{24\alpha_{\rm{at}} \sigma T^{2}_{0} \tau}{m_{\rm{at}} c \epsilon _{0}}\right)^{2}S_{\delta T}(\omega)\, .
\end{equation}
Using equation \ref{eq:anoise} we can then deduce that for rubidium atoms in a vacuum chamber at $ T_{0}=300$ K with a gradient of $\tau = 0.1$ K/m the temperature stability requirement at $1.7$ Hz is $\sqrt{S_{\delta T}}\le 2 \times 10^{-3} $ K/$\sqrt{\rm{Hz}}$.

\subsubsection{DC Stark force.}
A non homogeneous linearized electric field $E(t)=E_{0}(t)+\epsilon x$ induces an acceleration on atoms in the ground state $a_{x}(t)=\frac{\alpha_{at} E_{0}(t) \epsilon }{2m_{\rm{at}}}$ which results in a sensitivity to electric field fluctuations : 
\begin{equation}
    S_{a_{x}}(\omega)=\left(\frac{\alpha_{at} \epsilon }{2m_{\rm{at}}}\right)^{2}S_{ E_{0}}(\omega)\, .
\end{equation}
For an electric field gradient of $\epsilon = 0.1$ V/m$^{2}$, the electric field stability requirement is then $\sqrt{S_{E}(\omega)}\le 0.2 $ (V/m)/$\sqrt{\rm{Hz}}$ at $1.7$ Hz.

\subsection{Differential wavefront distortions}\label{sec_coupling_wavefront}

Wavefront distortions in light-pulse atom interferometry have been theoretically and experimentally studied in the context of inertial sensors \cite{fils_influence_2005, louchet-chauvet_influence_2011, schkolnik_effect_2015, trimeche_active_2017, karcher_improving_2018}. The sensitivity to wavefront distortions  stems from the coupling of the atomic sample's time-varying spatial extend due to its finite initial size and temperature with the laser beam's non-uniform phase front. This effect has usually been considered as static and is the dominant systematic effect for state-of-the-art atomic gravimeters. In these cases, the baseline of the atom interferometer is  on the order or below \SI{1}{\meter} so changes of the wavefront distortion along the direction of propagation of the laser beams is usually not considered. The effect of wavefront propagation has been studied theoretically in the context of dual-species atom interferometer of longer baselines for testing the weak equivalence principle \cite{hu_analysis_2017, schubert_differential_2013}. Wavefront aberrations have furthermore been discussed in the context of satellite-based atomic gravitational wave detectors \cite{dimopoulos_atomic_2008, bender_comment_2011, dimopoulos_reply_2011,Hogan2011}.
We give here an estimate of the impact of wavefront distortions in a simple yet representative case, in order to derive requirements on the initial position and velocity jitter of the atom source. 

\subsubsection{Effect of curvature.}
\label{subsec:wf_curvature}
We consider two atom interferometers separated in x-direction by a baseline $L$, one near the beamsplitter and one near the retro-reflecting mirror. Assuming a  waist of the incident beam of $w_0$=50 mm at the position of the beamsplitter, corresponding Rayleigh length 1s $x_R \simeq \,$10 km, the radius of curvature at position $x$ along the baseline is given as $R(x) = x \sqrt{1+ (x_R/x)^2}$. The wavefront of the beam in the direction transverse to the propagation axis ($z$) is given by $\phi(r,x)=k_{l}r^2/[2R(x)]$. The  atom interferometer at the position $z$ is driven  by two counter-propagating beams of different curvatures, corresponding to a relative wavefront $\Delta\varphi_{las}(r,x)=\varphi_{las}(r,x)-\varphi_{las}(r,2L_{T}-x)=k_{l}r^2/(2R_{\text{eff}})$ with  $R_{\text{eff}}^{-1} = R(x)^{-1} - R(2L_{T}-x)^{-1}$. We will consider the largest possible contribution of this effect to the gradiometer phase shift, which corresponds to the relative phase between the most distant interferometers located at $x\simeq 0$ (close to the input telescope where $R_{\rm{eff}}\simeq R(2L_{T})$) and $x\simeq L_{T}$ (close to the retro-mirror where $\Delta\varphi_{las}(r,L_{T})\simeq 0$). To simplify notations, we drop out the $x$ dependence below.
The phase shift of the 4-pulse interferometer in the limit of infinitely short interrogation pulses (see Eq.~1) is given by 
\begin{equation}
   \Delta\phi=\Delta\varphi_{las}(r(0))-2\Delta\varphi_{las}(r(T))+2\Delta\varphi_{las}(r(3T))-\Delta\varphi_{las}(r(4T))\, ,
\end{equation}
with $4T$=800 ms the total interrogation time and $r(t)=r_0+v_0 t$ the transverse position of the atom in the laser beam at the time $t$ of the light pulse. In the case where the curvature of the two (lower and upper) beams is the same and where the light pulses occur well at the center ($r=0$), the contribution to the interferometer phase shift vanishes. 
A phase shift can appear in two cases: \textit{(i)} if the center of the beams are not well aligned with the atom trajectory, i.e. if the atom interacts with the beams away from the center of curvature (we denote these offsets as $\epsilon_{1,2}$) ; \textit{(ii)} in the case of a different curvature between the two beams ($R_{\text{eff},1}=R_{\text{eff},2}+\Delta R$), due, for example, to different waists $w_{0,1}=w_{0,2}+\Delta w$. Here, the subscript $1$ ($2$) refers to the bottom (top) beam where the $\pi/2$ ($\pi$) transitions are driven at $t=0$ and $t=4T$ ($t=T$ and $t=3T$). Finally, the atomic phase shift associated to the wavefront curvature becomes:
\begin{equation}
\Delta\phi_{\rm{wf}} = 2nk_{l} \times \Big(-\frac{1}{R_{\text{eff},1}}[4 v_0 T(\epsilon_1+r_0)+8v_0^2 T^2] + \frac{1}{R_{\text{eff},2}}[4 v_0 T(\epsilon_2+r_0)+8v_0^2 T^2] \Big).
\end{equation}
As expected from the quadratic dependence of the wavefront, all the terms represent second-order corrections scaling as the product of either of two small parameters: $\epsilon_{1,2}$ (offsets from the center), $r_0$ (jitter in initial position) and $v_0$ (jitter in initial velocity).

For a numerical estimate with $2n=1000$, we calculate the effect of the coupling between centering error and velocity jitter (term  $\sim \epsilon_{1,2} v_0 T$).
In that case, constraining the differential phase noise at a level below $0.1\ \mu$rad/$\sqrt{\rm{Hz}}$ in the target frequency band then  requires  a control of atomic velocity jitter at the level of $v_0<0.8$ (nm/s)/$\sqrt{\rm{Hz}}$ for a centering error $\Delta\epsilon=|\epsilon_1 - \epsilon_2|=0.5$~mm and $w_{0,1}=w_{0,2}=50$~mm. The fine-tuning of a difference of waists $\Delta w$ might further reduce the total value of the phase shift $\Delta\phi_{\rm{wf}}$ and thus relax the requirement on atomic velocity jitter.

\subsubsection{Higher order effects.}
Higher order (than curvature) effects will occur as the laser beam reflects off a mirror of imperfect flatness, which causes a differential phase shift between the two distant atom interferometers as the  phase field evolves upon propagation.
Estimation of this effect highly depends on the phase map of the field after its reflection on the mirror \cite{karcher_improving_2018}, which is outside of the scope of this manuscript. We can nevertheless point towards a possible procedure to study this effect: \textit{(i)} characterization of the surface of the retro-mirror; \textit{(ii)} simulation of the propagation of the field with the phase map from the mirror imperfect planarity (e.g. with the angular spectrum method), to obtain the relative phase map at various locations along the detector baseline; \textit{(iii)} Monte-Carlo simulation of the sampling of this phase map by an atomic cloud, taking into account position and velocity initial jitters. The result of such a simulation would yield requirements on the initial position and velocity jitters as well as on the planarity of the retro-mirror.


\subsection{Scattered light and diffraction phase shifts}
\label{subsec:diff_phase_shifts}
Large momentum transfer atom interferometers are impacted by diffraction phases which originate from the non-resonant couplings between momentum states during Bragg diffraction \cite{Buechner2003,Estey2015}. These diffraction phases depend on the local effective Rabi frequency and therefore on the local laser  intensity. Inhomogeneities in intensity arise from the Gaussian profile of the laser beam, which can be mitigated by the use of a top-hat laser beam \cite{Mielec2018}. Still, intensity inhomogeneities remain because of scattered light in the process of beam shaping or upon propagation along the baseline \cite{Vinet1996coherent,Vinet1996stat}. The effect of light scattered off the walls of the vacuum tube or off the optics depends on the vibrations of these elements, which renders the evaluation of its impact on a particular detector a challenging task. Dedicated working groups tackle this challenge in the case of ground-based laser interferometers \cite{Ottaway2012,Canuel:13} or for the LISA mission \cite{Spector2012}.
Estimating the contribution of scattered light in the ELGAR detector scheme is therefore outside the scope of this paper.

Still, to provide an estimate of diffraction phase shifts in a simple configuration and hence derive a requirement for the intensity homogeneity of the laser beam, we numerically investigated diffraction phases for a Mach-Zehnder geometry and compared two different realizations of large momentum transfer beam splitter, implemented with either sequential Bragg transitions or combined Bragg transitions and Bloch oscillations. The numerical simulations were carried out in a position space approach solving Schr\"odinger's equation with the help of the split operator method \cite{javanainen2006symbolic}. The atom-light interaction describing Bragg diffraction as well as Bloch oscillations was modeled as described in Ref.~\cite{muller2008atom}.


We analyzed the influence of a change of the Rabi frequencies of the first $\pi$ pulse and the first Bloch sequence by $2\%$ and calculated the resulting phase shift by numerically performing a phase scan. For the sequential Bragg interferometer we found a phase shift of $\Delta \phi$=9.3 mrad and for the Bragg+Bloch geometry a shift of $\Delta \phi$=1.3 rad.
The larger Bragg+Bloch phase shift can be partly explained by the fact that in this geometry the atoms are trapped in the optical lattice for $1.5$ ms. This directly leads to an energy shift for the upper arm of the atoms, but not the lower one and therefore to an extra relative phase shift. 
This shows that when considering the effects of the diffraction phase the usage of sequential Bragg pulses is the preferable way to imprint momentum on the atoms. 

Scaling up the momentum splitting to the desired $1000 \, \hbar k$ (i.e. 500 sequential $2 \, \hbar k$ pulses), a coarse extrapolation leads to a phase bias $\Delta \phi$=4.6 rad (for a $2\%$ change of laser intensity). We emphasize that this number is very specific to the geometry studied. This estimate is anticipated to be  reduced (at least factor 10) by the integration over the atomic ensemble which averages the local laser intensity profile. When considering the differential phase between two distant AIs, the relative phase  bias will result from the difference in intensities of the common interrogation laser beam at the two positions along the baseline. Assuming a difference of $2\%$ in local laser intensities then yields a relative phase bias of the order of 460~mrad. The phase noise can then be estimated by studying the  mechanism  underlying relative laser intensity fluctuations (e.g. vibration noise on the walls of the vacuum chamber that affect the locally scattered light). Although we gave here a  coarse estimate, our simple calculation highlights the challenge in controlling diffraction phases at the $\mu$rad$\sqrt{\rm{Hz}}$ level  and conceiving mitigation schemes.

\subsection{Effect of inter-atomic interactions ($^{87}$Rb atoms)}
\label{subsec:atomic_interactions}

Following \cite{debs2011cold}, we estimate the phase noise resulting from interactions between the atoms and the fact that the beamsplitters in the proposed scheme are only characterized up to the shot noise level. This phase noise contribution is directly given by the mean-field energy term of the Gross-Pitaevskii equation averaged over atomic spatial distribution
\begin{equation}
    \Delta \phi_{\rm{MF}}=\frac{1}{\hbar} \int_0^{4T} dt \; \langle \Delta E_{\rm{MF}} \rangle = \frac{4T}{\hbar}\frac{\sqrt{N}U}{V_{\rm{atoms}}}\, ,
\end{equation}
where $U=\frac{4 \pi \hbar^2 a_s}{m_{at}}$, $a_s$ is the s-wave scattering length, and $V_{\rm{atoms}}=\frac{4}{3}\pi((\frac{7}{10})^{1/3}R_{\rm{TF}})^3$ is effective atomic volume with $R_{\rm{TF}}$ being the Thomas-Fermi radius of the BEC. Considering $4T=0.8$ s, $N=10^{12}$, $a_s = 95~a_0$ for a case of $^{87}$Rb ($a_0$ is Bohr radius) and $R_{\rm{TF}}=1 \ $mm we obtain an estimation of $\Delta \phi_{\rm{MF}} = 0.013$~rad.
A reduced phase noise can be reached by using bigger atomic clouds with smaller atom number, which would compromise the shot noise limited sensitivity. Alternatively, with 10 interleaved interferometers \cite{Savoie2018} and a squeezed atomic source by $20$ dB \cite{Hosten2016}, an atomic source with $N=10^9$ atoms would allow for equivalent sensitivity but significantly reduced phase noise.  An interaction-induced phase noise of $\Delta \phi_{\rm{MF}} =0.1 \ \mu\rm{rad}/\sqrt{Hz}$ would then correspond to a Thomas-Fermi radius of about $R_{\rm{TF}}=16$~mm. This requirement on the cloud size, while being challenging for BEC, seems feasible for a sufficiently cold thermal atomic ensemble with an expansion frozen by delta-kicked cooling technique.   

\externaldocument{noise_couplings}
\section{ELGAR sensitivity and data}\label{sec_Sensitivity_curves}
\subsection{ELGAR Sensitivity Curve and operating parameters}

\ctable[
	caption = {Parameters of the ELGAR detector to reach a strain sensitivity of \SI{3.3e-22}{\per \hertz \tothe{\sfrac{1}{2}}} at the peak frequency of 1.7~Hz limited by atom shot noise.},
	label = {tab:operating_parameters},
	doinside = {\setlength{\tabcolsep}{30pt} \renewcommand{\arraystretch}{1.2}},
	width = 0pt,
	pos = t
	]{l r}{
		\tnote[a]{\SI{1e10}{\per \second} + 20 dB squeezing (in variance) or \SI{1e12}{\per \second}.}
	    \tnote[b]{Assuming 10 interleaved interferometers, \SI{1e9}{} atoms and 20~dB squeezing, see section \ref{subsec:atomic_interactions}}
	}{\FL
	\textbf{Atomic source} & \ML
	Species	& \(\mathrm{^{87}Rb}\) \NN
	Loading source & 2D\(+\) MOT \NN
	Equivalent atomic flux\tmark[a] & \SI{1e12}{\per \second} \NN
	Ensemble type & ultracold source \NN
	Expansion velocity ($T_\text{eff} \approx 100$~pK) & \SI{100}{\micro \meter \per \second} \NN
	Vertical launching velocity & \SI{4}{\meter \per \second} \NN
	Cloud size\tmark[b] & \SI{16}{\milli \meter} \ML
	\textbf{Detector} & \ML
	Single gradiometer & \NN
	\hspace{0.2cm} Configuration & Double loop, four pulses \NN
	\hspace{0.2cm} Interrogation time  & $4T=800$~ms \NN
	\hspace{0.2cm} Atom optics & Sequential Bragg \NN
	\hspace{0.2cm} Momentum transfer & $2n=1000~\hbar k$ \NN
	\hspace{0.2cm} Baseline & $L=16.3$~km \NN
	\hspace{0.2cm} Peak strain sensitivity (at 1.7~Hz) & \SI{4.1e-21}{\per \hertz \tothe{\sfrac{1}{2}}} \NN
	Full detector& \NN
	\hspace{0.2cm} Number of gradiometers per arm & $N=80$\NN
	\hspace{0.2cm} Gradiometer separation & $\delta=200$~m \NN
	\hspace{0.2cm} Total baseline & $L_T= 32.1$~km \NN
	\hspace{0.2cm} Peak strain sensitivity (at 1.7~Hz) & \SI{3.3e-22}{\per \hertz \tothe{\sfrac{1}{2}}} \LL
	}
\ctable[
    caption = {Noise requirements for the ELGAR detector as discussed in Sec.~\ref{sec_other_noise_couplings} to reach the designed strain sensitivity.},
    label = {tab:stability_parameters},
    doinside = {\setlength{\tabcolsep}{5pt}\hspace*{-28pt}
    \renewcommand{\arraystretch}{1.4}},
    width = 0pt,
    pos = t,
    ]{l r}{
    \tnote[a]{Based on the equivalent atomic flux in table~\ref{tab:operating_parameters}.}
    \tnote[b]{Assuming a rotation rate of \SI{73}{\micro \radian \per \second}.}
    \tnote[c]{Assuming an initial velocity noise of the atoms of \SI{50}{\nano \meter \per \second \per \hertz \tothe{\sfrac{1}{2}}}.}
    \tnote[d]{\(\Gamma \approx\) \SI{1.5e-6}{\per \second \tothe{2}} and assuming a source velocity noise of \SI{50}{\nano \meter \per \second \per \hertz \tothe{\sfrac{1}{2}}}.}
    \tnote[e]{The laser beam radius of curvature is \SI{25}{\kilo \meter} (Gaussian beam waist of \SI{50}{\milli \meter}).}
	\tnote[f]{Assuming a magnetic field gradient of \SI{1}{\nano \tesla \per \meter}.}
	\tnote[g]{Assuming a temperature gradient of \SI{0.1}{\kelvin \per \meter}.}
	\tnote[h]{Assuming an electric field gradient of \SI{0.1}{\volt \per \meter \tothe{2}}.
}
}{\FL
	\textbf{Noise source} & \textbf{Constraint} \ML
	Phase noise from atom shot noise \tmark[a] & $1 \ \mu$rad.Hz$^{-1/2}$ \ML
	Velocity noise (coupled to static rotation) \tmark[b] (\ref{sec_coupling_rotation}) & \SI{50}{\nano \meter \per \second \per \hertz \tothe{\sfrac{1}{2}}} \NN
	Static relative beam alignment \tmark[c] (\ref{sec_coupling_beam_misalignment}) & \SI{0.32}{\nano \radian} \NN
	Compensation of static gravity gradient \tmark[d] (\ref{sec_coupling_grav_grad}) & \(1\%\) \NN
	Relative beam angle jitter (coupled to imperfect velocity)  (\ref{sec_coupling_g} (ii)) & \SI{4}{\pico \radian \per \hertz \tothe{\sfrac{1}{2}}} \NN
	Laser beam pointing jitter (coupled to static gravity difference) (\ref{sec_coupling_g} (iii)) & \SI{30}{\pico \radian \per \hertz \tothe{\sfrac{1}{2}}} \NN
	Wavefront curvature coupled to velocity noise \tmark[e] (\ref{sec_coupling_wavefront}) & \SI{0.8}{\nano \meter \per \second \per \hertz \tothe{\sfrac{1}{2}}} \ML
	Acceleration noise from (\ref{sec_mag_elect_temp}) & \NN
	\hspace{0.2cm} Magnetic field \tmark[f] & \SI{0.1}{\nano \tesla \per \hertz \tothe{\sfrac{1}{2}}} \NN
	\hspace{0.2cm} Blackbody radiation\tmark[g] & \SI{2}{\milli \kelvin \per \hertz \tothe{\sfrac{1}{2}}} \NN
	\hspace{0.2cm} DC Stark shift\tmark[h] & \SI{0.2}{\volt \per \meter \per \hertz \tothe{\sfrac{1}{2}}} \LL
}

\begin{figure}[hbt]
    \centering
    \includegraphics[keepaspectratio,width=10cm]{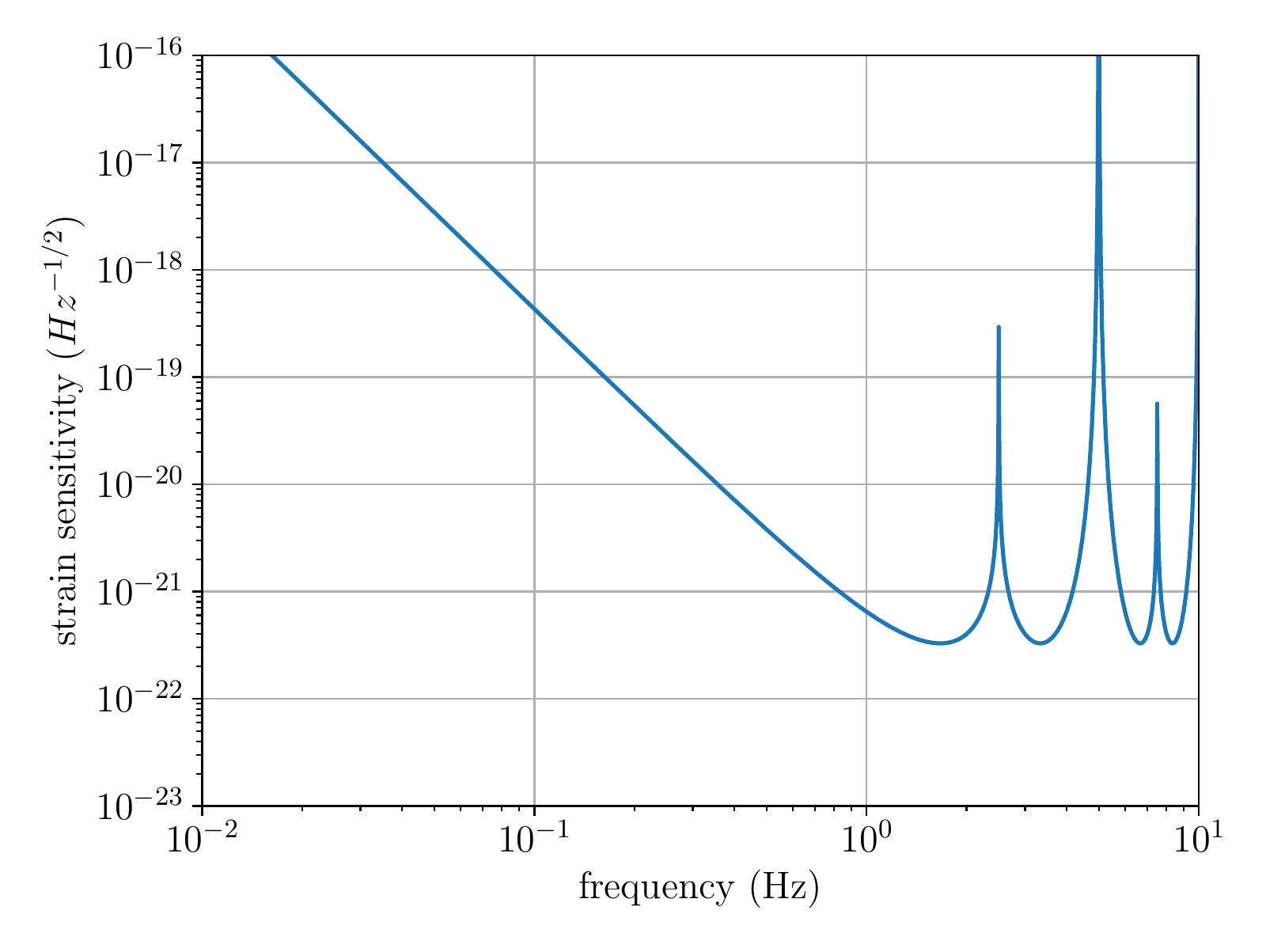}
    \caption{Strain sensitivity of ELGAR limited by atom shot noise, corresponding to a white phase noise of $1 \ \mu$rad.Hz$^{-1/2}$, with the parameters of table \ref{tab:operating_parameters}. The other noise sources are kept below atom shot noise according to the parameters of table \ref{tab:stability_parameters}.}
    \label{fig_strain_sensitivity}
\end{figure}

The configuration parameters  of the ELGAR detector and the noise requirements  are summarised in tables~\ref{tab:operating_parameters}~and~\ref{tab:stability_parameters}, respectively. The table~\ref{tab:stability_parameters} for the noise requirements is  subdivided into three parts.
The first part includes the atom shot noise that is, by design, the dominant noise contribution.  The second part accounts for the controllability of atomic motion and atom optics, the mutual couplings, and the couplings to the environment.
The third part covers the influence of the environment, i.e. static electric and magnetic field and the temperature, to the atoms.
The atom shot noise limited sensitivity curve of ELGAR, calculated with the parameters of table \ref{tab:operating_parameters}, is illustrated in Fig.~\ref{fig_strain_sensitivity}. 
Seismic and Newtonian noises are reviewed in sections~\ref{sec_seis} and \ref{sec_NN_reduction} respectively and the strategies adopted to mitigate them below the atom shot noise are  discussed therein. All the other noise sources are then kept below atom shot noise according to the parameters of table \ref{tab:stability_parameters}.

In Fig.~\ref{fig_strain_sensitivity}, the interrogation time $T$ sets the corner frequency of maximum strain sensitivity at 1.7~Hz.
The resonance peaks correspond to the windowing effect of the atom interferometer captured by the transfer function $\omega |G(\omega)|$~\cite{leveque}.
As discussed in section~\ref{sec_Atom_optics}, ELGAR can accommodate different geometries including single- and folded triple-loop to improve the sensitivity at lower frequency or better suppression of the spurious phase terms.
As shown in Fig.~\ref{fig:atom-optics-strain-sensitivities}, by varying slightly the interrogation time $T$, ELGAR can operate in broadband mode to detect unknown gravitational wave signals and later switch to resonance mode to enhance the signals at specific frequencies.

\subsection{ELGAR within the global framework of GW Detectors}

\begin{figure}
    \centering
    \includegraphics[keepaspectratio,width=10cm]{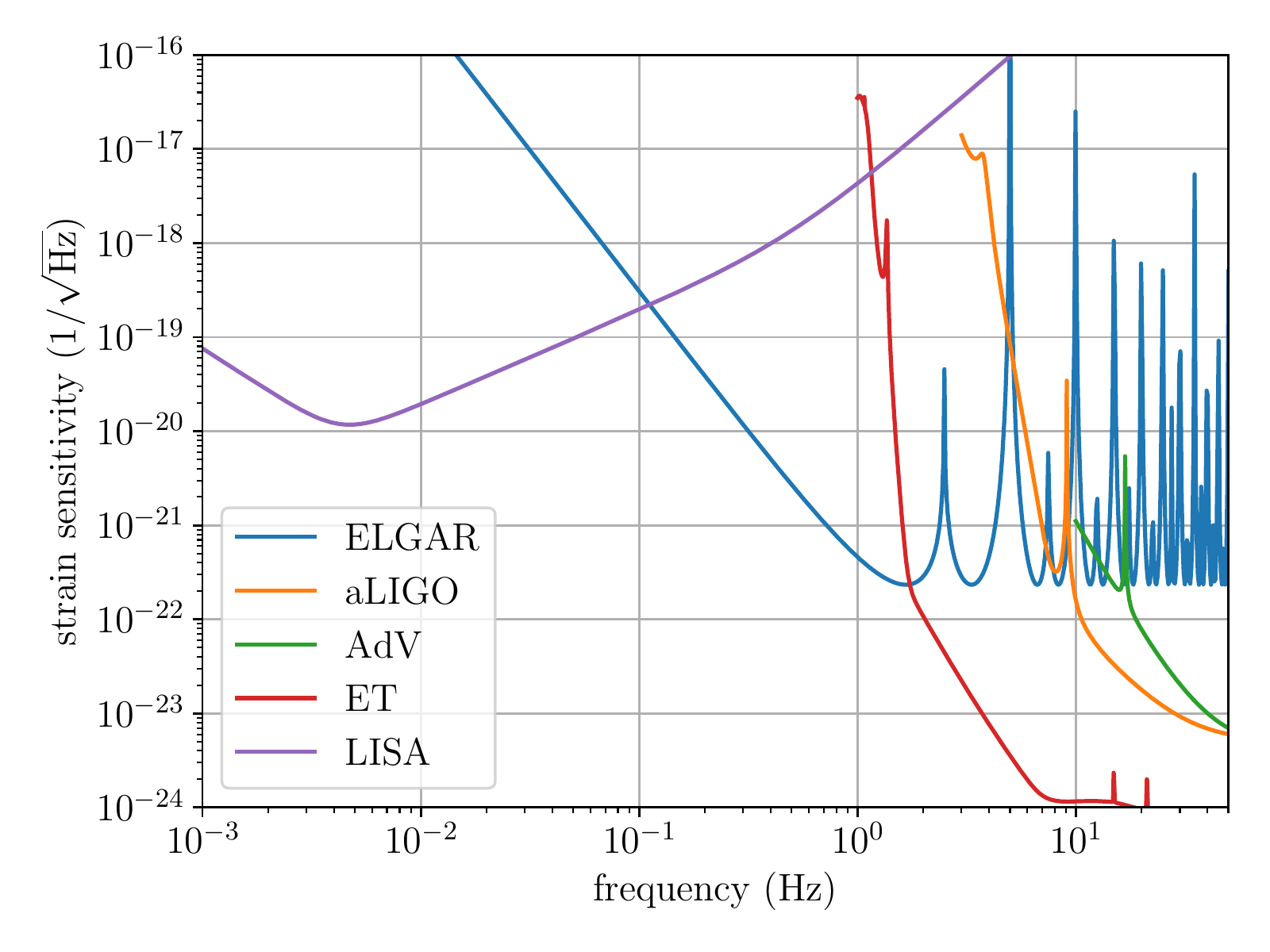}
    \caption{Strain sensitivities for different gravitational wave detectors, which include ELGAR (atom shot noise limit), aLIGO, ET and LISA, cover the frequency range from \SI{10}{\milli \hertz} to \SI{10}{\hertz}.~\label{fig_GWD_comparison}}
\end{figure}

ELGAR will complement the existing optical gravitational-wave instruments such as Advanced LIGO (aLIGO) and Advanced Virgo (AdV), and the future detectors such as Einstein Telescope (ET) and LISA, by covering a frequency band gap between the sensitivity curves of ground-based and space detectors.
The strain sensitivities for different detectors including ELGAR, aLIGO, AdV, ET and LISA are illustrated in Fig.~\ref{fig_GWD_comparison}.
As seen in the previous section, the resonance peaks in the ELGAR sensitivity curve correspond to the windowing effect of the atom interferometer and can be overcome by operating the antenna in a broadband mode (see Sec.~\ref{sec_operating_mode}).
The ground-based detectors such as aLIGO or AdV are limited by seismic and thermal noises for frequency below 10~Hz.
The space-based detectors including LISA can operate at a much lower frequency band from 0.1~Hz to mHz thanks to the large absence of Newtonian gravitational fluctuations in space.
ELGAR therefore offers a unique opportunity to explore GWs in the deciHertz band where an ambitious science program can be carried out. The new possibility offered by ELGAR for astrophysics, gravity and fundamental physics will be detailed in Sec.~\ref{sec_Sources_of_GWs}.

\subsection{Data Management for the GW and geophysical, atmospheric and environmental monitoring communities}
The ELGAR detector will generate two kinds of data: the atom interferometry strain sensor output, providing the time series $h(t)$ of the GW induced strain, and a large number of ancillary data from a whole range of environmental sensors.
The technical and administrative intricasies has been successfully tackled by the current network of optical GW detector operators.
The advantages of common and coordinated management of data among different detectors are well known.
Individual GW detectors are poorly directional, and source localization requires the combination of the outputs from multiple, simultaneous distant detectors.
Lower frequency sensors provide earlier data, which could be used to prime or fine-tune other detectors~\cite{Cannon2012TAJ}.
More generally, most GW searches require the use of a network of detectors.

ELGAR will produce strain data at much higher spatial resolution in a frequency band substantially lower than those of the existing detectors.
This offers the unique opportunity to compare this gravitational data with other local measurements.
Inversely the GW final strain data will benefit from the combination of existing and new data, to disentangle the effect of GW from that of classical Newtonian noise.
Thus, besides the interest for sharing the $h(t)$ strain data for GW observation, ELGAR will produce a massive amount of data with potential interest for geophysical, atmospheric and environmental monitoring, both in the gravitational information embedded in the atom interferometry signals and in the ancillary data from environmental sensors.
The exploration of the parameter space of gravitational-wave sources is a large-scale effort that involves many research groups and relies on interaction between gravitational-wave source modelling groups and the gravitational-wave data-analysis community.
The gravitational (and thus mass-transport) data also provide a completely novel input to the geophysical, atmospheric and environmental monitoring communities.

State-of-the-art methods seek to address these challenges by considering data encoded in large-scale matrices and employ tools like Principal Component Analyses (PCA), Singular Value Decomposition (SVD) and Non-Negative Matrix Factorization (NMF).
Despite their capabilities, a number of inherent limitations characterize these approaches, including the inability to encode high-dimensional observations or data from multiple sources/modalities.
Furthermore, these approaches are data-agnostic, which limits their potential for the specific setting.
Within ELGAR, novel mathematical frameworks will be exploited and developed, by modelling observation using high-dimensional data structures known as multi-way tensors, investigating approaches like robust tensor PCA, and low-rank tensor recovery.
Furthermore, we will consider exploiting available observations into a supervised machine learning paradigm in order to introduce methods like deep tensor neural networks for the optimal representation of measurements.

Specifically, the envisioned data management framework will offer a number of novel capabilities in terms of multi-modal high-dimensional data observation quality enhancement.
First, it will support the separation of the contribution of different noise sources from the observations by identifying the essential statistical characteristics, both in the time as well as the frequency domain for each signal category.
In addition, it will offer the ability to perform joint analysis of multiple time series of in-situ measurements from deployed sensing networks.
By developing a unified framework, a variety of signals from disparate origins, represented in different forms (e.g., timeseries and imagery) will be (jointly) processed in order to characterize their spatio-temporal evolution, and thereby facilitate the calculation of the geophysical computational models.
Last, it will act as the core framework for the imputation of missing measurement from different spatial locations and temporal instances, recovering lost measurements due to sensor and/or storage failures, exploiting properties like model sparsity and low-rank characteristics.

In addition to the enhancement of the quality of observations, the developed framework will also support the clustering and detection of anomalies through cutting-edge signal modeling and learning.
This objective will be achieved by autonomously generating the nominal data space, considering key operational characteristics parameters of each sensor category (e.g., nominal range of values, sensitivity, accuracy, and drift).
Based on this information, the sensing uncertainties will be automatically estimated, exploiting their representation in lower-dimensional spaces, which will both automatically update the nominal operational bounds, as well as indicate when specific regions of data exceed these bounds, due to the existence of short (e.g. outlier) or longer (e.g., hardware) failure of the sensing infrastructure.
To accommodate the expected increased volume of the data collected and the expectation of providing reduced, yet equivalently informational observations, we will employ cutting-edge machine learning algorithms, focusing on deep learning architectures, for inferring the optimal anomaly detection policies in a completely unsupervised way \cite{chalapathy2019deep}.
Furthermore, we will extend state-of-the-art by simultaneously considering observations from multiple modalities/sources and time-instances \cite{yang2017deep} into the anomaly detection policies.

For supporting the above goals, the data storage and access needs of ELGAR will be studied in order to propose the proper storage and computing infrastructure to meet them.
In order to estimate infrastructure requirements, the landscape of the ELGAR's data will be analyzed in four dimensions: a) their size, b) their access patterns, c) their processing needs, and d) the robustness of the proposed approach(es).
Then the serving model of the data will be analyzed along with the different computing infrastructure options that can better serve the data requirements.
User-friendly and customizable ways to access the data and formulate information needs will also be consider, mainly based on the Ontology-Based Data Access (OBDA) paradigm and exploratory approaches, allowing the integration and interoperation among different scientists and data repositories.
Finally, ELGAR will  ensure that the data it provides will implement the FAIR Data  principles \cite{wilkinson2016fair}, through the use of metadata catalogs that support the  discovery of digital assets. 
\externaldocument{Sec_08}

\section{Astronomy and physics with ELGAR}\label{sec_Sources_of_GWs}

The first few years of Gravitational Wave Astronomy, made possible by the large optical instruments LIGO~\cite{Aasi2015} and Virgo~\cite{Acernese2014}, have  significantly altered our knowledge and expectations about GW sources. The first detection, the famous GW150914 event~\cite{Abbott2016}, consisting of a pair of $36 M_\odot + 29 M_\odot$ black holes (BHs), merging into a final $62 M_\odot$ BH, and the subsequent multiple GW observations have demonstrated that binary black hole (BBH) signals are by a factor $\mathcal{O}(10)$ the dominant component of the observed mass spectrum~\cite{LIGOScientific:2018mvr} of GW sources.  Earlier, binary neutron star (BNS) coalescences were considered the most likely source of GWs: the period evolution in systems like the Hulse-Taylor Binary Pulsar (PSR B1913+16) with the predicted loss of orbital energy had provided indirect confirmation of the emission mechanism~\cite{Hulse:1974eb} (see also~\cite{Weisberg:2016jye}), and the observational evidence, by way of the pulsar phenomenon, of several BNS systems in our galaxy resulted in a credible estimate for an event rate~\cite{BNScensus:2005,RatesPaper:2010}. These predictions were actually confirmed when in the summer of 2017 LIGO and Virgo observed the first BNS event~\cite{GW170817:event}, yet BBH events remain prevalent.  The abundance of BBH events is {\it a posteriori} understandable: The GW signal amplitude scales roughly as $h \propto M^{5/3}$, hence a $\sim 30 + 30 M_\odot$ BBH system is roughly detectable $\mathcal{O}(10)$ farther away than a $10 + 10 M_\odot$ BBH system, previously taken as a benchmark. The observed volume scales approximately with the cube of the maximum observable distance, which explains the observed $\mathcal{O}(10^3)$ rate enhancement over the benchmark. We remark that even a single BNS observation has had an immense scientific value, also thanks to the association with a gamma-ray burst~\cite{GW170817:gw-grb}, confirming a long standing hypothesis about the GRB origin~\cite{BNS-GRB:1984}.  From the LIGO-Virgo observations during the two first observing runs~\cite{LIGOScientific:2018mvr}, O1 and O2, the  merger rate estimated for BNSs is $110 - 3840\,$Gpc${}^{-3}$yr${}^{-1}$ ($90\%$ confidence intervals), and for BBHs is $9.7 - 101\,$Gpc${}^{-3}$yr${}^{-1}$.  Given that there are no established detections for BH-NS mergers, only upper limits can be established and all of them (also $90\%$ confidence intervals) are below $610\,$Gpc${}^{-3}$yr${}^{-1}$.

The observations of LIGO and Virgo belong to the high-frequency band, between $1 - 10^{4}\,$Hz.  There are two other bands where there is significant progress towards the direct direction of GWs: (i) The low frequency band, between $10^{-4}-1\,$Hz, not accessible from ground due to seismic and gravity-gradient noises, where space-based detectors can operate. (ii) The very-low frequency band, between $10^{-9} - 10^{-6}\,$Hz, the realm of Pulsar Timing Arrays (PTAs).  The low-frequency band has already a space mission selected (on 2018), the Laser Interferometer Space Antenna (LISA)~\cite{2017arXiv170200786A}, the L3 mission of the European Space Agency with a launch date expected in 2034. LISA will consists in a triangular constellation of three spacecrafts exchanging laser beams and 2.5 million km of arm-length trailing the Earth on a heliocentric orbit. The required sensitivity is attained by suppressing the laser frequency noise below the secondary noises by  a combination of laser frequency locking and noise cancellation via Time-Delay Interferometry. The LISA Pathfinder mission~\cite{Armano:2016bkm,Armano:2018kix,Armano:2019dxs} has demonstrated, between December 3rd, 2015 and June 20th, 2017, the main technology for LISA. LISA  has a very wide science case as described in the white paper {\em The Gravitational Universe}~\cite{Seoane:2013qna}.  On the other hand, in the  very-low frequency band several consortia of radiotelescopes measure, over long time spans, the time of arrival of radio-pulses emitted by well-chosen set of (millisecond) pulsars.  By correlating the measurements of the different pulsars, deviations in the times of arrival of the radio-pulses due to passing gravitational waves are sought. There are three Pulsar Timing Array (PTA) collaborations: The European PTA(EPTA~\cite{Desvignes:2016yex}), the North American Nanohertz Observatory for Gravitational Waves (NANOGrav~\cite{Arzoumanian:2017puf}), and the Parkes PTA (PPTA~\cite{Reardon:2015kba}). They form the International PTA (IPTA~\cite{Perera:2019sca}) with the aim of enhancing the sensitivity by combining the data of the individual PTAs.  The sensitivity of these PTAs is already inside the predicted discovery space for GW backgrounds produced by the emission of inspiraling supermassive black hole binaries, with masses between $10^{8}-10^{10}\,M_{\odot}$. 
 
The decihertz GW band, between $0.1-10\,$Hz, and where we do not have any current detector (not even approved for construction), is very rich in GW sources which is reflected in the fact that it is the bridge between two distinctive GW bands in terms of sources, the high-frequency band, where second-generation ground based detectors have already observed a number of sources~\cite{LIGOScientific:2018mvr}, and the low-frequency band, where LISA~\cite{2017arXiv170200786A} will operate.  The main sources for this band are: {\em Compact Binaries}. The components of these binaries can be diverse: Mainly white dwarfs, NSs and BHs.  Among the BHs we have to differentiate them according to mass (stellar-mass BHs and intermediate-mass BHs) and origin (stellar origin, globular clusters, early universe). It is also important to understand to which stages of binary-black coalescence (inspiral, merger and ringdown) ELGAR is sensitive.  In principle, ELGAR should detect:  (i) The merger (and ringdown) of intermediate mass BBHs (with masses in the range $10^{2}-10^{4} M_{\odot}$). (ii) The inspiral phase and  of stellar-mass binaries, like BBH coalescence; BNS coalescence; neutron star-black hole coalescences; even binaries containing white dwarfs. (iii) Stochastic gravitational-wave backgrounds for ELGAR.  We have to distinguish backgrounds due to the emission of many compact binary inspirals  from those produced during the early-universe by means of high-energy processes.

\subsection{Extending the BBH spectrum}

The large number of BBHs detected motivates a considerable effort to better understand the origin of these systems, whether they result from the common evolution of pairs of massive stars, or form through capture mechanism in dense stellar environments~\cite{BBHs:O1O2properties}. To fully answer these questions, we want to characterise and extend the mass spectrum of these systems: are BBH pairs like GW150914 the most massive we should expect? Are there more massive systems to be detected, that we cannot just see yet? And it is important to measure accurately parameters like spin magnitudes and directions, which carry information about the past evolution of the system.

To address these questions meaningfully, it is necessary to enlarge the window of observation towards lower frequencies, since the maximum frequency of the GWs emitted roughly scales as $M^{-1}$; an event $30$ times more massive than GW150914 would be confined at frequencies below $\sim 10\,$Hz, where LIGO and Virgo are essentially blind because of seismic and suspension thermal noise walls.
In the long term, the ground-based Einstein Telescope (ET)~\cite{Punturo2010} will push the lower frequency limit down to $\sim 3\,$Hz, thus considerably widening the range of detectable masses~\cite{ETscience:2012}, whereas the space based LISA detector~\cite{2017arXiv170200786A} will open up the mHz - Hz range to observation, making possible to detect very massive systems, and extreme events like the infall of matter into supermassive BHs.

\begin{figure}
\centering
\includegraphics[width=0.8\textwidth]{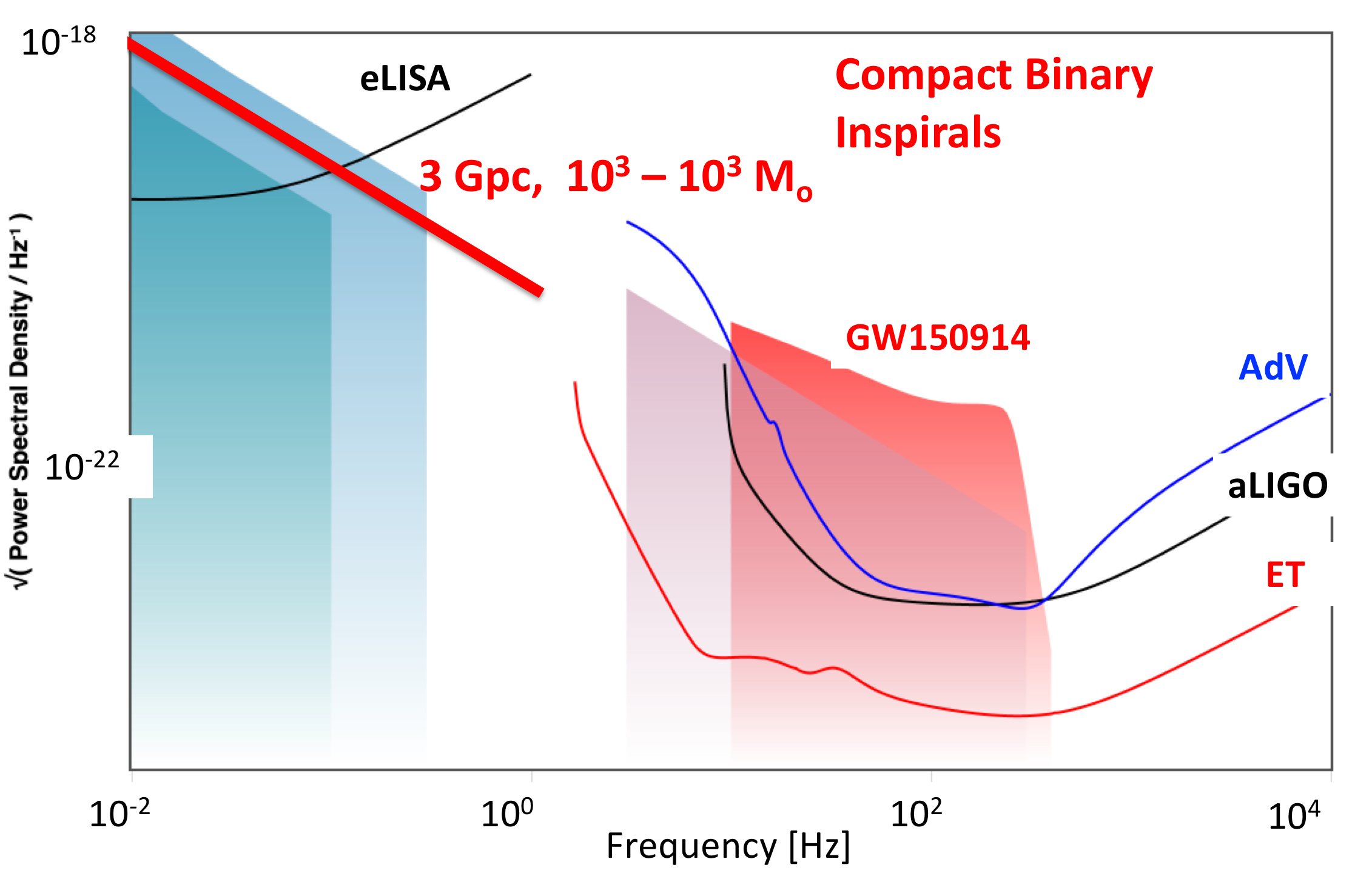}
\caption{\label{fig:binaries} Design sensitivity of the second-generation ground-based detectors LIGO and Virgo, the future third-generation ground-based detector ET, and of the future space-based detector LISA, along with the trace of the BBH event GW150914, of an hypothetical IMBH event at $3\,$Gpc, and spectra of BNS events. Figure obtained using the GWplotter tool~\cite{Moore2015CQG}.}
\end{figure}

However, we can see in Fig.~\ref{fig:binaries} that a gap will remain which could prevent, for instance, to directly observe the merger phase of the so-called Intermediate Mass Black Holes (IMBH), systems including BHs of $\mathcal{O}(10^3 M_\odot)$. Filling this gap is one of the purposes of an AI designed for the detection of GWs; evidence for such systems could shed light on the possible existence of a ladder of BH masses, from stellar mass to supermassive ones. An absence of evidence could indicate that entirely different formation mechanisms are at play in different mass ranges.

We can predict the waveforms emitted by massive BBHs; considering that the only relevant physical parameters are the Newton constant $G_N$ and the speed of light $c$, from dimensional analysis only we have that
\begin{equation}
h\left(t, d, M_1, M_2, \mathbf{S}_1, \mathbf{S}_2\right) = h\left(\frac{t}{G_N \mu c^{-3}}, \frac{d}{G_N \mu c^{-2}}, \frac{M_1}{\mu}, \frac{M_2}{\mu},\frac{\mathbf{S}_1}{G_N \mu^2 c^{-1}}, \frac{\mathbf{S}_2}{G_N \mu^2 c^{-1}}\right)\,,
\end{equation}
where $\mathbf{S}_{1,2}, M_{1,2}$ are the BHs spins and masses, and $\mu$ is an arbitrary mass scale. In a system (say) $\lambda$ times more massive than a reference one (could be GW150914 for instance), noticing that the waveform is independent of $\mu$, the equivalence holds
\begin{equation}
h\left(t, d, \lambda M_1, \lambda M_2, \lambda^2 \mathbf{S}_1, \lambda^2 \mathbf{S}_2\right) = h\left(\frac{t}{\lambda}, \frac{d}{\lambda}, M_1, M_2, \mathbf{S}_1, \mathbf{S}_2\right)\,.
\end{equation}
In the frequency domain
\begin{eqnarray}
\hat{h}(f, d, \lambda M_1, \lambda M_2) &\equiv& \int_{-\infty}^{+\infty} h(\frac{t}{\lambda}, \frac{d}{\lambda}, M_1, M_2) e^{-2\pi i f t} dt\\
&=& \lambda \int_{-\infty}^{+\infty} h(\tau, \frac{d}{\lambda}, M_1, M_2) e^{-2\pi i \lambda f \tau} d\tau = \lambda^2\,\hat{h}(\lambda f, d, M_1, M_2)\nonumber\ ,
\end{eqnarray}
where in the last equation we have exploited the $\frac{1}{d}$ dependency of the waveform on distance. Hence the frequency evolution $f_l \rightarrow f_u$ for a system with masses $\lambda M_{1,2}$ is mapped onto the evolution
$\lambda f_l \rightarrow \lambda f_u$ of a system with masses $M_{1,2}$ at the same distance, including a $\lambda^2$ factor on amplitude and $\lambda$ scaling in the frequency. It follows a scaling by $\lambda^{3/2}$ of the $SNR$

\begin{equation}
SNR \equiv 2\sqrt{\int_{f_l}^{f_u} \frac{\left|\hat{h}(f, d, \lambda M_1, \lambda M_2)\right|^2}{S_n(f)} df} = 2\lambda^{3/2} \sqrt{\int_{\lambda f_l}^{\lambda f_u} \frac{\left|\hat{h}(\nu, d, M_1, M_2\right|^2}{S_n({\frac{\nu}{\lambda})}} d\nu}\ .
\end{equation}

For instance, a $10^3 + 10^3 M_\odot$ BBH ($\lambda \simeq 30$ with respect to GW150914), located at 3 Gpc could be observed in the band $1 - 10\,$Hz with $SNR \simeq 5$ using a detector having a noise floor $\mathcal{O}(10^{-21} {\rm Hz}^{-1/2})$; even though this noise floor is $100$ times worse than Advanced LIGO or Virgo, the longer duration in band of the signal would allow to accumulate a comparable $SNR$.

\subsection{Binaries of intermediate-mass black holes}
\label{sec.IMBHbinaries}

The size of a massive black hole (MBH) in comparison with its host galaxy is
tiny. For instance, a MBH with a mass of $10^6\,M_{\odot}$ the difference spans
over ten orders of magnitude.  In spite of this, there is a link between the
properties of the host galaxy and the MBH. The mass of the MBH and the velocity
dispersion $\sigma$ of the spheroidal component is one of the best
examples~\cite{KormendyHo2013}. However, the low end of this ``mass-sigma''
correlation is uncertain because we lack data.  By extending this correlation
to smaller systems, we expect that dense stellar systems such as clusters
harbor intermediate-mass black holes, IMBHs, i.e. black holes with masses
$10^2-10^4,\,M_{\odot}$ (see~\cite{Mezcua2017} for a review).

For two IMBHs to form a binary emitting GWs, there are two different
possibilities.  First, (i) two host clusters born in a ``cluster of clusters'',
as those which we observe in the Antenn{\ae} galaxy (see introduction
of~\cite{ASF06}), can be gravitationally bound, and doomed to merge with each
other. In this process, the IMBHs sink to the center of the system due to
dynamical friction first, to form a binary (BBH), and later that binary shrinks
its semi-major axis due to interactions with stars. After each of such
interactions the star can take away a bit of the energy of the BBH. If there
are enough stars to interact with, the BBH evolves and eventually enters the
GW-dominated regime.  This means that the main driving mechanism in the
evolution of the BBH are GWs, so that the orbit circularizes. This process has
been described in the works of~\cite{ASF06,Amaro-SeoaneEtAl09a,AS10a}. A
space-based observatory such as LISA should detect these systems with an event
rate of $4-5~{\rm yr}^{-1}$.  Another possibility of a BBH to form is (ii) via
the runaway formation of two very massive stars in a single cluster with a
binary fraction of $10$\%, as described in~\cite{GFR06}.
In~\cite{FregeauEtAl06} it is estimated that LISA and Advanced LIGO should
detect tens of them, but this depends on the distribution of the masses of the
host cluster and also their central densities.

\begin{figure}
\resizebox{\hsize}{!}
          {\includegraphics[scale=1,clip]{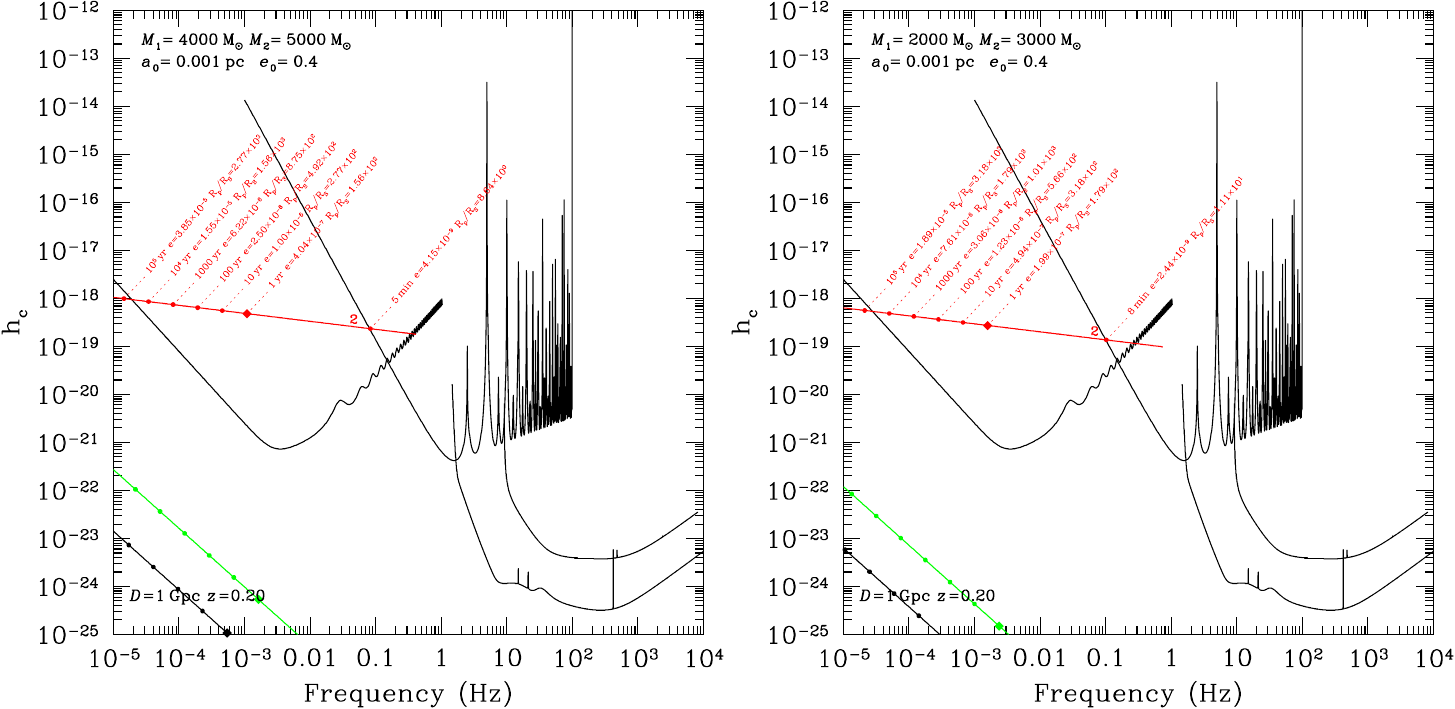}}
\caption
   {
\textit{Left panel: }Characteristic amplitude $h_c$ as a function of frequency for a BBH of masses
$4000\,M_{\odot}$ and $5000\,M_{\odot}$. We show the initial dynamical
parameters on the two left corners. The BBH initially has an eccentricity
of $e_0=0.4$, an initial semi-major axis of $a_0=10^{-3}\,{\textrm pc}$ 
and are is located at a distance of $D=1\,\textrm{Gpc}$. We show the
sensitivity curve for ELGAR starting at $10^{-3}\,{\textrm Hz}$ and 
$h_c \sim 10^{-14}$. We also show the curve for LISA starting at the lowest
frequency, the ET starting at $1\,{\textrm Hz}$ and LIGO, at $10\,{\textrm Hz}$.
We display the first harmonics of the GW in the approximation of~\cite{Peters64}
and show different moments in the evolution with circles 
on the top of the second one. The last diamond corresponds to one year before
the merger. At each of these, we additionally give
the eccentricity and the periapsis distance normalised to the Schwarzschild
radius. The source enters the ELGAR regime 5 minutes before merger. \textit{Right panel: }Same as the left panel but for slightly different
masses, and 8 minutes before merger.
   }
\label{fig.BBHs_ELGAR}
\end{figure}

In the work of~\cite{Amaro-SeoaneSantamaria10} it is shown that a BBH can spend
a significant amount of its evolution before the merger in the decihertz
regime. 
In Fig.~\ref{fig.BBHs_ELGAR} we show two BBH spending a significant amount of
its evolution before the merger in the ELGAR regime.  This means that these
systems are perfect targets for ELGAR. A joint detection of these BBHs with
ELGAR and the early inspiral with LISA would help to obtain enhanced
measurements of the parameters and also to potentially break degeneracies.  We
can see on the right panel that the system mergers out of the LISA window, and
spends 8 minutes in the ELGAR sensitivity window and merges at about
$1\,{\textrm Hz}$. This means that if an observatory such as LISA was operative
at the same time as ELGAR, it could give a warn of years in advance and seconds
of precision to it.

\subsection{Intermediate-mass ratio inspirals}
\label{sec.IMRIs}

Compact objects such as stellar-mass black holes, neutron stars or white dwarfs
can form a binary with a MBH and contrary to extended stars, approach closer
distances without suffering significant tidal stresses. In the process the
binary system emits gravitational radiation as orbital energy is lost.  Such
radiation provides us with detailed information about the binary and allows us
to test gravity in the strong regime.  Due to the mass ratio, we talk about
extreme-mass ratio inspirals because $q \gtrsim 10^4:1$ (EMRI,
see~\cite{Amaro-SeoaneLRR,Amaro-SeoaneGairPoundHughesSopuerta2015}). Such
EMRIs are one of the main objectives of LISA~\cite{2017arXiv170200786A} and
form in galactic nuclei.

Smaller mass ratios, of $q \sim 10^2-10^4:1$ correspond to ``intermediate-mass
ratio inspirals'' (IMRIs, see
e.g.~\cite{Amaro-SeoaneEtAl07,BrownEtAl2007,RodriguezEtAl2012}), and are
optimal sources of ELGAR. IMRIs form in smaller systems, in globular clusters,
with typically very high eccentricities, via two-body relaxation or the braking
mechanism, i.e. the parabolic formation of a binary, as described
in~\cite{Hansen1972,QuinlanShapiro1989,KocsisEtAl2006,MandelEtAl2008,OlearyEtAl09,LeeEtAl2010,HongLee2015}).
The work of~\cite{KonstantinidisEtAl2013} is the first numerical simulation
which shows the formation and evolution of an IMRI in a globular cluster. The
authors observe a binary of a stellar-mass black hole of mass $26\,M_{\odot}$
forming with the central IMBH, of mass $M_{\rm BH}=500\,M_{\odot}$ with an
initial semi-major axis of $a \sim 10^{-5}$ pc and an eccentricity of
$e=0.999$, which fits in the parabolic capture mechanism
of~\cite{QuinlanShapiro1989}. Due to the relativistic
recoil~\cite{CampanelliEtAl2006,BakerEtAl2006,GonzalezEtAl2007}, the merged
system leaves the whole cluster. After this initial paper, other independent
groups have found similar results, such as~\cite{LeighEtAl2014}
and~\cite{MacLeodEtAl2016}. In particular, the simulations
of~\cite{HasterEtAl2016} are basically a reproduction of the set-up
of~\cite{KonstantinidisEtAl2013} but with a different numerical scheme which
however remarkably lead to a very similar result.

\begin{figure}
\centering
\includegraphics[width=1.0\columnwidth]{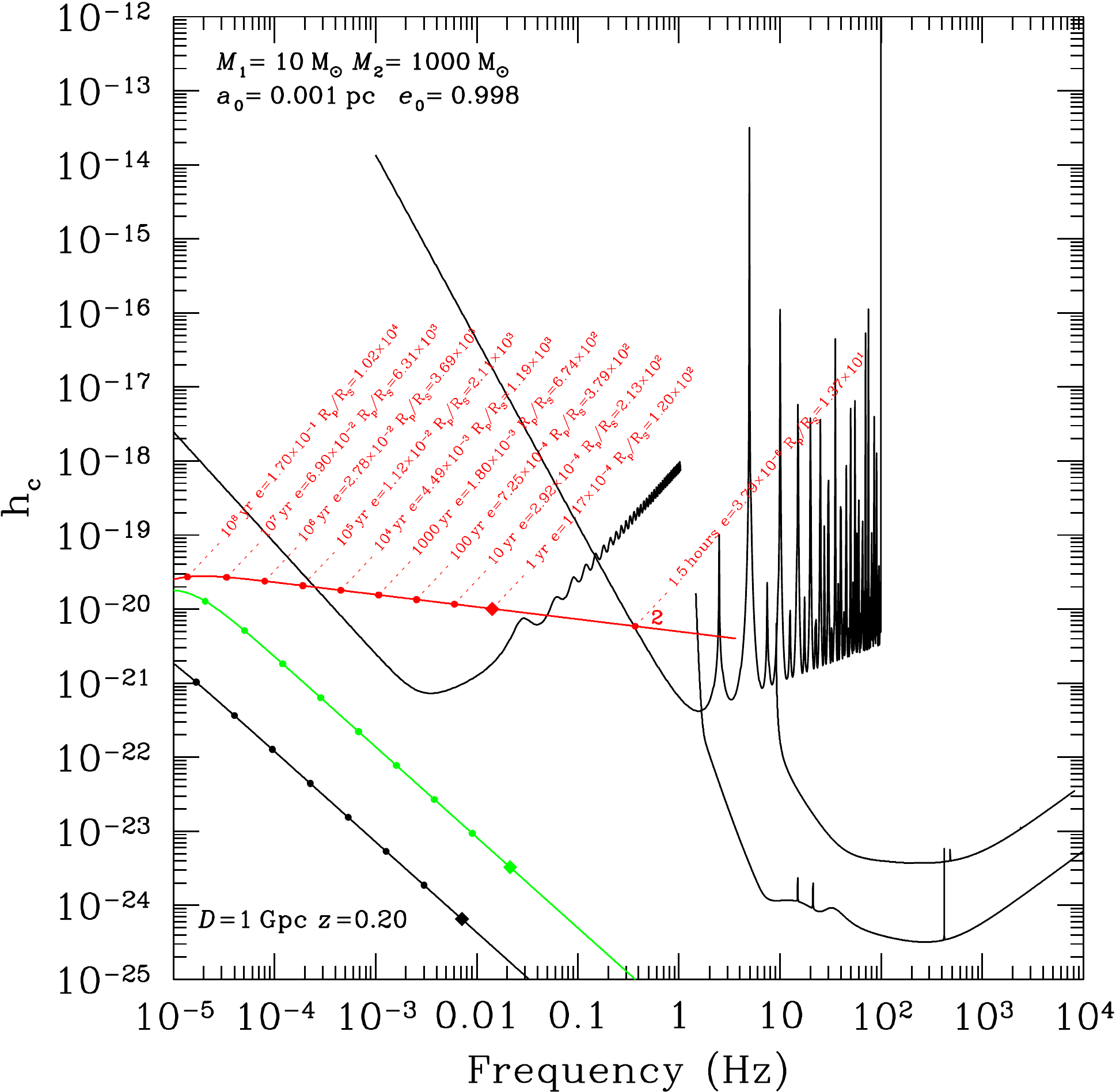}
\caption{
Same as Fig.~\ref{fig.BBHs_ELGAR} but for an IMRI of masses $1000\,M_{\odot}$ and $10\,M_{\odot}$.
The initial eccentricity is of $e_0=0.998$, as we expect from these systems (see \cite{Amaro-Seoane2018,Amaro-SeoaneLRR}),
and an initial semi-major axis of $a_0=10^{-3}\,\textrm{pc}$.
The IMRI spends 1.5 hours in the ELGAR regime. LISA can forewarn ELGAR years in advance and
with at least seconds of precision about the time when the source enters in its domain and provide
detailed information about all dynamical parameters of the system.
}   
\label{fig.ELGAR_IMRI}
\end{figure}

Ground-based detectors such as the proposed ET~\cite{Punturo2010,HildEtAl2011}
(with SNRs of up to $\sim 2000$, and are of typically a few hundred and of tens
for masses up to $2000\,M_{\odot}$) but also LIGO and Virgo, can detect the
final stages in the evolution of IMRIs (with SNRs of a few tens) and, depending
on the eccentricity, jointly with LISA (with SNRs of a few to tens), as shown
in~\cite{Amaro-Seoane2018}.  Due to the range of frequencies that these
sources have, a decihertz observatory such as ELGAR can also have an important
impact.  In Fig.~\ref{fig.ELGAR_IMRI} we show the characteristic amplitude in
function of the frequency of an IMRI, as well as a few
different moments in the evolution.

It has also been shown in the work of~\cite{Amaro-Seoane2018} that by adding a
decihertz observatory such as ELGAR, we can impose better constraints on the
system's parameters.  LISA and, later, ELGAR can measure the inspiral and hence
derive parameters such as e.g. the chirp mass. At more advanced stages in the
evolution, LIGO/Virgo can detect the merger and ringdown, which allows us to
derive the final mass and spin. Due to this combined detection, and contrary to
an individual one, one could break various degeneracies and obtain enhanced
measurements of the parameters.

\subsection{Stellar-mass binaries}
\label{sec.stellarmass}

The LIGO and Virgo observatories cannot discern the formation channel of
stellar-mass BH binaries (BBH).  In the work of~\cite{Amaro-SeoaneChen2016} is
shown that BBH formed either dynamically or in the field via stellar evolution
share identical properties, i.e. masses larger than the nominal
$10\,M_{\odot}$, a mass ratio ($q\equiv M_2/M_1$) of about $1$, low spin, and
nearly circular orbits. 

A compound detection of these detectors with LISA could allow us to to study
different moments in the evolution of BHBs on their way to
coalescence~\cite{2017arXiv170200786A}.  LISA can detect BHBs when the BHs are
still $10^2-10^3~R_S$ apart, years to weeks before they enter the LIGO/Virgo
band.  At such separations, the orbital eccentricity in principle bears the
imprint of the formation channel because (i) BHBs in dense stellar systems form
on systematically more eccentric orbits and (ii) the GW radiation at this stage
is too weak to circularize the orbits. Therefore, a priori detecting circular
binaries with LISA typically would imply a formation in the field, while a
detection of LISA of eccentric ones should have formed dynamically.

\begin{figure}
\centering
\includegraphics[width=0.8\columnwidth]{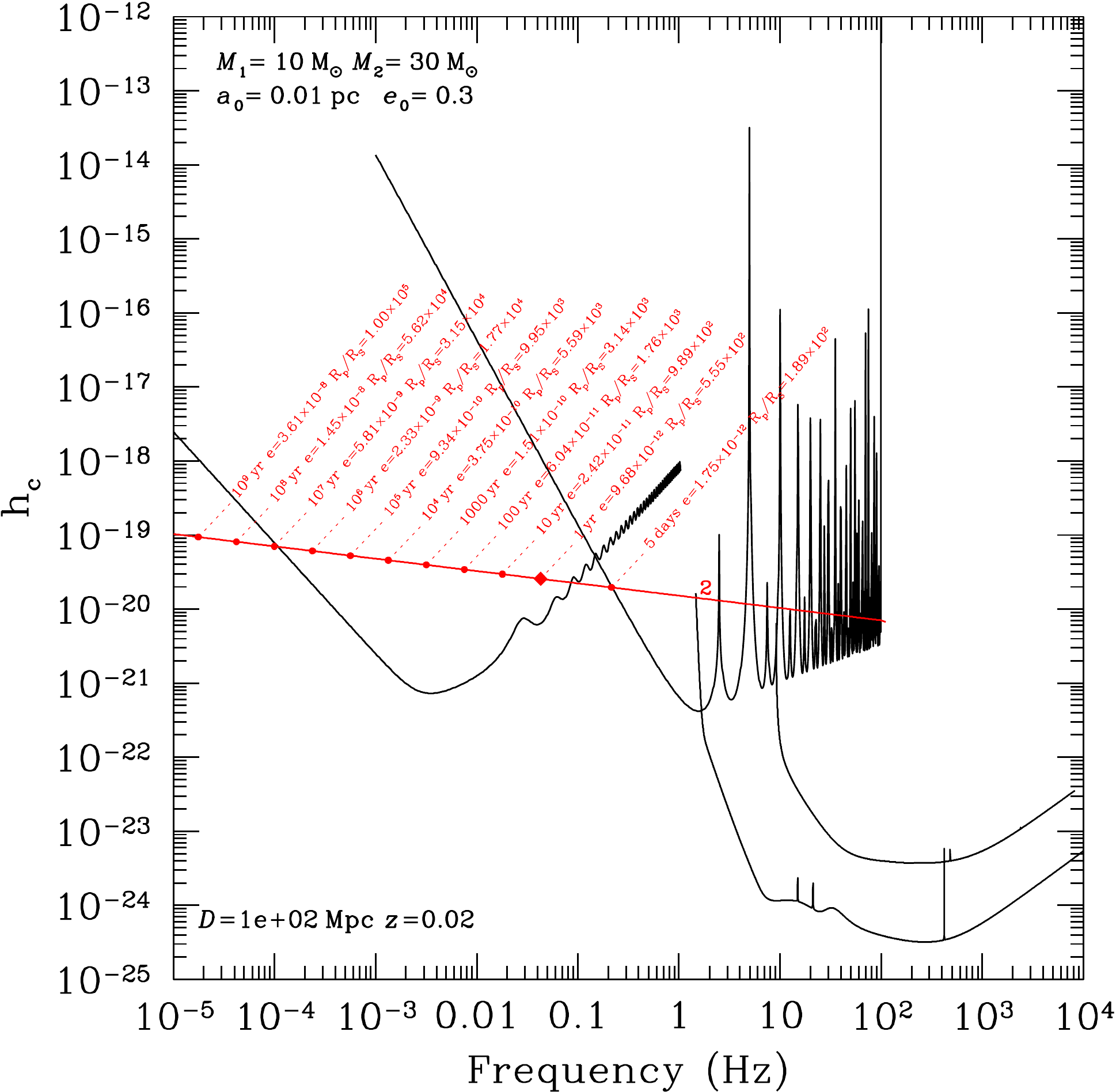}
\caption{
Same as Fig.~\ref{fig.BBHs_ELGAR} but for a BHB of masses $10\,M_{\odot}$ and
$30\,M_{\odot}$. The system enters the ELGAR sensitivity curve five days before merger,
and starting at about $0.3\,\textrm{Hz}$, crosses most of it in frequency.
\label{fig.10_30Msun_a1e-2_e3e-1}} 
\end{figure}

Nevertheless BHBs formed dynamically stand a big chance of eluding the LISA
band, as shown in~\cite{ChenAmaro-Seoane2017}.  Indeed, in their work the
authors show that only a very small subset of the detections would lead to a
joint detection with LISA and LIGO/Virgo. This means that, even if we
successfully detect BHBs with LISA and LIGO/Virgo, we still will not be able to
disentagle the nature of their formation.  This is a consequence of LISA not
being able to detect very eccentric BHBs, as shown in the work of~\cite{ChenAmaro-Seoane2017},
in particular their Fig. 1.
In~Fig.~\ref{fig.10_30Msun_a1e-2_e3e-1} we show a BHBs with an initial low eccentricity.
A decihertz detector such as ELGAR would contribute to recover
the information about the formation process of the binary.

\subsection{Stochastic Backgrounds of Gravitational Waves}

The incoherent sum of numerous unresolved gravitational wave signals results in a stochastic background of GWs~\cite{Christensen:2018iqi,Caprini:2018mtu}. We can distinguish two types of stochastic GW backgrounds: (i) Originated by high-energy processes in the early universe: from the inflationary epoch to different phase transitions; spacetime defects of cosmological scales such as cosmic strings, etc.  The energy scale of the processes associated to these physical processes is around $\sim 10^{3}\,$TeV (in contrast with LIGO/Virgo, around $\sim 10\,$TeV, and LIGO/Virgo, around $\sim 10^{5}\,$TeV). (ii) Backgrounds of astrophysical origin, produced by supernovae, magnetars, or the inspiral and merger of compact objects (NSs or BHs). Depending on the event rate of these astrophysical GW sources there may create a GW background overlapping the sensitivity band of ELGAR.

\subsection{Observing type Ia supernovae}
\begin{figure}
\centering
\includegraphics[width=0.8\textwidth]{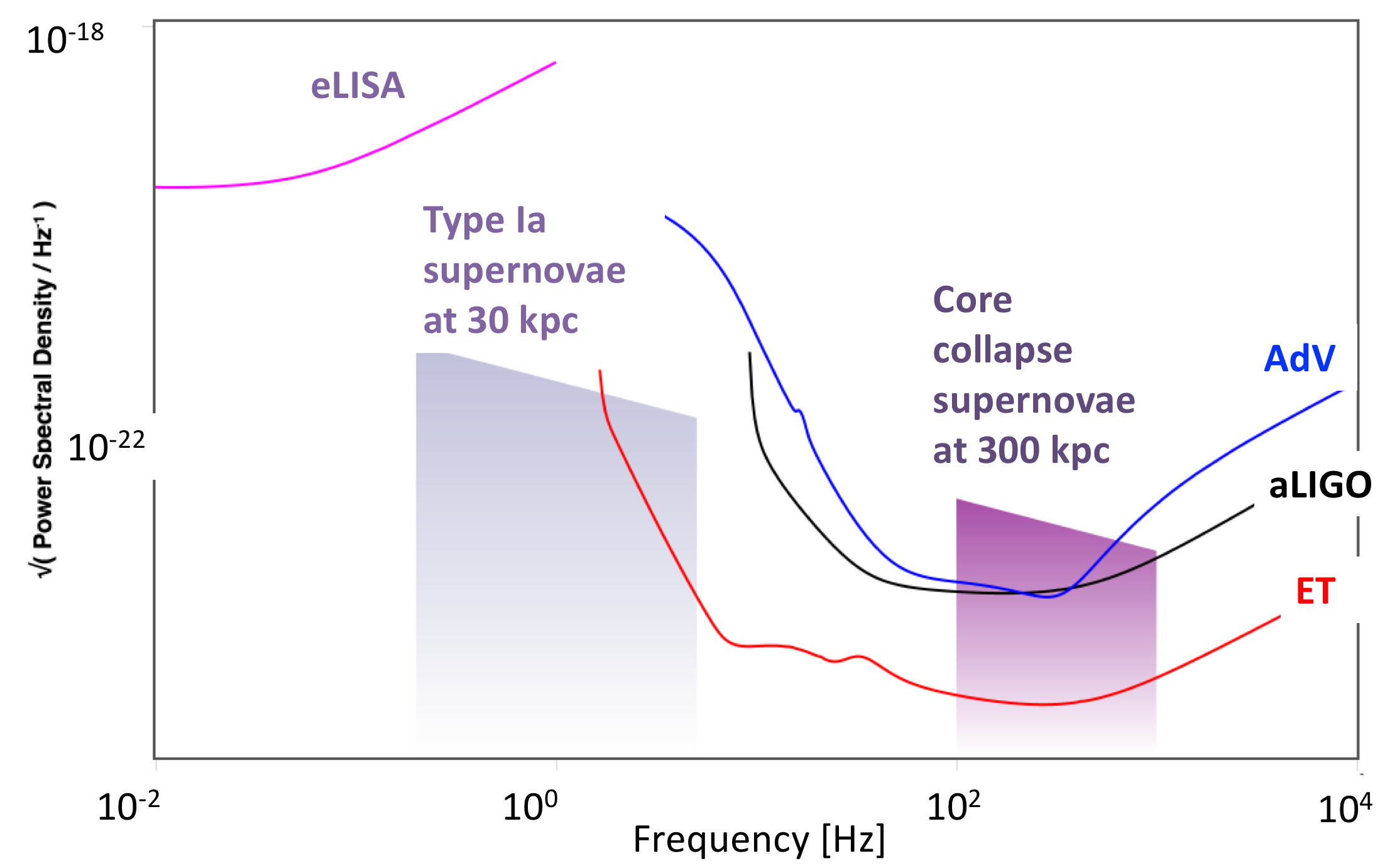}
\caption{\label{fig:supernovae} Type Ia and type II supernovae signal spectra, plotted against detector sensitivities. Figure obtained using the GWplotter tool~\cite{Moore2015CQG}.}
\end{figure}

Another source of interest at low frequencies is represented by the type Ia supernova events, expected to emit GWs and neutrinos~\cite{SnIa:2015} at frequencies in the gap between terrestrial large optical interferometers and the space detector LISA, as shown in Fig.~\ref{fig:supernovae}, where we plot the range of signal strengths for galactic sources. 

Again, these are signals potentially accessible for a detector with sensitivities close to $10^{-21} {\rm Hz}^{-1/2}$, exactly as type II (core-collapse) galactic supernovae are potentially accessible to LIGO and Virgo. Type Ia supernovae are about as frequent as core-collapse ones, with a frequency in our galaxy about one per hundred years; therefore the chances for detection are relatively modest, since the odds of detecting events in other galaxies are very unfavourable. Yet, like is the case for core-collapse supernovae, even the detection of a single event could provide a wealth of information on the detonation mechanism~\cite{SnIa:2015}.

%

\subsection{Multimessenger and Multiband GW Astronomy with ELGAR}

Following the detections of the LIGO and Virgo detectors, we expect that several astrophysical sources of GW transient signals will also emit electromagnetic (and possibly neutrino) counterpart signals~\cite{Sathyaprakash:2019rom,Fan:2015bia} that may detected by GW detectors, in particular by ELGAR. These counterparts (including cosmic rays) will be crucial to maximize the discovery potential of GW detectors, providing a complementary knowledge of astrophysical GW events and the emission mechanisms.  The main candidate for electromagnetic counterparts in the case of ELGAR will be binary systems containing a NS which may give rise to the emission of gamma-ray bursts (GRBs). In the case of ELGAR, BNSs will not merge inside the band, so we are looking at IMRIs containing a NS.

On the other hand, there are sources of GWs that can be observed in two different bands of the GW spectrum by two different observatories at different times. This has already been suggested between the high-frequency and low-frequency band~\cite{Sesana2016}, in the sense that the coalescence of binary black holes may be seen before by an observatory like LISA in the frequency band, specially if the masses of these Black Holes is at the high-end of what LIGO has been observing, as in the first detection of gravitational waves~\cite{Abbott2016}.  With LISA observations we can determine the time of coalescence and inform ground-based detectors operating at higher frequencies

Given that ELGAR is between the low- and high-frequency bands, we have two different possibilities of multiband detections where ELGAR could take part. ELGAR could detect binaries previously detected at the low-frequency band and also binaries that later will be detected by high-frequency detectors (either second-generation ones like LIGO~\cite{Aasi2015}, Virgo~\cite{Acernese2014}, and KAGRA or third-generation ones like ET~\cite{Punturo2010} or Cosmic Explorer (CE)~\cite{Reitze:2019iox}).

\subsection{Fundamental Physics with ELGAR} 

Gravitational waves offer many different ways of doing Fundamental Physics~\cite{Barack:2018yly,Sathyaprakash:2019yqt}.  We can distinguish different types of questions that we can address with gravitational waves depending on the system we are looking at and the assumptions we make about the systems itself.  For instance, if we distinguish between the generation and propagation of the gravitational waves we can see that different types of physics can influence these physical processes and hence they have a very different impact in the waveforms of the radiation emitted.  If we look at the propagation of gravitational waves coming from distance sources we can test different types of what we can call fundamental physics: Tests of possible violations of Lorentz invariance; test of the existence of extra polarizations of gravitational waves; to bounds on a hypothetical graviton mass (our best estimate, $m_{g} < 7.7 \times 10^{-23}$eV$/c^{2}$, comes from gravitational-wave observations~\cite{TheLIGOScientific:2016src,Abbott:2017vtc}); tests of parity violations, etc.  This type of new physics is usually associated with different theories that predict the presence of extra degrees of freedom, either coming from low-energy limit of quantum gravity theories or from phenomenological models inspired by quantum gravity. 

On the other hand, by looking at the late inspiral, merger and ringdown of compact binaries including Black Holes we can perform tests of the nature of compact objects (in particular of the no-hair conjecture for black holes) and also to perform tests of the most non-linear and dynamical (radiative) regime of General Relativity, not accessible to other types of observatories. We can also investigate whether certain modifications of General Relativity are correct and/or put bounds on them. The late inspiral and merger requires the use of Numerical Relativity or precise perturbative methods (like the effective-one-body approach that incorporates information from post-Newtonian theory, BH perturbation theory, and Numerical Relativity) and allows for tests of the nature of the compact objects and the theory of gravity. If these binary systems involve neutron stars we can also try to constraint the possible equation of state, which is currently not known, and which can tell us about properties of matter and high-energy physics interactions. In the case of black hole binaries we can also put bounds on exotic horizonless objects proposed as alternatives to BHs (see, e.g.~\cite{B2-48}), as for instance boson stars, gravastars, fuzzballs, etc.  Finally, the final ringdown when the deformed final BH from the coalescence relaxes to its final equilibrium state (a Kerr BH assuming General Relativity and the validity of the no-hair conjecture) via damped oscillations, the quasi-normal modes, can allow us to perform fine tests of the no hair conjecture.  The idea is simple, the quasi-normal modes are determined by the characteristic frequency of the oscillations and the characteristic damping time.  These in turn have been precisely computed in General Relativity (and in other alternative theories) for BHs and some alternatives to BHs.  Then, we can make as many tests as different quasinormal mode frequencies and damping times are we able to estimate from the gravitational wave signal.

Another tests of the nature of compact objects and gravity can come from the detections of intermediate-mass ratio inspirals with ELGAR.  This goes in the line of what has already been proposed for EMRIs and space-based detectors like LISA (see, e.g.~\cite{Babak:2017tow,Berry:2019wgg,Sopuerta:2010zy}).  There, it was estimated that during the last year before plunge an EMRI can spend of the order of $\sim q$ cycles ($10^{5}$ in the case of a $10\,M_{\odot}$ inspiraling into a $10^{6}\,M_{\odot}$ MBH).  It means that the GWs emitted carry a detailed map of the geometry of the MBH that allows for highly accurate estimations of the main EMRI parameters and even estimations the quadrupole and high-order multipoles~\cite{Babak:2017tow,Berry:2019wgg,Sopuerta:2010zy}.  For the Kerr metric, there are two sets of gravitational multipole moments, the mass and current (spin) moments, $\{{\cal M}_{\ell}\}_{\ell=0,\ldots,\infty}$ and $\{{\cal J}_{\ell}\}_{\ell=0,\ldots,\infty}$ respectively, that satisfy the following simple sets of relationships: ${\cal M}_{\ell} + i {\cal J}_{\ell} = M_{\rm BH} (i J_{\rm BH})^{\ell}$, where $M_{\rm BH}$ and $J_{\rm BH}$ are the mass and spin angular momentum of the MBH.  Therefore, if we are able to measure $N$ multipole moments from the gravitational wave signal we can perform $N-2$ tests of the geometry of the Kerr black holes.  We can extend this program to the case of IMRIs, where the number of cycles in the strong field regime should be of the order of $\sim 10^{2}-10^{4}$.  In this strong field regime, and in contrast with EMRIs, both the spin-orbit and spin-spin interactions are expected to be important, adding new time scales to the problem and increasing its complexity, both in terms of the structure of gravitational waveforms emitted but also in terms of the modeling required.  The EMRI modeling program is today an ongoing program based on the self-force approach that uses BH perturbation theory developments up to second order in the mass ratio.  In the case of IMRIs, the smaller mass ratios indicate that perturbation theory alone will not be sufficient and a combination of pertubative methods with numerical relativity will be needed.

Apart from tests on the nature of the most compact objects and the theory of gravity, ELGAR can also contribute to another important problem in Fundamental Physics: The nature of Dark Matter (DM).  There have been proposed different ways in which gravitational waves can provide information about DM~\cite{Bertone:2019irm}. We are going to mention here only the ones that are relevant for ELGAR. 
(i) {\em Primordial black holes} (PBHs).  They are many high-energy models with implications to early-Universe Cosmology predict the formation of BHs less than one second after the Big Bang in the mass range $10^{-16} - 10^{6} M_{\odot}$. PBHs have been proposed by different authors as a candidate for DM, some assuming assuming a single mass scenario and others with a certain mass distribution. 
Using different types of observations, a number of constraints on the fraction of DM that can be made out of PBHs (see, e.g.~\cite{Bertone:2019irm}) have been derived.  On the other hand, there are many different formation scenarios that may be distinguished by inferring the PBH mass spectrum.  Each formation scenario may lead to different gravitational-wave signatures. In this sense, it is important to mention that there are two cases of gravitational wave detections in which we can be sure that they are not due to BHs of astrophysical origin: The detection of nearby mergers of sub-solar mass BHs with ground-based detectors and binary black hole mergers at very high redshift ($z>40$) by future third-generation ground-based detectors like ET and CE (and similarly with space-based detectors like LISA).  Even in the case where PBHs do not constitute all of the DM, they can still be connected to DM by forming different types of particle DM mini-halos, whose annihilation may be enhanced giving rise, for instance, to gamma-ray emission.    
(ii) {\em Environmental effects}. There is the possibility that in the strong-gravity phase of the merger of black holes and/or neutron stars DM may play a role introducing small effects in the emitted gravitational waveforms. The question here is whether we can distinguish the DM environmental effect from other possible environmental effects~\cite{Barausse:2014tra}.
(iii) {\em Exotic Binary mergers}.  If DM is made out of some kind of particles, it is possible to conceive compact horizonless objects made out of them.  The coalescence of these objects may give rise to gravitational waveforms different from those corresponding to Black Holes and Neutron Stars.  This hypothetical scenario is clearly linked to the program of testing the nature of the most compact objects in the cosmos.
(iv) {\em  Non-perturbative DM dynamics}. DM particles may be produced non-perturbatively in the early universe giving rise to  characteristic stochastic gravitational wave backgrounds. The question in this case is whether these DM stochastic backgrounds can be distinguished from other stochastic backgrounds, something depends strongly on the shapes of the spectra.
(v) {\em Phase transitions}. They are one of the possible causes of stochastic backgrounds of gravitational waves of primordial original.  Given that the Standard Model of Particle Physics does not allow for first-order phase transition, detecting a stochastic GW background of this kind means new physics.  If these transition is associated with DM particles then we can infer some of the properties from the observations of the GW background.

\externaldocument{ELGAR-GW-Sources-and-Science-v6}

\section{Conclusions}
In this paper, we have proposed a European research infrastructure for the observation of gravitational waves in the deciHertz band, between $0.1-10\,$Hz.  
ELGAR will be based on the latest developments of quantum physics and will use a geometry based on an 2D array of atom interferometers. 
These techniques offer important perspectives to limit the impact of seismic noise and Newtonian Noise, which are expected to be major limitations for GW detectors at low frequency. 
Based on the preliminary design presented in this paper, an ELGAR detector tens of km long could achieve a sensitivity of $4.1 \times 10^{-22}/\sqrt{\rm{Hz}}$ at 1.7~Hz, assuming key developments in cold atom technologies mainly related to source parameters and atomic manipulation, but also improvements in Newtonian Noise reduction techniques.
The work presented here could be the base of a design study that will refine the requirements of the different systems of ELGAR, thanks to the development of a complete model of the antenna, from its metrology to its observable astrophysical sources. 
This work should also include a trade-off in terms of cost and performances for the realization of ELGAR on different candidate sites, such as the LSBB in France or the Sos Enattos and Seruci/Nuraxi sites in Italy. 
Together with the design study that will form the road map of ELGAR and identify the key experimental developments required, the realization of the antenna could be sustained from a growing number of national initiatives in Europe for the study of large scale atom interferometry, such as MIGA in France, VLBAI in Germany, or AION in United Kingdom. 
\par Opening GW detection in the in the deciHertz band, ELGAR will then fill the gap in the gravitational wave spectrum between ground-based detectors like Virgo and LIGO (high-frequency band) and space-based detectors like LISA (low-frequency band), where an ambitious science program can be carried out.
ELGAR will access a diverse variety of compact binary coalescence, both from their nature (white dwarfs, NSs, and BHs), their masses (from comparable mass to intermediate mass-ratio binaries), the orbital stage (inspiral, merger, ringdown), and even their presumable origin (stellar origin, globular clusters, early universe).
ELGAR will also be able to detect supernovae core collapse and stochastic backgrounds of gravitational waves, both from astrophysical and early-universe origin. 
ELGAR should thus be able to answer relevant questions in astrophysics and cosmology, concerning for example the existence of intermediate-mass BHs and their possible formation channels, and the existence of primordial BHs and the allowed mass ranges.
The possibility of detecting with ELGAR the merger and plunge of intermediate-mass ratio inspirals containing neutron stars would make the instrument a key resource for multi-band GW observations 
 and for multimessenger observations.
ELGAR will also have an important impact in fundamental physics, particularly for the detection of mergers of intermediate-mass BHs and/or intermediate-mass ratio inspirals with significant SNR. The precise parameter estimation for these sources will enable tests of the no-hair conjecture of BHs and even tests of General Relativity and alternative theories.
Additionally, ELGAR may also provide information about the nature of dark matter.
\section*{References}
\bibliographystyle{iopart-num}
\bibliography{ELGAR}
\end{document}